\documentclass[twocolumn,reprint,pra,aps,superscriptaddress,nofootinbib,9pt]{revtex4-1} 
\usepackage[title]{appendix}	% for appendix
\usepackage{etoolbox}	% to add colon in appendix numbering

\usepackage{epsf,graphicx}
\usepackage{amsmath,amssymb,amsfonts}
\usepackage{amstext} % for \text macro
\usepackage{multirow}
\usepackage{units}
\usepackage{enumitem}

\usepackage{booktabs}	% use \addlinespace[?pt] for rule-space
\usepackage[hypcap=true]{caption}
\usepackage{caption}
\usepackage{subfig}
\usepackage{hhline}
\usepackage{textcomp}
\usepackage{bm} % fonts 
\usepackage{float} % to stop tables floating here and there
\usepackage{rotating} % for sidewaytable
\usepackage{amsthm}
\usepackage{esvect} % for vectors: use vv{...}
\usepackage{gensymb} % to use \degree 
\usepackage{comment}		% to comment out text

\usepackage{arydshln} %--- \hdashline and \cdashline == dashed counterparts of \hline and \cline.

\usepackage{xparse}
\ExplSyntaxOn
\NewDocumentCommand{\mref}{m}{\quinn_mref:n {#1}}
\seq_new:N \l_quinn_mref_seq
\cs_new:Npn \quinn_mref:n #1
{
	\seq_set_split:Nnn \l_quinn_mref_seq { , } { #1 }
	\seq_pop_right:NN \l_quinn_mref_seq \l_tmpa_tl
	( % print the left parenthesis
	\seq_map_inline:Nn \l_quinn_mref_seq
	{ \ref{##1},\nobreakspace } % print the first references
	\exp_args:NV \ref \l_tmpa_tl % print the last or only one
	) % print the right parenthesis
}
\ExplSyntaxOff
\usepackage[
bookmarks=true,
colorlinks=true,
linkcolor=blue,
urlcolor=blue,
citecolor=blue,
pdftex,
bookmarks=true,
linktocpage=true, % makes the page number as hyperlink in table of content
hyperindex=true
]{hyperref}

\usepackage{bbm,bbold} %-- for \mathbbm{1}
\usepackage[mathscr]{euscript}
\DeclareSymbolFont{rsfs}{U}{rsfs}{m}{n}
\DeclareSymbolFontAlphabet{\mathscrsfs}{rsfs}
\usepackage{color}
\usepackage[usenames,dvipsnames,svgnames,table,colorspace]{xcolor}
\patchcmd{\appendices}{\quad}{: }{}{}	% to add colon in appendix numbering

\makeatletter
\renewcommand*{\@seccntformat}[1]{\csname the#1\endcsname.\hspace{.5em}}
\makeatother

\newcolumntype{C}{>{$}c<{$}}
\AtBeginDocument{
	\heavyrulewidth=.08em
	\lightrulewidth=.05em
	\cmidrulewidth=.03em
	\belowrulesep=.65ex
	\belowbottomsep=0pt
	\aboverulesep=.4ex
	\abovetopsep=0pt
	\cmidrulesep=\doublerulesep
	\cmidrulekern=.5em
	\defaultaddspace=.5em
}
\newcolumntype{L}{>{$}l<{$}}
\AtBeginDocument{
	\heavyrulewidth=.08em
	\lightrulewidth=.05em
	\cmidrulewidth=.03em
	\belowrulesep=.65ex
	\belowbottomsep=0pt
	\aboverulesep=.4ex
	\abovetopsep=0pt
	\cmidrulesep=\doublerulesep
	\cmidrulekern=.5em
	\defaultaddspace=.5em
}
\newcolumntype{R}{>{$}r<{$}}
\AtBeginDocument{
	\heavyrulewidth=.08em
	\lightrulewidth=.05em
	\cmidrulewidth=.03em
	\belowrulesep=.65ex
	\belowbottomsep=0pt
	\aboverulesep=.4ex
	\abovetopsep=0pt
	\cmidrulesep=\doublerulesep
	\cmidrulekern=.5em
	\defaultaddspace=.5em
}
\newcolumntype{?}[1]{!{\vrule width #1}}	% vertical rule with custom width
\theoremstyle{plain}
\newtheorem{theorem}{Theorem}

\newtheorem{lemma}{Lemma}

\newcommand*{\Scale}[2][4]{\scalebox{#1}{$#2$}} %--- reduce font size in math eqn
\newcommand*{\D}{\mathscrsfs{D}} % for var \mathcal{D}{+}
\newcommand*{\Dp}{\mathscrsfs{D}} % for var \mathcal{D}{+}
\newcommand*{\Dps}{\mathscrsfs{D}^{+}} % for var \mathcal{D}{+}
\newcommand{\overbar}[1]{\mkern 1.5mu\overline{\mkern-2.2mu#1\mkern-2.2mu}\mkern 1.5mu}
\newcommand*{\Dm}{ \overbar{\mathscrsfs{D}} } % for var \mathcal{D}{-}
\newcommand*{\Dms}{\mathscrsfs{D}^{-}} % for var \mathcal{D}{+}
\newcommand*{\Dpm}{\mathscrsfs{D}^{\pm}} % for var \mathcal{D}{+}
\newcommand*{\U}{\mathcal{U}} % for unitary operator
\newcommand*{\M}{\mathbf{M}} % for unitary operator
\newcommand*{\comp}{\mathcal{C}} % for computational basis
\newcommand\id{\ensuremath{\mathbbm{1}}} %---- \usepackage{bbm,bbold}
\newcommand\Omat{\ensuremath{\mathbbm{O}}} %---- \usepackage{bbm,bbold}
\newcommand*{\hada}{\mathbb{H}} % Hadamard
\newcommand*{\Hl}{\mathcal{H}} % Hilbert space

\newcommand*{\zb}{\bar{0}}
\newcommand*{\ob}{\bar{1}}
\newcommand*{\etal}{\textit{et al}}
\newcommand{\bpsp}{\textit{bpsp}}
\mathchardef\mhyphen="2D

\newcommand{\tab}{~~~~~~}
\newcommand{\half}{\frac{1}{2}} 
\newcommand{\invsqrttwo}{\frac{1}{\sqrt{2}}}
\renewcommand{\l}{\lambda}

\renewcommand{\b}{\beta}

\newcommand{\El}{E_{\l}}

\newcommand{\tensorprod}{\otimes}
\newcommand{\bra}[1]{\langle #1 |}
\newcommand{\ket}[1]{| #1 \rangle}
\newcommand{\braket}[2]{\langle #1 | #2 \rangle}
\newcommand{\ketbra}[2]{| #1 \rangle\langle #2 |}
\newcommand{\braopket}[3]{\langle #1 | #2 | #3 \rangle}

\newcommand{\Xlambdax}[2]{B_{#2}\tensorprod\sqrt{\El}~ \ket{#1}} 

\newcommand*{\concave}[1]{\sqrt{{#1}\!\left(1 \!-\! {#1} \right)}}

\newcommand*{\philog}[1]{\left(1+{#1}\right) \ln \left(1+{#1}\right) + \left(1-{#1}\right) \ln \left(1-{#1}\right)}
\def\spvecA#1;{\if;#1;\else #1\cr \expandafter \spvecA \fi}
\newcommand*{\colvecEightNobrc}[8]{\begin{matrix} {#1}\medskip\\{#2}\medskip\\{#3}\medskip\\{#4}\medskip\\{#5}\medskip\\{#6}\medskip\\{#7}\medskip\\{#8} \end{matrix}}

\newcommand*{\POVM}{\mathbf{M}}
\newcommand*{\matd}{\mathbb{D}}
\newcommand*{\vint}{\mathbf{a}}

\newcommand*{\ifft}{{\em iff}}

\newcommand*{\NEW}{\text{NEW}}
\newcommand*{\OLD}{\text{OLD}}

\newcommand{\thm}{{\sf  Thm.}}

\newcommand*{\IS}{\textbf{IS}} %-- denotes Initial State of Eve
\newcommand*{\IV}{\textbf{IV}} %-- denotes Initial Vector of Eve
\newcommand*{\W}{\mathbf{\mathcal{W}}} % for unitary operator

\newcommand*{\Ovecrow}{\vv{\mathbf{0}}} % for zero row-vec

\newcommand{\varA}[1]{{\operatorname{#1}}}			% for hyphenation in mathmode
\newcommand{\Poincare}{Poincar\'e}
\definecolor{darkpastelgreen}{rgb}{0.01, 0.75, 0.24} % =================	green
\definecolor{myorange}{rgb}{1.0, 0.4, 0.0}
\definecolor{byzantine}{rgb}{0.74, 0.2, 0.64}  % =================	 strong violet
\definecolor{violet}{rgb}{0.74, 0.2, 0.64}  % =================	 strong violet
\definecolor{carminered}{rgb}{1.0, 0.0, 0.22} % =================	red
\definecolor{brightpink}{rgb}{1.0, 0.0, 0.5}
\definecolor{mybrown}{rgb}{0.8, 0.34, 0.0}  % =================
\definecolor{mikadoyellow}{rgb}{1.0, 0.77, 0.05}

\colorlet{mybrownred}{RubineRed!90!black}   % for a text not wrong, but need modification
\colorlet{myyellow}{blue!5!black!30!Yellow} % for a text may be removed

\newcommand*{\UDeltaHada}{
	\begin{bmatrix} 
		\id_2^{E_2} & \cdot 		& \cdot 		& \cdot \\
		\cdot 		& \cdot 		& \cdot 		& \sigma_x^{E_2} \\
		\cdot 		& \cdot 		& \sigma_x^{E_2} & \cdot \\
		\cdot 		& \id_2^{E_2} 	& \cdot 		& \cdot \\ 
	\end{bmatrix}
}

\newcommand*{\HAuv}{
	\begin{bmatrix}
		\Dps_{uv} & -\Dms_{uv} \\
		\Dms_{uv} & \Dps_{uv}
	\end{bmatrix}
}
\newcommand*{\sigxHAuv}{
	\begin{bmatrix}
		\Dms_{uv} & \Dps_{uv} \\
		\Dps_{uv} & -\Dms_{uv}
	\end{bmatrix}
}

\newcommand*{\csigx}{
	\begin{bmatrix}
		\id_2 & \cdot \\
		\cdot & \sigma_x
	\end{bmatrix}
}
\newcommand*{\hadaTid}{
	\begin{bmatrix}
		\id_2 & \id_2 \\
		\id_2 & -\id_2
	\end{bmatrix}
}
\newcommand*{\Tef}{
	\begin{bmatrix}
		\id_2 & \id_2 \\
		\sigma_x & -\sigma_x
	\end{bmatrix}
}

\newcommand*{\UzeroExpanded}{  %--- without superscript $^{E_2}$ 
	\renewcommand{\arraystretch}{1.2}
	\begin{pmatrix}
		\sqrt{F_{xy}}~\HAuv & \sqrt{D_{xy}}~\HAuv & \Omat_2 & \Omat_2 \\
		\Omat_2 & \Omat_2 & \sqrt{D_{xy}}~\sigxHAuv & -\sqrt{F_{xy}}~\sigxHAuv \medskip \\
		\Omat_2 & \Omat_2 & \sqrt{F_{xy}}~\sigxHAuv & \sqrt{D_{xy}}~\sigxHAuv \\
		\sqrt{D_{xy}}~\HAuv & -\sqrt{F_{xy}}~\HAuv & \Omat_2 & \Omat_2 \\
	\end{pmatrix}
	\renewcommand{\arraystretch}{1}
}

\newcommand*{\UBellFuchs}{
	\renewcommand{\arraystretch}{1.4}
	\begin{pmatrix}
		\Dps_{xy}(\Dps_{uv}~~ -\Dms_{uv}) & \Dms_{xy}(-\Dms_{uv}~~ \Dps_{uv}) & \Ovecrow_2 & \Ovecrow_2 \\
		\Ovecrow_2 & \Ovecrow_2 & -\Dms_{xy}(\Dps_{uv}~~ -\Dms_{uv}) & \Dps_{xy}(-\Dms_{uv}~~ \Dps_{uv}) \\
		\Ovecrow_2 & \Ovecrow_2 & -\Dms_{xy}(\Dms_{uv}~~ \Dps_{uv}) & \Dps_{xy}(\Dps_{uv}~~ \Dms_{uv}) \\
		\Dps_{xy}(\Dms_{uv}~~ \Dps_{uv}) & \Dms_{xy}(\Dps_{uv}~~ \Dms_{uv}) & \Ovecrow_2 & \Ovecrow_2 \\
		\Ovecrow_2 & \Ovecrow_2 & \Dps_{xy}(\Dms_{uv}~~ \Dps_{uv}) & \Dms_{xy}(\Dps_{uv}~~ \Dms_{uv}) \\
		-\Dms_{xy}(\Dms_{uv}~~ \Dps_{uv}) & \Dps_{xy}(\Dps_{uv}~~ \Dms_{uv}) & \Ovecrow_2 & \Ovecrow_2 \\
		-\Dms_{xy}(\Dps_{uv}~~ -\Dms_{uv}) & \Dps_{xy}(-\Dms_{uv}~~ \Dps_{uv}) & \Ovecrow_2 & \Ovecrow_2 \\ 
		\Ovecrow_2 & \Ovecrow_2 & \Dps_{xy}(\Dps_{uv}~~ -\Dms_{uv}) & \Dms_{xy}(-\Dms_{uv}~~ \Dps_{uv}) 
	\end{pmatrix}
	\renewcommand{\arraystretch}{1}
}

\newcommand*{\Xcomp}
{
	\colvecEightNobrc
	{\sqrt{F_{xy}}~\Dps_{uv}}
	{\sqrt{F_{xy}}~\Dms_{uv}}
	{0}			%{\sqrt{1-D_{xy}}~\Ovec}
	{0}			%{\sqrt{1-D_{xy}}~\Ovec}
	{0}			%{\sqrt{D_{xy}}~\Ovec}
	{0}			%{\sqrt{D_{xy}}~\Ovec}
	{\sqrt{D_{xy}}~\Dps_{uv}}
	{\sqrt{D_{xy}}~\Dms_{uv}}
}
\newcommand*{\Ycomp}
{ 
	\colvecEightNobrc
	{0}			%{\sqrt{D_{xy}}~\Ovec}
	{0}			%{\sqrt{D_{xy}}~\Ovec}
	{\sqrt{D_{xy}}~\Dms_{uv}}
	{\sqrt{D_{xy}}~\Dps_{uv}}
	{\sqrt{F_{xy}}~\Dms_{uv}}
	{\sqrt{F_{xy}}~\Dps_{uv}}
	{0}			%{\sqrt{1-D_{xy}}~\Ovec}
	{0}		%{\sqrt{1-D_{xy}}~\Ovec}. 
}

\newcommand*{\XFuchs}
{ 
	\colvecEightNobrc
	{\sqrt{F_{xy}}~\Dps_{uv}}
	{0}			%{\sqrt{1-D_{xy}}~\Ovec}
	{0}			%{\sqrt{1-D_{xy}}~\Ovec}
	{\sqrt{F_{xy}}~\Dms_{uv}}
	{0}			%{\sqrt{D_{xy}}~\Ovec}
	{\sqrt{D_{xy}}~\Dms_{uv}}
	{\sqrt{D_{xy}}~\Dps_{uv}}
	{0}			%{\sqrt{D_{xy}}~\Ovec}	
}
\newcommand*{\YFuchs}
{
	\colvecEightNobrc
	{0}			%{\sqrt{D_{xy}}~\Ovec}
	{\sqrt{D_{xy}}~\Dps_{uv}}
	{\sqrt{D_{xy}}~\Dms_{uv}}
	{0}			%{\sqrt{D_{xy}}~\Ovec}
	{\sqrt{F_{xy}}~\Dms_{uv}}
	{0}			%{\sqrt{1-D_{xy}}~\Ovec}
	{0}			%{\sqrt{1-D_{xy}}~\Ovec}
	{\sqrt{F_{xy}}~\Dps_{uv}}	
}

	\usepackage{tikz}
	\usetikzlibrary{decorations.pathmorphing,patterns}
	\usetikzlibrary{matrix} % for block alignment
	\usetikzlibrary{arrows} % for arrow heads
	\usetikzlibrary{calc} % for manimulation of coordinates
	\usepackage{pgfplots}
	\usepackage{tikz-3dplot}
	\usepackage{contour}  % to use contour
	\usepackage{circuitikz} % for quantum circuits	
	
	\usetikzlibrary{intersections}  % intersection point of curves
	
\tikzstyle{startstop} = [rectangle, rounded corners, thick, minimum width=1cm, minimum height=.5cm,text centered, draw=black, fill=gray!6]
\tikzstyle{startstopdash} = [rectangle, rounded corners, dashed, thick, minimum width=1cm, minimum height=.5cm,text centered, draw=black, fill=gray!20]
\tikzstyle{startstop2} = [rectangle, rounded corners, minimum width=1cm, minimum height=.5cm,text centered, text width = 2 cm, draw=black, fill=yellow!30] %<text width fixed>
\tikzstyle{startstopflex} = [rectangle, rounded corners, minimum width=1cm, minimum height=.5cm,text centered, draw=black, fill=cyan!30] %<text width varied>
\tikzstyle{temp} = [rectangle, minimum width=1cm, minimum height=.5cm,text centered, draw=black, fill=white!30] %<text width fixed>
\tikzstyle{iff} = [implies-implies,double equal sign distance]
\tikzstyle{arrow} = [thick,->,>=stealth]
\tikzstyle{arrowDashed} = [dashed,->,>=stealth]
\tikzstyle{line} = [thick]
\tikzstyle{lineDashed} = [dashed,thin]
\tikzstyle{branch} = [circle,inner sep=0pt,minimum size=1mm,fill=black,draw=black]
\tikzstyle{block} = [draw,rectangle,thin,minimum height=2em,minimum width=2em]
\tikzstyle{blockdash} = [draw,rectangle,dashed,thin,minimum height=2em,minimum width=2em]
\tikzstyle{tinyblock} = [draw,rectangle,thin,minimum height=2em,minimum width=2em]
\tikzstyle{sum} = [draw, fill=blue!20, circle, node distance=1cm]

\pgfplotsset{%
	every tick label/.append style = {font=\tiny},
	every axis label/.append style = {font=\scriptsize}
}

\usepackage{tikz}
\usepackage{standalone}

\begin{document}
	\title{A complete characterization of the optimal unitary attacks in quantum cryptography \\
		with a refined optimality criteria involving the attacker's Hilbert space only
		}
	%\date{}
	\author{Atanu Acharyya}
	\email{pub.academy.15@gmail.com}
	\affiliation{Applied Statistics Unit, Indian Statistical Institute, Kolkata 700 108, India}
	\author{Goutam Paul}
	\email{goutam.paul@isical.ac.in}
%	%\affiliation{Cryptology and Security Research Unit, Indian Statistical Institute, Kolkata 700 108, India}
	\affiliation{Cryptology and Security Research Unit, R. C. Bose Centre for Cryptology and  Security, Indian Statistical Institute, Kolkata 700 108, India}

%A. Acharyya, and G. Paul,
%%Revisiting optimal eavesdropping in quantum cryptography: Optimal interaction is unique up to rotation of the underlying basis
%\hyperref{https://journals.aps.org/pra/abstract/10.1103/PhysRevA.95.022326}{category}{name}{Phys. Rev. A \textbf{95}, 022326 (2017)}.
	
\begin{abstract}
Fuchs \etal. 
[\hyperref{http://journals.aps.org/pra/abstract/10.1103/PhysRevA.56.1163}{category}{name}{Phys. Rev. A, 1997}] suggested an optimal attack on the BB84 protocol, where the necessary and sufficient condition for optimality involves the joint Hilbert space of the sender and the attacker. 
%via indirect learning that can tolerate up to 14.64\% channel error. 
In this work, we propose a refined optimality criteria involving the Hilbert space of the attacker only. It reveals that the optimal (non-zero) overlaps between the attackers post-interactions states must be equal and numerically same as the difference between the fidelity and the disturbance at the receiving end. That amount turns out to be same as the reduction (factor) in Bell violation when estimated for the equivalent entanglement-based protocol. Further, a series of necessary and sufficient conditions unveil the structure of the optimal states which therefore are the only and all possible optimal interactions. %They are shown to be isomorphic to the ones found  by Acharyya  \etal.[\hyperref{http://journals.aps.org/pra/abstract/10.1103/PhysRevA.95.022326}{category}{name}{Phys. Rev. A, 2017}]. 
We show that these optimal states are same as the outputs of an optimal phase-covariant cloner.
We also demonstrate various methods to derive optimal unitary evolutions that an eavesdropper is interested to know in order to mount an optimal attack.
%	An optimal attack on the BB84 protocol via indirect learning was prescribed by Fuchs \etal. 
%	[\hyperref{http://journals.aps.org/pra/abstract/10.1103/PhysRevA.56.1163}{category}{name}{Phys. Rev. A, 1997}] which exhibited an optimal interaction that attains the maximum possible knowledge-gain by an attacker. A necessary and sufficient condition on the joint Hilbert space of the sender and the attacker was suggested there for an attack to be optimal. Here we propose a more efficient condition restricted on the Hilbert space of the attacker only. It exhibits an intriguing connection with an optimal Bell violation alongside the optimal shrinking of Bloch vectors. Further, we move through a series of necessary and sufficient conditions to derive infinitely many optimal interactions which therefore are the only and all possible optimal interactions. Those are shown to be same (one-to-one) as the ones derived by Acharyya  \etal.[\hyperref{http://journals.aps.org/pra/abstract/10.1103/PhysRevA.95.022326}{category}{name}{Phys. Rev. A, 2017}]. We also demonstrate various methods to derive optimal unitary evolutions that an eavesdropper is interested to know in order to mount an optimal attack.
\end{abstract}
	\maketitle

\section{Introduction}
\label{intro}
The BB84 protocol~\cite{bb84} can establish an information-theoretically secure secret key between two distant parties. Alice encodes a stream of classical bits (\textit{cbits}) into an ensemble of quantum bits (\textit{qubits}) using two \textit{mutually unbiased bases} (MUBs). 
%One of them is the \textit{computational basis} $\{\ket{x}, \ket{y}\}$, aka $xy$ basis, and the other one is the \textit{Hadamard basis} $\{\ket{u}, \ket{v}\}$, aka $uv$ basis. She encodes the cbit $0$ into a qubit with the state $\ket{x}$ or $\ket{u}$, and encodes $1$ into $\ket{y}$ or $\ket{v}$. \clrR{Thus, Alice encodes a cbit $a$ by a basis $\b$ into a qubit in state $\ket{a^{\b}} = \hada\ket{a^0}$.}
She then transmits the qubits one-by-one over a quantum channel. 
Bob, at the receiving end, measures individually in one of the encoding bases, chosen randomly. Later they  reconcile bases publicly over an authenticated classical channel to filtrate a \textit{sifted key}.
 
A third party (Eve) is allowed to tamper the quantum channel. However, any approach to learn the state of the qubit introduces an error which is further detectable by the recipient. The legitimate parties can estimate the \textit{quantum bit error rate} (QBER) by discussing over the public channel on a part of the sifted key. Within a threshold value QBER$^\star$, a classical post-processing (CPP) is faithful to filter a shared secret on which Eve has virtually no information.
%The \textit{quantum bit error rate} (QBER) estimated by the legitimate parties cannot exceed a threshold value QBER$^\star$ to ensure feasibility of a classical post-processing (CPP) in order to filter a shared secret on which Eve has virtually no information.

An advanced eavesdropping model~\cite{fuchs97} is to extract the information of a transmitted qubit via an ancilla qubit by interacting unitarily. Given that the attacker is allowed to defer her measurement until after basis reconciliation, %In this scenario, it is known~\cite{fuchs97} that 
an one-way (OW) CPP is faithful if the estimated QBER remains below the critical value 0.1464 where the secret key-rate becomes zero. The authors could estimate 
the maximum \textit{knowledge gain} (KG) by an attacker that eventually appeared a tight bound due to an witness interaction. Nonetheless, there could be infinitely many such saturating candidates (interactions) which are unitarily equivalent~\cite{ach17}. % although they are unitarily equivalent.
In that attack model, a candidate interaction must pass a formal verification of optimality, viz., a \textit{necessary and sufficient condition}(NSC)~\cite{fuchs97} involving the joint Hilbert space of the sender and the attacker. 

\begin{em}
	We suggest here a necessary and sufficient condition for optimality that involves the Hilbert space of the attacker only. The verification is easier to perform than that in~\cite{fuchs97}. This new criteria explicitly depicts the geometry of the optimal states. We find its direct connection with the equivalent entanglement-based protocol and with optimal phase-covariant (pc) cloner~\cite{BCAM2K}. 
	%other modes of attack like cloning, entanglement-based protocols etc.
	%more counterintuitive, and has deeper connections as follows. 
\end{em}
% which is more counterintuitive
%{\em As an eavesdropper is confined in her own Hilbert space, we suggest here another necessary and sufficient condition for optimality which is more counterintuitive and easier to verify within the Hilbert space of the attacker.}

To be precise, an optimal attack is characterized by the non-zero overlaps between various post-interaction states of Eve's ancilla. The optimal overlap must equate the fidelity less the disturbance incurred at Bob's end. We show that the amount is same as the reduction (factor) in the CHSH sum~\cite{chsh69, cirelson80} for an equivalent entanglement-based scheme. Geometrically speaking, it amounts to the contraction in the Bloch vectors associated with Bob's states.

%The geometry of Eve's optimal states and that of the receivers states has an interesting connection: the non-zero overlap in the former case is same as the contraction in the Bloch vectors for the later. 
%It vindicates that  the overlap between the disturbed states is same as the overlap between the non-disturbed states of an optimal attacker. %interaction vectors 
%The optimal overlap, is in one hand same as the fraction of reduction in the Bloch vectors  
%at the receiving end, and on the other hand it equates the fraction of reduction in the CHSH sum~\cite{chsh69, cirelson80} for an entanglement-based \textit{E-91} scheme~\cite{e91}.

We carry on through a chain of NSCs to derive infinitely many optimal interactions, and therefore without ambiguity, these are the only and all possible optimal interactions. They are unitarily the same as the optimal states derived earlier by Acharyya \etal. in~\cite{ach17}. An optimal \textit{post-interaction joint state} (PIJS) clearly exhibits an one-to-one correspondence with the optimal measurement of Eve. Thus, Eves measurement setup determines her interaction and vice versa. Relation between Eve's optimal measurements for two MUBs are established. The optimal PIJSs are in sync with the outputs obtained by an optimal pc-cloner~\cite{BCAM2K}. 

%Further, we show how a measurement setup for an encoding basis determines the measurement setup in its conjugate basis.
%the way the optimal \textit{interaction vectors} (IVs) of Eve in the conjugate bases are interrelated, 
%the corresponding optimal POVMs in the conjugate bases are also interrelated that we find here.  

We then consider the task of characterizing the optimal unitary attacks, i.e., to derive the optimal unitary operators. First, we describe the basic approach to find an optimal unitary for a given optimal PIJSs. We discuss the limitations of that method when it comes to work for the optimal PIJSs described in an arbitrary measurement basis for Eve. To bypass this hurdle, we start with the PIJSs by fixing Eve's measurement basis as the computational basis, and develop a quick hack to get the simplest form of an optimal unitary that must accompany a specific \textit{initial state} (IS) of the ancilla. However, given a PIJS, there could be infinitely many optimal unitary interactions: we discuss methods to find any and all of them from the prior knowledge of an already derived optimal unitary. Once an optimal unitary is found for an IS, one can leverage this knowledge to find an optimal unitary for any other IS just by finding a transformation rule between the initial states. Finally, we figure out the transformation rule to get an optimal unitary in one measurement basis from an already derived optimal unitary in some other measurement basis. We demonstrate these methods in place for a few chosen states to understand the other intricacies. Essentially we have characterized the whole space of optimal unitary attacks.

First, we discuss in Sec.~\ref{sec:recap} the framework of optimal eavesdropping~\cite{fuchs97}, the generic optimal interactions~\cite{ach17}, and the interrelation between Eves optimal measurements 
%in each of the conjugate bases
across the two MUBs~\footnote{Although a new addition, these interrelations are better fitted in this recapitulation section only.}. Then we discuss briefly the results in Sec.~\ref{sec:iff}-\ref{sec:unitaryEvol} and defer their derivations and illustrations until in Sec.~\ref{sec:details}. We conclude by summarizing the new findings and also discuss further scopes to explore.

%\ref{sec:unitaryEvol}
%\ref{sec:connections}

\section{Elements of Optimal Eavesdropping} \label{sec:recap}
Here we brief the attack model, the objective functions to be optimized and their bounds, and the optimal states after an interaction. We exhibit some direct connections that a practical attack has with Bell violation and with an optimal pc-cloner.
 
\subsection{Alice's encoding}
%\begin{table*}[htbp!]
%	\begin{tabular}{ccc}
%		Basis-1 & z & $\{\ket{z}, \ket{-z}\}$ \\
%		Basis-2 & x & $\{\ket{x}, \ket{-x}\}$ 
%	\end{tabular}
%\end{table*}
For encoding, Alice uses two orthonormal bases conjugate to each other: the \textit{computational basis}, and the \textit{Hadamard basis}. %, indexed by 0 and 1, respectively. 
The basis states correspond to the eigenstates of the phase-flip operator $\sigma_z$ and bit-flip operator $\sigma_x$, respectively. The following notations for the bases and their states are used interchangeably throughout the paper. % depending on the context.
%\renewcommand{\arraystretch}{1.4}
%\begin{equation*}
%	\begin{array}{cc c cc}
%%		\multicolumn{2}{c}{\text{Basis-}1}
%%		&& \multicolumn{2}{c}{\text{Basis-}2}  
%%		\\ \hline \hline
%		\text{Basis}  & \text{States}
%		&& \text{Basis}  & \text{States}  
%		\\ \hline \hline
%		+ & \{\ket{0}, \ket{1}\} 
%		&& \times & \{\frac{\ket{0}+\ket{1}}{\sqrt{2}}, \frac{\ket{0}-\ket{1}}{\sqrt{2}}\} 
%		%		\\ \hline
%		%		0 & \{\ket{0}, \ket{1}\} 
%		%		&& \bar{0} & \{\ket{\bar{0}}, \ket{\bar{1}}\}
%		\\ \hline 
%		xy & \{\ket{x}, \ket{y}\} 
%		&& uv & \{\ket{u}, \ket{v}\}
%		\\ \hline
%		0 & \{\ket{0}^{0}, \ket{1}^{0}\} 
%		&& 1 & \{\ket{0}^{1}, \ket{1}^{1}\}    
%		\\ \hline
%		Z & \{\ket{z}, \ket{-z}\} 
%		&& X & \{\ket{x}, \ket{-x}\}    
%		\\ \hline \hline
%	\end{array}
%\end{equation*}

{ \setlength{\tabcolsep}{0.5em} % for column spacing
	\renewcommand{\arraystretch}{1.2}% for row spacing
\begin{table}[htbp!]
\begin{tabular}{C|CC?{0.4mm}CC|C}	
	\toprule
\multicolumn{2}{c}{Computational basis} &&& \multicolumn{2}{c}{Hadamard basis}
	\\ 		
	\hline	 %\cmidrule{1-2} \cmidrule{5-6}  %\midrule
\text{Basis}  & \text{States}
	&&& \text{Basis}  & \text{States}  
	\\ 
	\specialrule{0.8pt}{1pt}{1pt}	\midrule		% \hline \hline
+ & \{\ket{0}, \ket{1}\} 
	&&& \times & \{\frac{\ket{0}+\ket{1}}{\sqrt{2}}, \frac{\ket{0}-\ket{1}}{\sqrt{2}}\} 
		%		\\ \hline
		%		0 & \{\ket{0}, \ket{1}\} 
		%		&& \bar{0} & \{\ket{\bar{0}}, \ket{\bar{1}}\}
		\\ 
		\midrule[0.8pt] 
\multicolumn{6}{c}{Various labellings used}
	\\
	\midrule[0.8pt]
xy & \{\ket{x}, \ket{y}\} 
		&&& uv & \{\ket{u}, \ket{v}\}
		\\ 
		\hline
0 & \{\ket{0}^{0}, \ket{1}^{0}\} 
		&&& 1 & \{\ket{0}^{1}, \ket{1}^{1}\}    
		\\ 
		\hline
Z & \{\ket{+z}, \ket{-z}\} 
		&&& X & \{\ket{+x}, \ket{-x}\} 
	\\ \bottomrule
\end{tabular}
\end{table}
}

$\bar{\b}$ denotes the conjugate of a basis $\b$. 
%The conjugate of a basis $\b$ is denoted sometimes as $\bar{\b}$. 
%When a basis is $\b$, its conjugate basis is $\bar{\b}$.
The Hadamard transform $\hada := \invsqrttwo\left(\sigma_z+\sigma_x\right)$ flips the bases ($\hada : \b \mapsto \bar{\b}$) while the basis states can be written with respect to the computational basis elements as $\ket{a}^{\b}=\hada^{\b}\ket{a}$ for $a=0,1$.
%In an encoding basis, 
The orthogonal counterpart of a state $\ket{a}$ is denoted by $\ket{a\oplus 1}$ or $\ket{\bar{a}}$.
%The states in bases $Z$ and $X$ are the eigenstates of $\sigma_z$ and $\sigma_x$, respectively.
Alice encodes the cbit $0$ into a qubit in state $\ket{x}$ or $\ket{u}$, and encodes $1$ into $\ket{y}$ or $\ket{v}$.

%\section{A Circuit diagram}
\begin{figure*}[htbp]
	\begin{center}
		\caption{	\underline{A circuit diagram for an optimal eavesdropping on BB84 protocol.} \\ 
			Alice uses one of the two MUBs, $\beta$, to encode a \textit{cbit} `$a$' into a \textit{qubit} $\ket{a}^{\beta}$. Eve attaches an ancilla $\ket{e}$ and evolves the joint system unitarily ($\mathcal{U}_e$) that creates an entangled state $\ket{S_a}^{\beta}$. Bob measures the received qubit in basis $\beta^{\prime}$ to get the \textit{cbit} $b$, and keeps it if the bases are matched. After basis reconciliation, Eve measures her ancilla in the POVM basis $\{\ket{M_{\lambda}}^{\beta}\}$. She interprets her outcome $\lambda$ by a strategy and bet for $a_{\lambda}$ to guess Alice's \textit{cbit}. When Eves choices for the unitary and the measurement are optimal, she guesses the key best while not forcing to abort the protocol. 	
			%Two encoding bases for the sender: the computational basis $\{\ket{0}, \ket{1}\}$, and the Hadamard basis $\{\frac{\ket{0}+\ket{1}}{\sqrt{2}}, \frac{\ket{0}-\ket{1}}{\sqrt{2}}\}$. 
		}
		\label{fig::crckt}
	%%%%%%%%%%%%%%%%%%%%%%%%%%%%%%%%%%%%%%%%%%%%%%%%%%%%
%% Author: Atanu Acharyya
%% Senior Research Fellow
%% Applied Statistics Unit
%% Indian Statistical Institute
%% 203 B T Road, Kolkata 700 108, India
%% Email: pub.academy.15@gmail.com
%%%%%%%%%%%%%%%%%%%%%%%%%%%%%%%%%%%%%%%%%%%%%%%%%%%%%

%===============================
%	quantum MEATER
%===============================

%------- define METER -----------------
\tikzset{
	meter/.append style={
		draw, inner sep=10, rectangle, font=\vphantom{A}, minimum width=30, line width=.8,
		path picture={
			\draw[black] 
				( [shift={(.1,.3)}] path picture bounding box.south west) 
				to [bend left=50] 
				( [shift={(-.1,.3)}] path picture bounding box.south east);
			\draw[black,-latex] 
				([shift={(0,.1)}]path picture bounding box.south) 
				-- 
				([shift={(.3,-.1)}]path picture bounding box.north);
		}
	}
}

%\begin{tikzpicture}
%\node[meter] (meter) at (10,5) {};
%\end{tikzpicture}

\begin{tikzpicture}	
%\draw[thick,->,dashed] (-1,2) node[left=1pt]{\footnotesize $a$}--(0,2);
%\draw[thick,->,dashed] (-1,2.5) node[left=1pt]{\footnotesize $\beta$}--(0,2.5);
%\draw[thick,->,dashed] (0,2.5)--(0,2)--(0,1.3);
%\draw[thick,decorate,decoration={brace},xshift=-4pt,yshift=0pt]
%	(-1.4,2) -- (-1.4,2.5) node [black,midway,xshift=-1cm] {Alice encodes};	
%\draw (-1,5) to node[currarrow,sloped] {} (-1,4);  %==== midway arrow
\draw[dashed] (-1,3) node[above=1pt]{$\stackrel{cbit}{  a\in\{0,1\} }$ } to node[currarrow,sloped] {} (0,2.4);
\draw[dashed] (+1,3) node[above=1pt]{$\stackrel{basis}{\beta\in\{0,1\}}$} to node[currarrow,sloped,allow upside down] {} (0,2.4);
%\draw[thick,->,dashed] (-1,3) node[above=1pt]{\footnotesize $a\in\{0,1\}$}--(0,2.4);
%\draw[thick,->,dashed] (+1,3) node[above=1pt]{\footnotesize $\beta\in\{0,1\}$}--(0,2.4);
\node [] at (0,2.2){Alice encodes};
	\draw[thick,->,dashed] (0,2)--(0,1.3);
\node[] (A) at (0,1) {$\ket{a}^{\beta}$}; \node [left of=A,xshift=-0.4cm] {Alice sends};
\node[] (E) at (0,0) {$\ket{e}$}; \node [left of=E,xshift=-0.4cm] {Eves ancilla};

\draw[thick,->] (0.4,1)--(2,1);
\draw[thick,->] (0.4,0)--(2,0);
\draw[thick] (2,-0.5) rectangle (3,1.5) node [pos=.5]{\large $\mathcal{U}_e$};
\draw[line,dashed] (3.5,-0.5) -- (3.5,2) node[above=0.3]{$\ket{S_a}^{\beta}$};
\node[meter,scale=0.6] (meterB) at (4.3,1) {};
\node[above of=meterB, yshift=-0.5cm] (){~~~~$\beta^{\prime}\in\{0,1\}$};
\draw[thick,->] (3,1)--(meterB);
\draw[thick,->] (meterB)--(5.5,1) node[right=0.2](cbitBob){$b$};
\draw[thick,->] (5.9,1)--(7,1) node[right=0.2](){sifted key};
%\draw(cbitBob.10) -- ++(6,1); \draw(cbitBob.-10) -- ++(6,1);
\draw[line,dashed] (6.5,-0.5) -- (6.5,2) node[above=0.3]{Basis reconciliation};

\node[meter] (meterE) at (7.5,0) {};
\node[below of=meterE, yshift=0.2cm] (){$\{\ket{M_{\lambda}}\}^{\beta}$};
\draw[thick,->] (3,0)--(meterE);
\draw[thick,->] (meterE)--(8.5,0) node[right=0.2]{$\lambda$} (9,0) --node[midway,above]{\footnotesize Strategy}(10.5,0) node[right=0.2]{$a_{\lambda} = \begin{cases}
	 0, \text{~~if~~} \lambda=0,2 \\ 
	 1, \text{~~if~~} \lambda=1,3
	\end{cases}$};
%\draw[thick,decorate,decoration={brace,mirror},xshift=-4pt,yshift=0pt]
%	(9.5,0.5) -- (9.5,-0.5) node [black,midway,xshift=-1cm] {};
\end{tikzpicture} 
	\end{center}
\end{figure*}

\subsection{The attack model}
Eve attacks the quantum channel with an intention to indirectly learn the transmitted qubits one-by-one. She attaches a probe in state $\ket{e}\in\Hl_{E}$ to Alice's qubit that was transmitted in state $\ket{a^{\b}}\in\Hl_{A}$. She evolves the joint system unitarily ($\U$) from the pre-interaction joint state $\ket{a^{\b}}\ket{e}\in\Hl_{A}\tensorprod\Hl_{E}$ to the post-interaction joint state
%	Let, Alice prepares her qubit in a state $\ket{a} \in \{\ket{x}, \ket{y}, \ket{u}, \ket{v}\}$, drawn with \textit{prior probability} $p_a$, such that $p_x+p_y=1=p_u+p_v$. In order to learn the qubit-state indirectly, Eve attaches a probe having initial state $\ket{e}\in\Hl_{E}$, and evolve unitarily ($\U$) the joint system from the pre-interaction joint state $\ket{a}\ket{e}\in\Hl_{A}\tensorprod\Hl_{E}$ to the post-interaction joint state (PIJS) 	  
%	\begin{eqnarray} \label{eq:post-int-joint-state}
%		\ket{S} =  \U\left(\ket{s}\ket{\psi_0}\right).
%	\end{eqnarray} 
	\begin{eqnarray} \label{eq:post-int-joint-state}
		\ket{S_a^{\b}} =  \U~\ket{a^{\b}}\ket{e}.
	\end{eqnarray}
	%	For Alices' symbol $a^{\b} \in \{x,y,u,v\}$, denote the PIJS symbol $S_a^{\b}$ as $X,Y,U,V$ respectively.		
	The interaction entangles Eve's probe with Alice's qubit, and the joint state possess the following Schmidt decomposition:
	%. The PIJS is a superposition of two states: one retains Alice's qubit-state and the other one flips it. %\footnote{When $a=x$, $\bar{a}=y$ and vice versa. Similarly, in $uv$ basis.}. 
	%It has the following Schmidt decomposition:	
	\begin{eqnarray}  \label{eq:entangledExpr}
	\ket{S_a}^{\beta} &=& \sqrt{F_{\b}}~\ket{a}^{\b}\ket{\xi_a}^{\b} + \sqrt{D_{\b}}~\ket{\bar{a}}^{\b}\ket{\zeta_a}^{\b}. 
	\end{eqnarray}
%	As Bob measures, even for correct choice of the basis $\b$, 
	Bob receives the qubit intact or flipped, with frequency $F_{\b}$ or $D_{\b}$, respectively\footnote{We use $F_{\b}$ and $1-D_{\b}$ interchangeably.}.	Consequently, he finds the channel producing a QBER $D_{\b}$ for an encoding basis $\beta$. 	
	%That error rate (or disturbance) plays a significant role in quantifying the maximum KG by Eve.
	%  $D_{ss^{\prime}}$ in $ss^{\prime}$ basis.
	
%	In case the state $\ket{S}$ gets entangled post interaction, a convenient description of the state of Eve's probe can be considered the density operator $\rho_s := \tr_A\left(\ketbra{S}{S}\right)$, partially traced over Alice's qubit. %where $\tr_A$ represents partial trace over Alice's qubit.
	%	\begin{eqnarray} \label{eq:rho_s} 
	%	\rho_s &:=& \tr_A\left(\ketbra{S}{S}\right),
	%	\end{eqnarray}	
	%In the Smith decomposition, 
	The corresponding states of Eve's ancilla: the \textit{fidelity state} $\ket{\xi_a}^{\b}$ and the \textit{disturbed state} $\ket{\zeta_a}^{\b}$ are mutually orthogonal.	
	%The fidelity state has zero overlap with its disturbed counterpart 	
	% An optimal PIJS in $xy$ basis has the following Schmidt decomposition:
	%	\begin{eqnarray}  \label{eq:entangledExprXY}
	%	\ket{X} &=& \entangledExpr{xy}{x}{x}{y}{x}, \nonumber \\
	%	\ket{Y} &=& \entangledExpr{xy}{y}{y}{x}{y}. 
	%	\end{eqnarray}	
	In a fixed basis, when these four {\em interaction vectors} (IVs) share real-valued inner products, they can be grouped into two mutually orthogonal sets: %one with the two fidelity states 
	the fidelity set $ \{\ket{\xi_a},\ket{\xi_{\bar{a}}}\}$, and the disturbed set $ \{\ket{\zeta_a},\ket{\zeta_{\bar{a}}}\}$. 		
%	In an encoding basis, if the inner products between the IVs are assumed to be real, those four states can be grouped into two mutually orthogonal sets, and therefore could be distinguished by Eve. For instance, in the $xy$ encoding basis, the two such groupings of $IV_{xy}:= \{\ket{\xi_x}, \ket{\xi_y}, \ket{\zeta_x}, \ket{\zeta_y}\}$ are: the \emph{fidelity states} $\ket{\xi_x},\ket{\xi_y}$ and the \emph{disturbed states} $\ket{\zeta_x}, \ket{\zeta_y}$. 
	%	\begin{eqnarray*} \label{rel:perp_opt}
	%	\{\ket{\xi_x},\ket{\xi_y}\} \perp \{\ket{\zeta_x}, \ket{\zeta_y}\}. 
	%	%\nonumber \\ \{\ket{\xi_u},\ket{\xi_v}\} \perp \{\ket{\zeta_u}, \ket{\zeta_v}\}. 
	%	\end{eqnarray*} 
	Clearly, a two-qubit probe suffices to describe Eve's four-dimensional Hilbert space $\Hl_{E}=\Hl_{2}^{\tensorprod2}$.
	%	and similar restriction for $uv$ basis. It depicts that Eve's probe lives in a Hilbert space $\Hl_{E}$ of dimension at most four and therefore 2 qubits suffice. The computational basis $\{\ket{00}, \ket{01}, \ket{10}, \ket{11}\}$ can be used 	\footnote{$\Hl_d$ denotes $d$-dimensional Hilbert space. We sometimes use the notations $\ket{0}$, $\ket{1}$, $\ket{\bar{0}}$ and $\ket{\bar{1}}$ instead of $\ket{x}$, $\ket{y}$, $\ket{u}$ and $\ket{v}$ respectively. %Otherwise, we follow the notations of Fuchs \etal.~\cite{fuchs97} for easy understanding. 	} for mathematical description of the elements in the four-dimensional Hilbert space $\Hl_{E}=\Hl_{2}^{\tensorprod2}$. 
	
	%	The optimal PIJSs in $uv$ basis have the similar expressions:
	%	\begin{eqnarray} \label{eq:entangledExprUV}
	%	\ket{U} &=& \entangledExpr{uv}{u}{u}{v}{u}, \nonumber \\
	%	\ket{V} &=& \entangledExpr{uv}{v}{v}{u}{v}. 
	%	\end{eqnarray}
	%with similar orthogonality restriction on the IVs.	
To distinguish these four states, she needs to incorporate a generalized measurement with four outcomes.   
Thus, her measurement is considered to be a \textit{positive operator-valued measure} (POVM): a resolution of unity into non-negative Hermitian operators~\cite{fuchs96, qNC02}. 
Denote her POVMs $\{E_\l\}$ or $\{F_\l\}$ depending on whether Alice encodes in $xy$ or $uv$ basis. Denote them commonly as $\{M_\l\}^{\b}$, where an outcome is labeled by $\l\in\{0,1,2,3\}$. She then interprets the outcome following a {\em strategy} which is a rule for Eve to assign a guess for the state of the signal sent by Alice. 	
	%The strategy is chosen to be the one that maximizes her KG on Alice's symbol. 

\subsection{Functions to be optimized}
After the measurement by Eve and Bob, each of the three parties is left with a classical {\em random variable} (r.v.), denoted here as $A,B$, and $E$, for Alice, Bob, and Eve, respectively. For a permissible QBER, all the legitimate parties are concerned about, is the {\em secret key-rate} (SKR) which is the ratio of the length of the final secret key and the sifted key. No analytic expression is known for the SKR, except a lower bound~\cite{CK78} which depends on the bipartite {\em mutual informations} (MI): $MI_{AE}$ and $MI_{EB}$. Minimizing SKR amounts to maximizing $MI_{AE}$ %which is the MI among the r.v.s A and E. 
which in turn is an appropriate candidate to estimate Eve's knowledge gain from measurement outcomes as it captures the reduction of entropy in Alice's random variable due to Eve's knowledge from outcomes. A closely related, but easier to estimate quantifier is her {\em information gain} (IG)~\cite{fuchs97}. The optimal MI is found to be a concave function of the optimal IG.

An {\em optimal interaction} is the one that can maximize KG in both the bases. However, acquiring the maximum knowledge depends on the right choice of the measurement, called \textit{optimal measurement}.

For BB84 protocol, for equal prior, both IG and MI is a function of three parameters: two density operators $\rho_a^{\b}, \rho_{\bar{a}}^{\b}$ and the POVM $\{M_{\lambda}\}^{\b}$.
For a fixed QBER $D_{\beta}$ in each bases, a {\em global maxima} exists for each of the functions IG and MI in each of the bases %~\footnote{When $\beta=xy$, $\bar{\beta}=uv$ and vice versa. Sometimes, we use the indices 0, 1 instead of the labels $xy, uv$ respectively.} 
%$\b\in\{xy,uv\}$ 
and is attainable~\cite{fuchs97}. 
\begin{equation*}
\begin{array}{rclcrcl}
IG_{\beta}^{\star} & = & 2 \concave{D_{\bar{\beta}}}, ~~&~~MI_{\beta}^{\star} &=& \half~\phi\left(IG_{\beta}^{\star}\right), 
%\\[1.5ex]
%IG_{uv}^{\star} & = & 2 \concave{D_{xy}}, ~~&~~MI_{uv}^{\star} &=& \half~\phi\left(IG_{uv}^{\star}\right), 
\end{array}
\end{equation*}
for the concave function
\begin{eqnarray*}
	\phi(z) &:=& \philog{z}.
\end{eqnarray*}	
%	Notice that maximum KG in an encoding basis is a function of the QBER in the conjugate basis. 
%	\newcolumntype{R}{>{$}r<{$}}  % for math entries
%	\newcolumntype{L}{>{$}l<{$}}
%	\newcolumntype{C}{>{$}c<{$}}
%	\renewcommand{\arraystretch}{1.8}
%	\begin{table}[htbp]  % half-sided table
%		%\small
%		\centering  
%		\caption{Optimal bounds for IG and MI in each bases}
%		\label{tab:bounds}
%		\begin{tabular}{RCL|C|RCL}  %R@{\extracolsep{\fill}}
%			\toprule \rule{0pt}{4.5ex}		
%			IG_{xy}^{\star} & = & 2 \concave{D_{uv}} ~~&~~~&~~IG_{uv}^{\star} & = & 2 \concave{D_{xy}} ~ \\[1.5ex]
%			\hline
%			MI_{xy}^{\star} &=& \half~\phi\left(IG_{xy}^{\star}\right) ~~&~~~&~~ MI_{uv}^{\star} &=& \half~\phi\left(IG_{uv}^{\star}\right)~ \\[1.5ex]
%			\hline
%			~\phi(z) &:=& \multicolumn{5}{C}{\philog{z}}~ \\[1.5ex]
%			\hline\hline
%		\end{tabular}
%	\end{table}
%	\renewcommand{\arraystretch}{1}
The upper bounds for both IG and MI is attainable in each of the bases for independently chosen error rates $D_{xy}$ and $D_{uv}$.	
%Both the bounds IG$_{xy}^{\star}$, and IG$_{uv}^{\star}$ could simultaneously be achieved by Eve while fixing $D_{xy}$ and $D_{uv}$ independently. The same pair of interactions saturate the bounds for MI in the respective bases. %This is so because, given a fixed basis, the set of conditions to saturate the bounds for IG and MI remain same~\cite{ach17}.
	 
Finding an optimal POVM for such IVs correspond to a rather easier optimization problem: maximize IG over all POVMs~\cite{fuchs96}.	
%	A simpler version to get started with could be the following: given an interaction (i.e., the parameters $\rho_x, \rho_y$ get fixed), maximize information gain IG$_{xy}$ over all measurements $E_{\Lambda}$, which essentially finds the optimal measurement for a given interaction~\cite{fuchs96}.
%		\begin{center}
%			maximize IG$_{xy}$ over all POVMs $\{E_{\Lambda}\}$. 
%		\end{center}
An upper bound exists and is achievable in each of the encoding bases. In $xy$ basis, the maximum IG is attained by the orthonormal eigenprojectors $\{E_{\lambda} := \ketbra{E_\l}{E_\l}\}$ of the Hermitian $\rho_x - \rho_y$.  %It essentially finds the optimal measurement for a given interaction.	
%	\begin{equation} 
%	\begin{aligned} \label{Gxy:Gamma}
%	 \Gamma_{xy}~:=~\rho_x - \rho_y. %\label{eq:gamma} &
%	\end{aligned}	
%	\end{equation}
For equal prior (and not necessarily for unequal prior), the same measurement optimizes both IG and MI for an optimal interaction.  
	
%	An optimal PIJS in a basis $s\sprp$, has the following Schmidt decomposition:
%	\begin{eqnarray}  \label{eq:entangledExprXY}
%		\ket{S} &=& \entangledExpr{\sS}{s}{s}{\sprp}{s}. 
%	\end{eqnarray}
%	with $\ket{\xi_s} \perp \ket{\zeta_s}$ which are called {\em interaction vectors} representing the state of the ancilla used by Eve. 
	%While $s=x$ fixes $s^{\prime}=y$ and $S=X$, for $s=y$ one gets $S=Y$.
%	\begin{equation} 
%	\begin{aligned} \label{rel:perp_s}
%	& \ && \ket{S} = \entangledExpr{\sS}{s}{s}{\sprp}{s}, & \\
%	& \text{with} && \ket{\xi_s} \perp \ket{\zeta_s}. %\label{eq:gamma} &
%	\end{aligned}	
%	\end{equation}
\subsection{The optimal states after an interaction}
 An {\em optimal interaction} induces a restriction %on the PIJSs by restricting the nature of the IVs.
 on the IVs. Optimal IVs must satisfy some necessary and sufficient conditions~\cite{fuchs97}. Deriving optimal IVs from these conditions remained a harder task [done in Sec.~\ref{sec:iff}]. Nevertheless, a judicious bet on a specific choice of IVs passed the verification~\cite{fuchs97}. %Therefore, it fulfilled the primary goal to show that the upper bound on IG and MI is attainable in each of the bases for independently chosen error rates $D_{xy},D_{uv}$.  

Although, there could be various other choices~\cite{ach17}, infinitely many for each of the encoding bases, they are unitarily equivalent. 	
%To avoid such intelligent guesswork, the authors in~\cite{ach17} derived the optimal PIJSs from the first principle. It depicted infinitely many optimal interactions in each bases, which has unique form when expressed in the (optimal) measurement basis. Each of the optimal IVs could be represented as a superposition of two of the measurement directions having amplitudes $\Dps_{uv}$ and $\Dms_{uv}$ defined as follows~\cite{ach17}: %\clrR{The corresponding optimal POVM could then be given by the eigenbasis of the Hermitian $\Gamma_{xy}$ of Eq.~\eqref{eq:Gamma_xy:intetactionVectors}.} 
%When Alice chooses $xy$ basis for encoding, Eve can choose her measurement setup as any four-dimensional orthonormal basis $\{\ket{E_\l}\}$, for which the optimal IVs can be expressed as follows:
In $xy$ basis, the optimal IVs of Eve can be expressed in her orthonormal measurement basis $\{\ket{E_\l}\}$ as follows:
	\begin{eqnarray} \label{eq:interaction-vecs_opt_xy}
	\ket{\xi_x^{\star}} \!=\! \Dps_{uv}\ket{E_0} \!+\! \Dms_{uv}\ket{E_1}, &~~&
	\ket{\xi_y^{\star}} \!=\! \Dms_{uv}\ket{E_0} \!+\! \Dps_{uv}\ket{E_1}, \tab \nonumber\\
	\ket{\zeta_x^{\star}} \!=\! \Dps_{uv}\ket{E_2} \!+\! \Dms_{uv}\ket{E_3},  &~~& 
	\ket{\zeta_y^{\star}} \!=\! \Dms_{uv}\ket{E_2} \!+\! \Dps_{uv}\ket{E_3}. \tab \nonumber\\
	\end{eqnarray}	
Note that, an optimal IV is a superposition of two measurement directions having amplitudes $\Dps_{uv}$ and $\Dms_{uv}$ defined as	
%	Following the notations~\cite{ach17} $\Dps_{uv}, \Dms_{uv}$ defined as
%	\begin{eqnarray}  \label{rel:DpDm}
%	\Dps_{uv} := \invsqrttwo(\sqrt{1-D_{uv}}+\sqrt{D_{uv}}), &&
%	\Dms_{uv} := \invsqrttwo(\sqrt{1-D_{uv}}-\sqrt{D_{uv}}). 
%	\end{eqnarray}
\begin{eqnarray}  \label{rel:DpDm}
	\Dpm_{\b} &:=& \frac{\sqrt{1\!-\!D_{\b}}\pm\sqrt{D_{\b}}}{\sqrt{2}}. 
\end{eqnarray}
%\begin{eqnarray}  \label{rel:DpDm}
%\Scale[0.98]{	\Dps_{uv} := \frac{\sqrt{1\!-\!D_{uv}}+\sqrt{D_{uv}}}{\sqrt{2}}, ~~
%	\Dms_{uv} := \frac{\sqrt{1\!-\!D_{uv}}-\sqrt{D_{uv}}}{\sqrt{2}}. }\tab 
%\end{eqnarray}
%	\begin{eqnarray}\label{rel:calD_uv} 
%	\Dps_{uv}\cdot\Dms_{uv} & = & \half\left(1-2D_{uv}\right),\nonumber\\
%	\Dps_{uv}^2+\Dms_{uv}^2 & = & 1, \nonumber\\
%	\Dps_{uv}^2-\Dms_{uv}^2 & = & 2\concave{D_{uv}}.
%	\end{eqnarray} 

Similarly, the general expression representing the optimal IVs in $uv$ basis are as follows:
%a generic form for optimal interaction vectors was established: % Eq.~\mref{eq:opt_alpha_rel}
\begin{eqnarray} \label{eq:interaction-vecs_opt_uv}
\ket{\xi_u^{\star}} \!=\! \Dps_{xy}\ket{F_0} \!+\! \Dms_{xy}\ket{F_1}, &~~&
\ket{\xi_v^{\star}} \!=\! \Dms_{xy}\ket{F_0} \!+\! \Dps_{xy}\ket{F_1}, \tab \nonumber\\
\ket{\zeta_u^{\star}} \!=\! \Dps_{xy}\ket{F_2} \!+\! \Dms_{xy}\ket{F_3},  &~~& 
\ket{\zeta_v^{\star}} \!=\! \Dms_{xy}\ket{F_2} \!+\! \Dps_{xy}\ket{F_3}. \tab \nonumber\\
\end{eqnarray}

Any specification of the orthonormal basis $\{\ket{E_{\lambda}}\}$ (or $\{\ket{F_{\lambda}}\}$) provides a specific instance of optimal IVs in computational basis, e.g., the optimal IVs due to Fuchs \etal.~\cite{fuchs97}. Due to varied choices of the eigenbasis, there are infinitely many setups of the optimal IVs when expressed in computational basis. %\footnote{\clrR{For equal prior, one can simply replace $D_{xy}, D_{uv}$ by $D$ and get the similar results~\cite{ach17}.-- Make a para: put a practical attack must be symmetric.}} 
%Maximum information can be extracted by measuring with orthonormal eigenprojectors $\{\ketbra{E_{\lambda}}{E_{\lambda}}\}$.
A one-to-one correspondence between the optimal IVs in each basis can be established (Sec.~\ref{sec:details}) since the optimal measurement directions $\{\ket{E_\l}\}$ in $xy$ basis are interrelated to the optimal measurement directions $\{\ket{F_\l}\}$ in $uv$ basis as follows:
\begin{eqnarray} \label{reln:optimalPOVMs}
2\ket{F_0} &=& \ket{E_0} + \ket{E_1} + \ket{E_2} + \ket{E_3},  \nonumber \\ 
2\ket{F_1} &=& \ket{E_0} + \ket{E_1} - \ket{E_2} - \ket{E_3},  \nonumber \\
2\ket{F_2} &=& \ket{E_0} - \ket{E_1} - \ket{E_2} + \ket{E_3},  \nonumber \\
2\ket{F_3} &=& \ket{E_0} - \ket{E_1} + \ket{E_2} - \ket{E_3}.
\end{eqnarray}
For instance, the measurement basis $\{\ket{E_\l}\} = \{\ket{00},\ket{11},\ket{10},\ket{01}\}$ fixes the measurement basis $\{\ket{F_\l}\} = \{\ket{\zb\zb},\ket{\ob\ob},\ket{\ob\zb},\ket{\zb\ob}\}$ for Eve.
%~\footnote{We sometimes use the notations $\ket{0}$, $\ket{1}$, $\ket{\bar{0}}$ and $\ket{\bar{1}}$ instead of $\ket{x}$, $\ket{y}$, $\ket{u}$ and $\ket{v}$ respectively.}.
These were the optimal measurement bases for an optimal Eve described by Fuchs \etal.~\cite{fuchs97}. Note that the ordering of the basis elements is retained in that case. \\
%in the optimal instantiation~\cite[Eqs.~(50,~51)]{fuchs97} of the interaction vectors.

%	\begin{eqnarray} \label{eq:interaction-vecs_opt_xy}
%	\ket{\xi_x} &=& \Dps_{uv}~\ket{E_0} + \Dms_{uv}~\ket{E_1}, \nonumber\\
%	\ket{\xi_y} &=& \Dms_{uv}~\ket{E_0} + \Dps_{uv}~\ket{E_1}, \nonumber\\
%	\ket{\zeta_x} &=& \Dps_{uv}~\ket{E_2} + \Dms_{uv}~\ket{E_3}, \nonumber\\
%	\ket{\zeta_y} &=& \Dms_{uv}~\ket{E_2} + \Dps_{uv}~\ket{E_3},
%	\end{eqnarray}

%Strategy of Eve can be determined by evaluating the sign parameters $\sgn(Q_{x\l}-Q_{y\l})$: a $+ve$ value indicates higher likelihood for $x$, while a $-ve$ value indicates $y$ is more likely. For the optimal IVs~\mref{eq:interaction-vecs_opt_xy} in $xy$ basis, these values are $+1, -1, +1, -1$ for $\l=0,1,2,3$ respectively~\cite{ach17}. Thus Eves strategy will be to guess $x$ for eigenvalues $\l=0,2$ (i.e., eigenstates $\ket{E_0}, \ket{E_2}$) and to guess $y$ for eigenvalues $\l=1,3$ (i.e., eigenstates $\ket{E_1}, \ket{E_3}$).

%\subsection{Optimal strategy}
\textbf{Optimal strategy: }
Strategy of Eve can now be determined as follows. 
As Alice declares her basis to be $\b\in\{0,1\}$, Eve measures her ancilla in basis $\{\ket{M_\l}^{\b}\}_{\l\in\{0,1,2,3\}}$ and interprets her measurement outcome in terms of a guess on Alice's bit. For $+$ve outcome, which occurs for $\l=0,2$, she bets on 0, whereas, for $-$ve outcome, which occurs for $\l=1,3$, she bets on 1. 

%Following this strategy, Eve can optimally learn $\half\phi\left(2\sqrt{D_{\bar{\b}}(1-D_{\bar{\b}})}\right)$ per bit of the transmission with fidelity $\left(\half + \sqrt{D_{\bar{\b}}(1-D_{\bar{\b}})}\right)$ in lieu of introducing an error rate $D$ within the channel between the honest parties. The distinguishing advantage for an optimal attack is $\sqrt{D(1-D)}$.
% the maximum (possible) information about the transmission.

To mount an optimal attack, Eve performs a suitable interaction (the allowed unitaries can be found in Sec.~\ref{sec:unitaryEvol}), measures accordingly after basis reconciliation, and finally guesses the signal applying her strategy. %Her KG is maximum \ifft~her IVs are optimal and her measurement is also optimal.
Fig.~\ref{fig::crckt} provides a schematic view of the attack model.
%Remember that an optimal measurements $E_{\l}$ corresponds to the projectors $\ketbra{E_{\l}}{E_{\l}}$ (\clrR{??}).
%An interaction is optimal whenever Eve's PIJS-IVs and the associated measurements are optimal in each basis. Such attacks that optimize Eve's KG are known as {\em optimal eavesdropping}.  

%\section{Key-rate: IG, MI(AB,AE)}
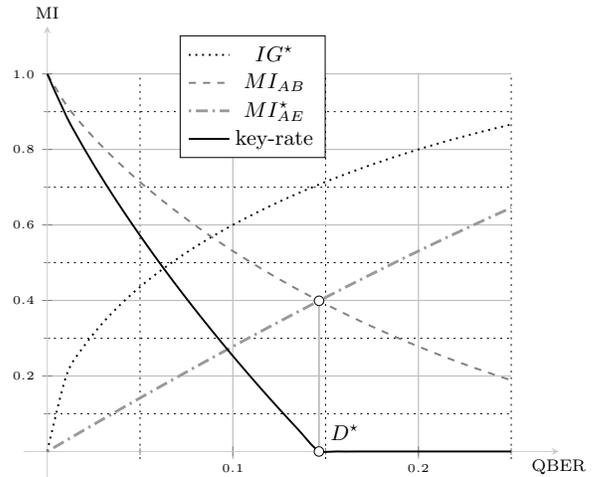
\begin{figure}[htbp]
%	\scale{0.8}
	%\begin{center}
%		\caption{ \underline{Key-rate for one-way classical post-processing.} \\
%			Plotted: optimal Information Gain by Eve, Mutual Informations %$MI_{AB}$ and $MI_{AE}$, 
%			between Alice-Bob and Alice-Eve, 
%			and the secret key-rate. %\\
%			The graph of $MI_{AE}$ reveals the information-disturbance trade-off.
%			For QBER $D^{\star}=0.1464$, $MI_{AB}$ and $MI_{AE}$ %two curves $MI_{AB}$ and $MI_{AE}$ intersect
%			coincides, and the key-rate drops to zero. Below this error rate, an OW-CPP is faithful.
%		} 
		\caption{ \underline{Key-rate for one-way classical post-processing.} \\
			Plotted: optimal Information Gain, bipartite Mutual Informations, %$MI_{AB}$ and $MI_{AE}$, 
			and the secret key-rate. %\\
			The graph of $MI_{AE}$ reveals the information-disturbance trade-off.
			For QBER $D^{\star}=0.1464$, $MI_{AB}$ and $MI_{AE}$ %two curves $MI_{AB}$ and $MI_{AE}$ intersect
			coincides, and the key-rate drops to zero. Below this error rate, an OW-CPP is faithful.
		} 
		\label{fig::keyRate}
	\begin{center}	
		%%%%%%%%%%%%%%%%%%%%%%%%%%%%%%%%%%%%%%%%%%%%%%%%%%%%
%% Author: Atanu Acharyya
%% Senior Research Fellow
%% Applied Statistics Unit
%% Indian Statistical Institute
%% 203 B T Road, Kolkata 700 108, India
%% Email: pub.academy.15@gmail.com
%%%%%%%%%%%%%%%%%%%%%%%%%%%%%%%%%%%%%%%%%%%%%%%%%%%%%

%-------------	Key rate lower bound_____Mutual Info I_{AB}-I_{AE}	------------
\begin{tikzpicture}[
scale=0.9,	%,xscale=1.5,yscale=1.5
declare function={
	C(\x) = (2*\x);
	D(\x) = ln(C(\x));
	ig(\x) = 2.0*sqrt(\x*(1-\x)); 	%------------ ig-opt: 2\sqrt{D(1-D)}
	%miAB(\x) = 1 + \x*\lg(\x) + (1-\x)*\lg(1-\x);
	miAB(\x) = 1 + \x*log2(\x) + (1-\x)*log2(1-\x); %------- MI_{AB} in base 2, i.e., in bits
	phiFunc(\x) = (1+\x)*log2(1+\x) + (1-\x)*log2(1-\x); 	%-----	phi-function
	miAE(\x) = 0.5*phiFunc(ig(\x));	%------- MI_{AE} in base 2, i.e., in bits
	keyRate(\x) = max(0,miAB(\x) - miAE(\x));
}
]

\def \xmin{0} 	\def \xmax{0.25}
\def \ymin{-0.3}	\def \ymax{1}

\begin{axis}[
xmin=0,xmax=0.25,
ymin=-0.02,ymax=1,
grid=both,
major grid style={line width=.5pt,draw=gray!50},
minor grid style={thick, dotted,line width=0.1pt,draw=black},
axis lines=middle,
axis line style={shorten >=-20pt, shorten <=-10pt,draw=gray!40},
%x axis line style={name path=error},
%y axis line style={name path=mi},
%enlargelimits={abs=0.5},
x tick label style = {
	/pgf/number format/.cd,
		fixed, fixed zerofill, precision = 1, 
	/tikz/.cd
},
y tick label style = {
	/pgf/number format/.cd,
	fixed, fixed zerofill, precision = 1, 
	/tikz/.cd
},
minor x tick num=1, % <--- added
minor y tick num=1, % <--- added
xtick={0.1,0.2},
minor xtick={0.05,0.15,0.25},
ytick={0.2,0.4,...,1},
minor ytick={0.1,0.3,...,0.9},
xlabel style={
	anchor=north,
	at={(ticklabel* cs:1.0)},
	xshift=20pt
}, xlabel=QBER,
ylabel style={
	anchor=south,
	at={(ticklabel* cs:1.0)},
	yshift=20pt
}, ylabel=MI,
legend style={at={(0.6,1.1)},anchor=north east,align=left}
%legend pos=north east
]

%\draw[step=0.245cm,gray!50,very thin,opacity=0.9] (axis cs: 0,0) grid (axis cs: 10,10);
%\node at (axis cs: -0.15, -0.3) {$O$};
%\draw[->,thick,name path=Daxis] (\xmin-0.1,0)--(\xmax,0) node[right=1pt] {QBER};
%\draw[->,thick,name path=mi] (0,\ymin)--(0,\ymax) node[left=1pt] {MI};

\addplot[dotted, line width=0.8pt, domain=0:\xmax, smooth]{ig(x)};
\addplot[dashed, line width=0.8pt, draw=gray, domain=0:\xmax, smooth, name path=miAB]{miAB(x)};
%\addplot[thick, \red, domain=0:\xmax, smooth]{phiFunc(x)};
\addplot[thick, line width=1.2pt, 
dash pattern=on 1pt off 2pt on 5pt off 2pt, mark options={solid},draw=gray!80, domain=0:\xmax, smooth, name path=miAE]{miAE(x)};
\addplot[thick,  domain=0:\xmax, smooth, name path=keyRate]{keyRate(x)};

\addlegendentry{$IG^{\star}$}
\addlegendentry{$MI_{AB}$}
\addlegendentry{$MI_{AE}^{\star}$}
\addlegendentry{key-rate}

%\node[thick,right=4pt,scale=0.8] at (axis cs: 0.15,0.8){}; % {$IG^{\star}$}; % function label
%\node[thick,\blue,right=4pt,scale=0.8] at (axis cs: 0.2,0.18){}; % {$MI_{AB}$}; % function label
%\node[thick,\red,right=4pt,scale=0.8] at (axis cs: 0.18,0.62){}; % {$MI_{AE}^{\star}$}; % function label
%\node[thick,\grn,scale=0.8] at (axis cs: 0.14,0.2){}; % {key-rate}; % function label

% axes
%\draw[->,name path=error] (\xmin,0)--(\xmax,0) node[right=1pt] {QBER};
%\draw[->,thick,name path=mi] (0,\ymin)--(0,\ymax) node[left=1pt] {MI};
\node[circle,draw=black, fill=white,inner sep=0pt,minimum size=4pt,	label=above right: $D^{\star}$,] (QBERmax) at (axis cs: 0.1464,0) {};
%\filldraw[fill=black,draw=black] (0.1464,0) circle(0.6);
%\coordinate[thick,scale=0.8,label=above right: $D^{\star}$,] (QBERmax) at (axis cs: 0.1464,0);
\path [name intersections={of=miAB and miAE, by={ordinate}}];
\draw[thick,  gray!50] (QBERmax) -- (ordinate);
\node[circle,draw=black, fill=white, inner sep=0pt,minimum size=4pt] at (ordinate) {};

\end{axis}
\end{tikzpicture}
	\end{center} 
\end{figure}

\subsection{Practical eavesdropping: the secure zone. \\ Connecting Bell violation and cloning.}
%{Practical eavesdropping: the secure zone. \\ Connecting Bell violation and Bloch-vector shrink} \label{sec:connections}
A practical eavesdropping should ideally leave the error rate symmetric across the two basses, i.e., $D_{xy} = D_{uv} = D$. Otherwise, the legitimate parties can detect the difference during the error-estimation phase,  and thereby detect the presence of a malevolent party. For a QBER $=D$, the maximum amount of the IG in both the bases reaches $2\concave{D}$, and is achievable~\cite{fuchs97}. 
%For a symmetric attack, the MI between Alice-Eve is same as the MI between Bob-Eve.

Due to symmetric eavesdropping, the quantum channel between Alice-Bob and that between Alice-Eve can be interpreted as a binary symmetric channel with data-flipping rate $D$ and $D_E=\half - \sqrt{D(1-D)}$, respectively. Thus, at error-rate $D$, the respective bipartite mutual informations become
\begin{eqnarray*}
	MI_{AB} &=& 1-H(D) =  \half~\phi\left(1-2D\right), \\
	MI_{AE} &=& 1-H(D_E) =  \half~\phi\left(2\concave{D}\right),
\end{eqnarray*} 
when expressed in \textit{bits per sifted-photon} (\bpsp).
%$I_{AB} = \half~\phi\left(1-2D\right)$ and $I_{AE} = \half~\phi\left(2\concave{D}\right)$ 
%which coincides when $1-2D = 2\concave{D}$,
%	i.e., for the \emph{critical value} $D=D^{\star}$ of Eq.~\eqref{QBER_max}.
	
%The amount of QBER characterizes the severity of the attack. 
The \textit{secret key-rate} $K$ %is bounded below by
%, which is the ratio of the length of the secret key and that of the \clrR{raw key}, 
is bounded below by the difference $MI_{AB} - MI_{AE}$. 
For a QBER $D$, it amounts to $K_D = H(D_E) - H(D)$ \bpsp. 
%, where $H(p)$ denotes the entropy with probability $p$. 
% between the mutual informations.
%$MI_{AB}$ and $MI_{AE}$, and thereby depends on the QBER. 
It decreases with growing QBER, and vanishes when the two MIs coincide which happens at the threshold [Fig.~\ref{fig::keyRate}]
%As the error-rate increases, the key-rate decreases and becomes zero for $D=D^{\star}$. For \clrR{OW-CPP}, it assumes the value 
\begin{eqnarray} \label{QBER_max}
D^{\star} &=& \half\left(1-\frac{1}{\sqrt{2}}\right) \approx 0.1464. 
\end{eqnarray}
Beyond this tolerable rate, an OW-CPP may not guarantee to filtrate a secure key. 
Within the \textit{secure zone} $D\in [0,D^{\star})$, key-filtration is guaranteed because Bob possess more information on Alice's bit than Eve does.
%$I_{AB}$ remains more than $I_{AE} (=I_{EB})$. 
%Fig.~\ref{fig::keyRate} consolidates the scenario. 
%It cannot cross a specific limit in order to ensure a classical post-processing feasible.

%Following the optimal strategy, Eve can glean $\half\phi\left(2\sqrt{D(1-D)}\right)$ bits per sifted-photon of the transmission.  She incurs an error rate $D_E = \half - \sqrt{D(1-D)}$ in her measurement of Alice's information, while Bob experiences an error rate $D$ to judge Alice's bits.
Following the optimal strategy, Eve can glean $(1-H(D_E))$ bits per sifted-photon of the transmission with fidelity $1-D_E$ in lieu of introducing an error-rate $D$ at Bob's end.
	%with fidelity $\left(\half + \sqrt{D(1-D)}\right)$ in lieu of introducing an error rate $D$ within the channel between the honest parties. 
The distinguishing advantage for an optimal attack is $\sqrt{D(1-D)}$.

An optimal attack on the prepare-and-measure (\textit{p$\&$m}) scheme that we considered here has some interesting connections with the optimal attack on its entanglement-based (\textit{eb}) counterpart as well with optimal cloning mechanisms.  

In the  \textit{eb} protocol, the legitimate parties observe a Bell violation so far the estimated QBER remains in the secure zone of the \textit{p$\&$m} scheme.
%While Alice-Bob find themselves in the secure zone in the prepare-and-measure scheme, they observe Bell violation in its entanglement-based counterpart. 
An optimal attack with QBER $D$ reduces the CHSH correlation co-efficient to $\eta_D2\sqrt{2}$ for  $\eta_D:=1-2D$. An optimal attack also leaves Bob with the Bloch vectors contracted by a factor of $\eta_D$. 

An optimal attack on the \textit{p$\&$m} scheme can also be achieved via an optimal phase-covariant cloner~\cite{BCAM2K}. 
The cloner is asymmetric since it creates two clones of the senders state: a degraded copy for her own with fidelity $(\half + \sqrt{D(1-D)})$, and a superior copy for Bob with fidelity $1-D$. At the threshold QBER, both the fidelity for Bob and Eve reaches the maximum of $1-D^{\star}=\half\left(1+\frac{1}{\sqrt{2}}\right)$ i.e., $85.36 \%$, both in cloning and in \textit{p$\&$m} scheme. Moreover, the optimal PIJSs agrees with the outputs of an optimal pc-cloner. To be specific, for Eve's measurement basis (Fuchs basis) $\{\ket{E_0},\ket{E_1},\ket{E_2},\ket{E_3}\} = \{\ket{00},\ket{11},\ket{01},\ket{10}\}$, the optimal PIJSs are same as those in \cite[Eq.~(36)]{BCAM2K}.

\section{A necessary and sufficient condition for optimality and deriving optimal interaction vectors} \label{sec:iff}
%\textbf{Notations: } $E_{\l}$ is the POVM for $xy$ basis, i.e., basis-0. $F_{\l}$ is the POVM for $uv$ basis, i.e., basis-1. Here, the optimal measurements $E_{\l}$ corresponds to the projectors $\ketbra{E_{\l}}{E_{\l}}$.
%What is the certificate for an interaction by an eavesdropper to be optimal? 
Optimality of an interaction require a certificate, e.g., a necessary and sufficient condition~\cite{fuchs97}. The verification involves the PIJSs in the joint Hilbert space. Here we suggest a refined NSC involving the states of Eve only that makes the verification easier. The journey also leads to derive the optimal IVs which are unitarily equivalent to those derived in~\cite{ach17}.
%--[Eqs.~(38,39)]

%
\subsection{A Necessary and sufficient condition \\ due to Fuchs \etal.~\cite{fuchs97}}
Consider the optimality of the post-interaction states~\eqref{eq:entangledExpr}. For  Alices symbol $a^{\b} \in \{x,y,u,v\}$, denote the PIJS symbol $S_a^{\b}$ as $X,Y,U,V$, respectively. The NSC~\cite[Eqs.~(38,39)]{fuchs97} for optimality in $xy$ basis involves the following four states defined over the joint Hilbert space of Bob and Eve.
%\clrV{The derivation in~\cite{fuchs97} for the upper bound on IG (MI) passes through a chain of inequalities. The set of conditions to achieve the upper bound in each of these inequalities were cooked together to come up with the NSC~\cite[Eqs.~(38,39)]{fuchs97} to check optimality. To verify optimality in $xy$ basis, consider the following four states in the joint Hilbert space of Bob and Eve.}
%Once Bob receives the signals from Alice, he measures them one-by-one in a randomly chosen basis. Let, his Von Neumann POVMs are $\{B_x, B_y\}$ in $xy$ basis and $\{B_u, B_v\}$ in $uv$ basis, where $B_a$ stands for the projector $\ketbra{a}{a}$ for $a\in\{x,y,u,v\}$.  
%The essential part of the \ifft~condition is to establish mutual parallelism among some vectors. For optimality of IV$_{xy}$, those four vectors are as defined below: 
%		\begin{eqnarray} \label{eq:Ulu}
%		\ket{U_{\l u}} = \Xlambdax{U}{u}, &~~& \ket{V_{\l u}} = \Xlambdax{V}{u}, \tab \nonumber\\ 
%		\ket{U_{\l v}} = \Xlambdax{U}{v}, &~~& \ket{V_{\l v}} = \Xlambdax{V}{v}. \tab
%		%B_u = \ketbra{u}{u}, ~~~ B_v = \ketbra{v}{v}, ~~~\text{ s.t., }~~ B_u + B_v = \mathbbm{1}.\nonumber\\
%		\end{eqnarray} 
\begin{eqnarray} \label{eq:Ulu}
\ket{W_{\l a}} := \Xlambdax{W}{a}, %&~~& \text{with~} W\in\{U,V\} \text{~and~} s\in\{u,v\}
%B_u = \ketbra{u}{u}, ~~~ B_v = \ketbra{v}{v}, ~~~\text{ s.t., }~~ B_u + B_v = \mathbbm{1}.\nonumber\\
\end{eqnarray} 
with $W\in\{U,V\}$ and $a\in\{u,v\}$; Bob uses the von Neumann POVMs $B_a:=\ketbra{a}{a}$.

For optimal $KG_{xy}$, the inner products $\braket{U_{\l u}}{V_{\l u}}$ and $\braket{U_{\l v}}{V_{\l v}}$ must be real and have the same sign \footnote{Henceforth, we use the notations $\l^{\b}$ and $\varepsilon_{\l}^{\b}$ to denote the eigenvalues and their signs~\cite{ach17} in a basis $\b$.}
 $\varepsilon_{\l}^{0}\in\pm1$.  Checking optimality is essentially to check the following parallelism:
%	the first two vectors are parallel to each other and similarly for the last two, i.e.,
$$\ket{U_{\l u}}\parallel\ket{V_{\l u}} \text{~~and~~} \ket{U_{\l v}}\parallel\ket{V_{\l v}}.$$

%\begin{proposition} \label{propn:iff_Fuchs}
	The PIJSs $\ket{X}, \ket{Y}$ are optimal for Eve with a POVM $\{E_{\lambda}\}$ \ifft~the following conditions are satisfied:
%	the conjugate states $\ket{U}, \ket{V}$ satisfy the following conditions:
%	\begin{equation} \label{eq:iff_Fuchs}
%	\begin{aligned} 
%	& ~~\phantom{and} & \sqrt{D_{uv}}~\ket{U_{\l u}} = \varepsilon_\l \sqrt{1\!-\!D_{uv}}~\ket{V_{\l u}},  \tab \nonumber \\
%	& ~~\text{and} & \sqrt{D_{uv}}~\ket{V_{\l v}} = \varepsilon_\l \sqrt{1\!-\!D_{uv}}~\ket{U_{\l v}}. \tab 
%	\end{aligned}	
%	\end{equation}
%	\clrR{eq. num not displaying?}\\
\begin{subequations} \label{eq:iff_Fuchs}
	\begin{align}
		&& \sqrt{D_{uv}}~\ket{U_{\l u}} &= \varepsilon_\l \sqrt{1\!-\!D_{uv}}~\ket{V_{\l u}},  \tab 
				\tag{\ref{eq:iff_Fuchs}.u} \label{eq:iff_Fuchs.u}
		\\
		&& \sqrt{D_{uv}}~\ket{V_{\l v}} &= \varepsilon_\l \sqrt{1\!-\!D_{uv}}~\ket{U_{\l v}}. \tab 
				\tag{\ref{eq:iff_Fuchs}.v} \label{eq:iff_Fuchs.v}
	\end{align}
\end{subequations}	
%	\begin{eqnarray} \label{eq:iff_Fuchs}
%	\sqrt{D_{uv}}~\ket{U_{\l u}} &=& \varepsilon_\l \sqrt{1\!-\!D_{uv}}~\ket{V_{\l u}},  \tab \nonumber \\
%	\sqrt{D_{uv}}~\ket{V_{\l v}} &=& \varepsilon_\l \sqrt{1\!-\!D_{uv}}~\ket{U_{\l v}}. \tab 
%	\end{eqnarray}
	Similarly, analogous conditions hold for the optimality of the PIJSs in $uv$ basis.
%\end{proposition}

%	For a signal sent in $\xy$ basis, to check optimality of a given interaction fixed by the postinteraction states $\ket{X},\ket{Y}$ and a given measurement $E_{\Lambda}$, the underlying vectors are defined as  
%	\begin{eqnarray} \label{eq:Ulu}
%	\ket{U_{\l u}} = \Xlambdax{U}{u}, &~~& \ket{V_{\l u}} = \Xlambdax{V}{u}, \tab \nonumber\\ 
%	\ket{U_{\l v}} = \Xlambdax{U}{v}, &~~& \ket{V_{\l v}} = \Xlambdax{V}{v}, \tab
%	%B_u = \ketbra{u}{u}, ~~~ B_v = \ketbra{v}{v}, ~~~\text{ s.t., }~~ B_u + B_v = \mathbbm{1}.\nonumber\\
%	\end{eqnarray}
%	where the operators $B_u,B_v$ are the projectors $\ketbra{u}{u},\ketbra{v}{v}$ respectively following the completeness relation $B_u + B_v = \mathbbm{1}$.
%	

%	\begin{eqnarray} \label{eq:condn_optG_1}
%	\ket{V_{\l u}} &=& \varepsilon_\l~\factor{uv}~ ~\ket{U_{\l u}} 
%	\end{eqnarray}
%	and 
%	\begin{eqnarray} \label{eq:condn_optG_2}
%	\ket{U_{\l v}} &=& \varepsilon_\l~\factor{uv}~ ~\ket{V_{\l v}}, 
%	\end{eqnarray}

\subsection{A new necessary and sufficient condition towards \\ completely characterizing Eve's optimal states}
Now, we move from this NSC to derive a refined one. In this pursuit, we move through a series of \ifft~conditions that eventually derives the optimal IVs in terms of the optimal measurement basis.

The following observation is going to help finding a refined certificate for optimality.
\begin{lemma} \label{lem:interreln:parity-overlaps}
	The post-interaction states of Eve exhibit an interrelation involving the overlap between the two undisturbed states and that between the two disturbed states.
	\begin{equation*}
		\left(1-D_{xy}\right)\braket{\xi_x}{\xi_y} + D_{xy}\braket{\zeta_x}{\zeta_y} = 2 \Dp_{uv}\Dm_{uv}.
	\end{equation*} 
\end{lemma}
The result follows by considering the inter-relations~\eqref{eq:interaction-vecs_uv_to_xy} between the IVs across the bases, while imposing the normalization constraint on $\ket{\xi_{u}}$. % in the first equation.

%We start from the necessary and sufficient conditions for optimality as described in Eq.~\eqref{eq:iff_Fuchs}, and derive a series of \ifft~conditions for optimality that eventually derives the optimal interaction vectors in terms of the optimal measurement basis. As a byproduct, we derive a necessary and sufficient condition for optimality which is more easier than that of Fuchs \etal., when it comes to verify whether a set of IVs is optimal or not.  %when it comes to verify whether a set of IVs is optimal. 

Here we derive a series of \ifft~conditions for an interaction to be optimal. The following conditions are equivalent. 

\begin{theorem} \label{thm:equivalent-iff-condns} 
	The set of interaction vectors $IV_{xy}$ is optimal along with the projectors $E_{\l}:=\ketbra{E_{\l}}{E_{\l}}$ for measurement iff any of the following conditions hold:
	\begin{enumerate}
		\item \label{iff:overlapIV_uv-mes_xy}
		The overlap between the measurement direction $\ket{E_{\l}}$ in $xy$ basis and the IVs in $uv$ basis are related in the following way:
		\begin{eqnarray} \label{rel:ElXiuZetav_ElXivZetau}
		\braket{E_{\l}}{\xi_u} &=& \varepsilon_{\l}^{0}~\braket{E_{\l}}{\zeta_v}, \nonumber \\
		\braket{E_{\l}}{\xi_v} &=& \varepsilon_{\l}^{0}~\braket{E_{\l}}{\zeta_u}.
		\end{eqnarray}
		%		Note that while the overlaps in Eq.~\eqref{rel:ElXiuZetav_ElXivZetau} consider the measurement directions in the $xy$ basis, the IVs are referred in the $uv$ basis.
		\begin{description}
			\item[Corollary \ref{iff:overlapIV_uv-mes_xy}]	 \label{iff:IV-overlaps} %\tag{\ref{iff:overlapIV_uv-mes_xy}} 
			The overlap between the IVs in $xy$ basis satisfy the following condition:
			\begin{equation} \label{rel:iff}
			\braket{\xi_x}{\xi_y} = \braket{\zeta_x}{\zeta_y} = 1-2D_{uv}. %2 \Dp_{uv}\Dm_{uv}.
			\end{equation}
			%Analogus condition applies to the conjugate basis.
		\end{description}		
		\item \label{iff:overlap-ratio}	
		The overlaps between the measurement direction $\ket{E_{\l}}$ in $xy$ basis and the IVs in the same basis must maintain the following ratio:	
		\begin{eqnarray} \label{rel:ElXixXiy_ZetaxZetay}
		\frac{\braket{E_{\l}}{\xi_x}}{\braket{E_{\l}}{\xi_y}} = \frac{\braket{E_{\l}}{\zeta_x}}{\braket{E_{\l}}{\zeta_y}} = \frac{\D_{uv}^{(+,\varepsilon_{\l}^{0})}}{\D_{uv}^{(-,\varepsilon_{\l}^{0})}} = \left(\frac{\Dps_{uv}}{\Dms_{uv}}\right)^{\varepsilon_{\l}^{0}}.
		\end{eqnarray}
		Here, we improvise to the following notation
		\begin{eqnarray} \label{notation:curly-D-epsa} 
		\D_{uv}^{(\sigma,\varepsilon_{\l}^{0})} &=& \invsqrttwo\left(\sqrt{1-D_{uv}}+\sigma\varepsilon_{\l}^{0}\sqrt{D_{uv}}\right), %\nonumber \\
		\end{eqnarray} 
		with the sign parameter $\sigma=\pm1$. It becomes $\D_{uv}^{+}$ or $\D_{uv}^{-}$, depending on whether the product $\sigma\varepsilon_{\l}^{0}$ becomes plus or minus, respectively.

		\item \label{iff:optIVs}
		The interaction vectors in the $xy$ basis can be expressed in an orthonormal basis $\{\ket{E_{\l\xi}^{+}}, \ket{E_{\l\xi}^{-}}, \ket{E_{\l\zeta}^{+}}, \ket{E_{\l\zeta}^{-}}\}$ as follows:
		\begin{eqnarray} \label{eq:opt-IVs:new}
		\ket{\xi_x} &=& \Dps_{uv}\ket{E_{\l\xi}^{+}} + \Dms_{uv}\ket{E_{\l\xi}^{-}}, \nonumber \\ 
		\ket{\xi_y} &=& \Dms_{uv}\ket{E_{\l\xi}^{+}} + \Dps_{uv}\ket{E_{\l\xi}^{-}}, \nonumber \\ 
		\ket{\zeta_x} &=& \Dps_{uv}\ket{E_{\l\zeta}^{+}} + \Dms_{uv}\ket{E_{\l\zeta}^{-}}, \nonumber \\ 
		\ket{\zeta_y} &=& \Dms_{uv}\ket{E_{\l\zeta}^{+}} + \Dps_{uv}\ket{E_{\l\zeta}^{-}}. 
		\end{eqnarray}
		The basis vectors $\ket{E_{\l\xi}^{\pm}}, \ket{E_{\l\zeta}^{\pm}}$ correspond to some unitary transform $\mathbf{R}^{\pm}$ of those two measurement directions $\ket{E_{\l}}$ that provide $\pm$ve outcomes.
%		While the basis vectors $\ket{E_{\l\xi}^{+}}, \ket{E_{\l\zeta}^{+}}$ correspond to some unitary transform $\mathbf{R}^{+}$ of those two measurement directions $\ket{E_{\l}}$ that provide $+$ve measurement outcomes, the remaining basis vectors $\ket{E_{\l\xi}^{-}}, \ket{E_{\l\zeta}^{-}}$ correspond to some unitary transform $\mathbf{R}^{-}$ of those two measurement directions $\ket{E_{\l}}$ that provide $-$ve measurement outcomes. 			
		%			The basis vectors $\ket{E_{\l\xi}^{\pm}}, \ket{E_{\l\zeta}^{\pm}}$ belong to the plane of  those two measurement directions $\ket{E_{\l}}$ that provide $\pm$ve measurement outcomes. The earlier vectors corresponds to some unitary transform $\mathbf{K}^{\pm}$ of the later vectors, and therefore are unitarily equivalent. Once viewed in this way, the optimal states~\eqref{eq:opt-IVs:new} are unitarily equivalent to those in~\cite[Eq.(38)]{ach17}.
		%While the remaining two vectors $\ket{E_{\l\xi}^{-}}, \ket{E_{\l\zeta}^{-}}$ belong to the span of those two measurement directions which corresponds to negative measurement outcomes.	
	\end{enumerate}	
\end{theorem} 
 The above four \ifft~conditions in Thm.~\ref{thm:equivalent-iff-condns} are equivalent, in the sense that any of them can be derived [see Sec.~\ref{sec:details}] from the other one, directly, or via some of the remaining conditions as sketched below. 

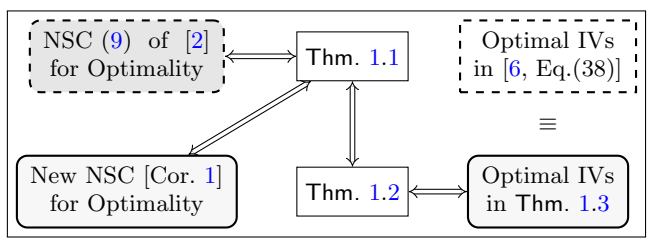
\begin{table}[h!]
	%\caption{Workflow for optimal IVs}. 
	%\label{fig::IV}
	\begin{tabular}{cccl} %{c|c|c|l}
		\boxed{%%%%%%%%%%%%%%%%%%%%%%%%%%%%%%%%%%%%%%%%%%%%%%%%%%%%
%% Author: Atanu Acharyya
%% Senior Research Fellow
%% Applied Statistics Unit
%% Indian Statistical Institute
%% 203 B T Road, Kolkata 700 108, India
%% Email: pub.academy.15@gmail.com
%%%%%%%%%%%%%%%%%%%%%%%%%%%%%%%%%%%%%%%%%%%%%%%%%%%%%

	\begin{tikzpicture}[node distance=1.0cm]
%	\node (A11) [block, draw=black] {
%		\begin{tabular}{c} 
%		\end{tabular}
%	};
	\node (A11) [startstopdash, draw=black] {
		\begin{tabular}{c}
		%Fuchs
				NSC~\eqref{eq:iff_Fuchs} ~of~ \cite{fuchs97}  \\
				for~Optimality
		\end{tabular}
	};
	\node (A12) [block, draw=black, right of=A11, xshift=2cm] {
		%\begin{tabular}{l}
		\thm~\ref{thm:equivalent-iff-condns}.\ref{iff:overlapIV_uv-mes_xy}
		%\end{tabular}
	};
	\node (A13) [blockdash, draw=black, thick, right of=A12, xshift=1.6cm] {
		\begin{tabular}{c}
		Optimal~IVs  \\
		in~\cite[Eq.(38)]{ach17} 
		\end{tabular}
	};
	\node (A21) [startstop, draw=black, below of=A11, yshift=-0.8cm]  {
		\begin{tabular}{c}
		New NSC [Cor.~\ref{iff:IV-overlaps}]  \\
		for~Optimality 
		\end{tabular}
	};
	\node (A22) [block, draw=black, right of=A21, xshift=2cm] {
		%\begin{tabular}{c}
		%T1.2  \\
		\thm~\ref{thm:equivalent-iff-condns}.\ref{iff:overlap-ratio} 
		%\end{tabular}
	};
	\node (A23) [startstop, draw=black, right of=A22, xshift=1.6cm] {
		\begin{tabular}{c}
		Optimal~IVs  \\
		in~\thm~\ref{thm:equivalent-iff-condns}.\ref{iff:optIVs} 
		\end{tabular}
	};

	% iff connections
	\draw [iff] (A11) -- 
	node[anchor=center, above] {} 
	(A12);  
	
	\draw [iff] (A12) -- 
			node[anchor=center, left] {} 
	(A21); 	
	
	\draw [iff] (A12) -- 
	node[anchor=center, right] {} 
	(A22); 
	
	\draw [iff] (A22) -- 
	node[anchor=center, below] {} 
	(A23); 
	
	\draw [] (A23)  			%========= place equiv sign
	node[anchor=center, above, yshift=0.7cm] {$\equiv$} 
	(A13);

	%	\draw [arrow] (A11) -- node[] {} (A21); % arrow
	%	\draw [arrowDashed] (A11) -- node[] {} (A21); % arrow dashed
	\end{tikzpicture}} 
	\end{tabular}
\end{table}

It's interesting to notice the change of basis while describing the overlap between Eve's measurement directions and the IVs. While Eve's measurements are considered in $xy$ basis, the IVs are counted for $uv$ basis and for $xy$ basis in Eq.~\eqref{rel:ElXiuZetav_ElXivZetau} and Eq.~\eqref{rel:ElXixXiy_ZetaxZetay}, respectively.

%In Eqs.~\mref{rel:ElXiuZetav_ElXivZetau, rel:ElXixXiy_ZetaxZetay} the  (measurement, IVs) basis-combination remains ($xy$, $uv$) and ($xy$, $xy$) respectively.
%In the first \ifft~condition,  Eq.~\eqref{rel:ElXiuZetav_ElXivZetau} considers the measurement directions in the $xy$ basis, whereas the IVs are referred in the $uv$ basis. In the byproduct \ifft~condition, Eq.~\eqref{rel:iff} captures the overlaps between the IVs in the $xy$ basis.
%In the second \ifft~condition, Eq.~\eqref{rel:ElXixXiy_ZetaxZetay} considers both the measurement direction and the IVs in the $xy$ basis. The final condition eventually provides the optimal IVs.

\subsection{Explaining the \ifft~conditions} \label{sec:interpret}
Let's explain the essence of the four \ifft~conditions described in Thm.~\ref{thm:equivalent-iff-condns} involving the optimality of the four interaction vectors in the $xy$ basis.

The \ref{iff:overlapIV_uv-mes_xy}$^{st}$ \ifft~condition says that the overlap between a measurement direction $\ket{E_{\l}}$ and a fidelity state corresponding to Alice's signal $u$ (or $v$) is same in magnitude as the overlap between that measurement direction and the disturbed state corresponding to Alice's signal $v$ (or $u$), except that they differ in sign $\varepsilon_{\l}^{0}$. 

%The \ifft~condition in Corollary~\ref{iff:IV-overlaps}, which is a byproduct of the \ref{iff:overlapIV_uv-mes_xy}$^{st}$ \ifft~condition of Thm.~\ref{thm:equivalent-iff-condns}, { \em can be used as a working formula to verify whether a given set of IVs is optimal or not, and this verification is far more easier than the verification criteria~\eqref{eq:iff_Fuchs} of \cite{fuchs97}. %provided by Fuchs \etal.
%It vindicates that, for an optimal asymmetric attack, in $xy$ basis, the overlap between the two fidelity states is same as the overlap between the two disturbed states and is equal to $(1-2D_{uv})$. The importance of this result lies in its connection with Bell violation and shrinking of Bloch vectors and is described elaborately in Sec.~\ref{sec:connections}. }
%For an optimal symmetric attack, this overlap exhibits an intriguing connection with the Bell violation as well as with the shrinking factor of Bloch vectors at the receiving end. To be specific, the amount of overlap $(1-2D)$ is in one hand equal to the reduction factor for the correlation signature in the Bell-CHSH inequality,
%(\clrR{ref Fuchs, and ref own section number where it's discussed}), 
%and in the other hand it's same as the amount of shrinking of the Bloch vectors for Bob that corresponds to the optimal joint states with Eve (\clrR{ref Gisin}).
%However, further probe is required to fully understand this connection between optimal eavesdropping and the Bell-CHSH correlation, which remained a long-standing mystery as of date.

The \ref{iff:overlap-ratio}$^{nd}$ \ifft~condition says that
the ratio of the overlaps between a measurement direction and the undisturbed states are same as the ratio of the overlaps between the measurement direction and the disturbed states. The ratio becomes $\nicefrac{\Dps_{uv}}{\Dms_{uv}}$ or its inverse depending on whether the measurement outcome is positive or negative in sign.

%The lem.~\mref{lem:overlap_IV_mes-dir$\cdot$lem:overlapIV_uv-mes_xy} an lem.~\ref{lem:overlap-ratio}

The \ref{iff:optIVs}$^{rd}$ \ifft~condition provides the optimal interaction vectors, and therefore are the only and all possible optimal IVs. They are unitarily equivalent to those in~\cite[Eq.(38)]{ach17} [see Sec.~\ref{sec:equivalence}]. 
%Thus, we are safe to do further work with those in Eq.~\eqref{eq:interaction-vecs_opt_xy}.

The \ifft~condition in Corollary~\ref{iff:IV-overlaps}, which is a byproduct of the \ref{iff:overlapIV_uv-mes_xy}$^{st}$ \ifft~condition of Thm.~\ref{thm:equivalent-iff-condns}, 
restricts Eve's optimal states to have a specific orientation in the four-dimensional Hilbert space. To be more specific, when Alice encodes is $xy$ basis, the overlap between the two fidelity states must be same as the overlap between the two disturbed states and is equal to $(1-2D_{uv})$. \\

%\begin{paragraph} {The new NSC and its significance: }
\textbf{The new NSC and its significance: }% \label{sec:interpret}
The necessary and sufficient condition in Corollary~\ref{iff:IV-overlaps} can be used as a working formula to verify whether a given set of IVs is optimal or not. 
%Apart from its simplicity and easy verification, the new NSC involves the IVs of Eve than the joint Hilbert space. 
It's efficient due to easy verification, it's simple as it involves Eve's states only than the joint Hilbert space as in~\cite{fuchs97}, it's intuitive as it demands a specific configuration of the states in Eve's Hilbert space. 
%This new verification criteria looks more appealing than the verification criteria~\eqref{eq:iff_Fuchs} of \cite{fuchs97}, because the new criteria involves only the attackers Hilbert space unlike the later one (by Fuchs \etal.) which involves the Hilbert space of both the attacker and the receiver. 

An optimal attack is essentially characterized by the optimal overlap, called here as \textit{optimal syndrome}, that amounts to $1-2D$ for a symmetric attack. It exhibits interesting links between various other approaches for eavesdropping. 
Although the connection between Bell violation and optimal state discrimination is known~\cite{fuchs97}, we find the connection more explicit here with respect to the optimal syndrome. For a specific error-rate $D$, the fraction of reduction in the optimal CHSH-sum in an \textit{eb} scheme is precisely the optimal syndrome in the \textit{p$\&$m} scheme. %[see Sec.~\ref{sec:connections} for details]. 
The Bloch vector at the receiving end shrinks by the same factor. 

%The importance of the newly derived NSC lies in its physical significance, viz., its connection with Bell violation and Bloch vectors contraction. For an  optimal symmetric attack that introduces a channel error $D$, both the pair of parity (fidelity and disturbed) states must share an overlap of $(1-2D)$ which in turn is the factor by which  the CHSH-sum estimated by the legitimate parties is reduced by the same factor while compared to a noise-free channel [see Sec.~\ref{sec:connections} for details]. The Bloch vector at the receiving end shrinks by the same factor.

%\clrV{ Keeping aside this mathematical criteria, all a practical eavesdropper needs to verify optimality is to tally her fidelity versus the estimated channel error between the legitimate parties: %(\clrR{looking at the frequency of her outcomes 2,3}). 
%for QBER $D$ her fidelity must be $\half + \sqrt{D(1-D)}$. }
%\end{paragraph}

%\clearpage

\section{Deriving optimal unitary evolutions}
\label{sec:unitaryEvol}
Given the optimal PIJSs $\ket{X^{\star}}, \ket{Y^{\star}}$, we wish to find an optimal unitary for a suitable initial state $\ket{\psi_0}$ of Eve's ancilla.
Mathematically speaking, the task is to solve the following equations. 
\begin{eqnarray}  \label{eq:post-interaction-joint-states_XY}
\U_{\psi_0}^{AE} \ket{0}_{A} \ket{\psi_0}_{E} = \ket{X^{\star}}, &~~~& \U_{\psi_0}^{AE} \ket{1}_{A} \ket{\psi_0}_{E} = \ket{Y^\star}. \tab
\end{eqnarray}
Although the same unitary serves the purpose in the conjugate basis, the measurement setup generally differs.
Given a specific reconciled basis, Eve's measurement basis is in one-to-one correspondence with the PIJSs: 
different measurement bases correspond to different PIJSs. Thus, we parameterize the PIJSs as $\ket{X}^{\M}, \ket{Y}^{\M}$ on a measurement setup that corresponds to an unitary transform $\mathbf{M}_{xy} := \left[\ket{E_0}, \ket{E_1}, \ket{E_2}, \ket{E_3}\right]$ of the computational basis. %(A choice for $\mathbf{M}_{xy}$ automatically fixes its conjugate counterpart $\M_{uv}$).
%defining a matrix $\M$ (e.g., $\M_{xy}$).
%Remember, that the PIJSs $\ket{X^{\star}}, \ket{Y^{\star}}$ are in one-to-one correspondence with the measurement directions fixed by $\M_{xy}$ which we call as $\M$ when unambiguous in the context. 
%Thus, an optimal unitary $\U$ depends on the two parameters $\ket{\psi_0}, \M$ and possibly on some other free parameters as we'll see shortly.
%To work with the optimal IVs, we prefer Eq.~\eqref{eq:interaction-vecs_opt_xy} over those in Eq.~\eqref{eq:opt-IVs:new} due to its simpler form.
  
\subsection{Optimal PIJSs for different measurement bases} \label{sec:mes-basis-vs-optPIJS}
The PIJSs described in Eq.~\eqref{eq:entangledExpr} live in eight dimensional Hilbert space as the attacker uses a two-qubit probe. The PIJSs are optimal whenever Eve's IVs are optimal for which we prefer Eq.~\eqref{eq:interaction-vecs_opt_xy} over Eq.~\eqref{eq:opt-IVs:new} for simplicity. 

For two-qubit probe of Eve, a PIJS live in eight dimensional Hilbert space. Assume, Alice's qubit is prepared in $xy$ basis. 
Let's consider two different choices of Eve's Measurement basis $\{\ket{E_{\l}}\}$: %$\{\ket{E_0},\ket{E_1},\ket{E_2},\ket{E_3}\}$: 
\begin{enumerate}
	\item the computational basis = $\{\ket{00},\ket{01},\ket{10},\ket{11}\}$, 
	\item the Fuchs basis = $\{\ket{00},\ket{11},\ket{10},\ket{01}\}$.
\end{enumerate}
The optimal states are denoted by $\ket{X^{\star}}^{\mathcal{C}}, \ket{Y^{\star}}^{\mathcal{C}}$ in the first case, and by $\ket{X^{\star}}^{\mathcal{F}}, \ket{Y^{\star}}^{\mathcal{F}}$ in the second case. Their vector form are tabulated below.
%Note that, $\ket{X}^{\mathcal{F}} = (\id_2 \otimes \M_{xy})\ket{X}^{\mathcal{C}}$, where $\M_{xy} = [\ket{00}, \ket{11}, \ket{10}, \ket{01}]$.
%The optimal PIJS in computational basis and Fuchs basis are given below:
%\begin{equation*}
%\begin{array}{ccccc}
%\hline \\ 
%\ket{X}^{\mathcal{C}} & \ket{Y}^{\mathcal{C}} &  & \ket{X}^{\mathcal{F}}  & \ket{Y}^{\mathcal{F}} 
%\medskip \\
%\hline \\
%\Xcomp & \Ycomp &  & \XFuchs  & \YFuchs \medskip \\	
%\hline	\\
%\end{array}
%\end{equation*}
\begin{table}[htbp!]
	\begin{tabular}{C|CC?{0.4mm}CC|C}	
		\toprule
\ket{X}^{\mathcal{C}} & \ket{Y}^{\mathcal{C}} &  && \ket{X}^{\mathcal{F}}  & \ket{Y}^{\mathcal{F}} 
		\\ \addlinespace
		\specialrule{0.8pt}{1pt}{1pt}	\midrule
\Xcomp & \Ycomp &  && \XFuchs  & \YFuchs  	
		\\ 
		\bottomrule
	\end{tabular}
\end{table}
One may consider any other measurement basis, where each measurement direction perhaps include all the computational basis states in its superposition. In that case, the corresponding PIJS may not have any zero entries in its co-ordinated form. We'll show how such \textit{`complicated' measurements} can easily be tackled by our approach surpassing the difficulty of the rudimentary basis-completion method as discussed below. 

%The $uv$-counterparts of the optimal PIJSs can easily be found by $\ket{U} = \frac{\ket{X}+\ket{Y}}{\sqrt{2}}$ and $\ket{V} = \frac{\ket{X}-\ket{Y}}{\sqrt{2}}$.

\subsection{Basis completion method to get an optimal unitary, and its shortcomings.} \label{easySol}
%for a given initial state
Following is the basic mathematical approach to solve Eq.~\eqref{eq:post-interaction-joint-states_XY}.\\

%\begin{theorem}
By introducing some \textit{auxiliary states}, an unitary evolution $\U_{\psi_0}$ can be viewed as a linear transformation that maps an orthonormal basis $\{\ket{0}_{A} \ket{\psi_i}_{E}, \ket{1}_{A} \ket{\psi_i}_{E}\}_{i\in\{0,1,2,3\}}$ to the orthonormal basis $\{\ket{X_i},\ket{Y_i}\}_{i\in\{0,1,2,3\}}$, where $\ket{X_0}=\ket{X^{\star}},\ket{Y_0}=\ket{Y^{\star}}$. %That is to say,
\begin{eqnarray*} \label{eq:alpha-beta}
\U_{\psi_0} \ket{0}_{A}\ket{\psi_i}_{E} = \ket{X_i}, &\tab& \U_{\psi_0} \ket{1}_{A}\ket{\psi_i}_{E} = \ket{Y_i}, \nonumber \\
\ && ~~\forall i\in\{0,1,2,3\}.
\end{eqnarray*}	
Then a solution for the optimal unitary can be given by
\begin{eqnarray} \label{sol:unitary:gen}
\U_{\psi_0} &=& \sum\limits_{i=0}^{3} \left(\ket{X_i}\bra{0_{A}} + \ket{Y_i}\bra{1_{A}}\right)\bra{\psi_i}_{E}.
\end{eqnarray}
%\end{theorem}
which can further be factored [see Sec.~\ref{sec:factorization}] in two unitaries as
%\begin{corollary}\label{cor:unitary:factorize:local}
%Note that, the solution unitary in Eq.~\eqref{sol:unitary:gen} above can be factored in two unitaries in the following way:
\begin{eqnarray} \label{unitary:factorize:local}
\U_{\psi_0} &=& 
\U_{X,Y}^{AE}~(\id_2^{A} \tensorprod \W_{\psi_0}^{E\dagger}).
\end{eqnarray}	
The first unitary $\U_{X,Y}$, that depends on the PIJSs $\ket{X^{\star}},\ket{Y^{\star}}$, is defined as 
\begin{eqnarray} %\label{sol:unitary:wrt:local}
\U_{X,Y} &:=& \sum\limits_{i=0}^{3} \ket{X_i}\bra{0_{A}}\bra{i_{E}} + \ket{Y_i}\bra{1_{A}}\bra{i_{E}},	
%\ \\ &=& 
%	 \left[~\ket{X_0}, \ket{X_1}, \ket{X_2}, \ket{X_3}, \ket{Y_0}, \ket{Y_1}, \ket{Y_2}, \ket{Y_3}~\right];
\end{eqnarray}
which has the following matrix representation
\begin{eqnarray*} %\label{sol:unitary:wrt:local}
\left[~\ket{X_0}, \ket{X_1}, \ket{X_2}, \ket{X_3}, \ket{Y_0}, \ket{Y_1}, \ket{Y_2}, \ket{Y_3}~\right].
\end{eqnarray*}
%\clrR{Is comma separated expression correct way to write?}\\
The local unitary $\W$, that depends on the initial state $\ket{\psi_0}$, is defined as follows
\begin{eqnarray} %\label{sol:unitary:wrt:local}
\W_{\psi_0}^{E} &=& \sum\limits_{i=0}^{3} \ket{\psi_i}\bra{i_{E}}
%~~=~~ \left[~\ket{\psi_0}~~\ket{\psi_1}~~\ket{\psi_2}~~\ket{\psi_3} ~\right].
\end{eqnarray}
which has the following matrix representation
\begin{eqnarray*} %\label{sol:unitary:wrt:local}
\left[~\ket{\psi_0}~~\ket{\psi_1}~~\ket{\psi_2}~~\ket{\psi_3} ~\right].
\end{eqnarray*}
%Note that, the first unitary $\U_{X,Y}$ depends on the \PIJSt~ $\ket{X^{\star}},\ket{Y^{\star}}$, while the second unitary $\W$ depends on the initial state $\ket{\psi_0}$. 
%\end{corollary}
%The factorization is detailed in Sec.~\ref{sec:factorization}.
%\clrV{PROBLEM:: getting the $\{\ket{X_i},\ket{Y_i}\}_{i\in\{1,2,3\}}$'s is difficult in general. A bypass method is employed.}

\underline{How does an optimal unitary works}: % functions
We observe from Eq.~\eqref{unitary:factorize:local} that an optimal unitary is a product of two unitaries. In order to evolve the joint system from the initial state $\ket{a}\ket{e}$, the part of it first transforms Eve's initial state to $\ket{00}$ leaving Alice's part invariant, and then the second part creates the required entanglement between Alice and Eve's states.

\underline{Infinitely many solutions}:
Note that, given the optimal PIJSs $\ket{X^{\star}}, \ket{Y^{\star}}$ and an initial state $\ket{\psi_0}$, a solution of Eq.~\eqref{eq:post-interaction-joint-states_XY} for the unitary $\U_{\psi_0}$ is not unique. There are infinitely many solutions: each of $\U_{X,Y}$ and $\W_{\psi_0}$ represent an infinite family of unitaries.  
Thus, the arbitration of an optimal unitary $\U$ is two-fold: 
\begin{enumerate}
	\item arbitration of $\U_{X,Y}$, which depends on the various choices of the auxiliary states $\{\ket{X_i}, \ket{Y_i}\}_{i=1,2,3}$.
	\item arbitration of $\W_{\psi_0}$ due to various choices of the free variables $\ket{\psi_i}_{i=1,2,3}$ required to complete the orthonormalization.
\end{enumerate}
 
\underline{Shortcoming with that approach}:
Getting a specific optimal unitary $\U_{\psi_0}$ corresponds to the problem of basis completion: once in the eight dimensional space of optimal PIJSs, and once in the four dimensional space of the initial state. To complete an orthonormal basis is not always straightforward, a trial and error approach may work following some calculation-intensive efforts. For instance, one may try it for the two measurement bases stated in Sec.~\ref{sec:mes-basis-vs-optPIJS}. But, for \textit{`complicated' measurements}, 
%directions having all non-zero overlaps with the computational basis states leads to PIJSs having no zero entries in its co-ordinated form. In such cases, 
basis completion is really a challenging task. 
On the other hand, different ordering of the auxiliary states lead to different optimal unitaries following a fresh computation. To pinpoint the simplest one (canonical form) among the infinite zoo of optimal unitaries is not immediate, which otherwise may be important from practical designing perspective. 
%On the other hand, different ordering of the auxiliary states needs be handled separately following fresh computations to produce different optimal unitaries. 
%However, as we'll see shortly, one of them actually has the simplest form, possibly helpful in designing an unitary operator for real implementations. 

\underline{How do we overcome}:
To overcome all these practical shortcomings with the approach of basis completion, we suggest henceforth a series of methodologies to obtain all possible optimal unitaries more easily. As a first step, we propose the following hack to adopt a completely new approach which is surprisingly easy and natural that provides an optimal unitary along with a specific initial state. The resulted optimal unitary is the simplest one (as we'll see shortly) among its all other alternative siblings, possibly helpful in designing an unitary operator for real implementations. Once we get an optimal unitary, it's easier to find its siblings by exploiting the factorization property explained in the earlier subsection.

%finding its alternates is quite easier, as explained in further subsections. Nevertheless, we'll find a nice connection between the two approaches.
 %to complete the basis in the space of PIJSs and it leads to some $\U_{XY}$. Similarly, given an initial state $\ket{\psi_0}$, one can somehow complete an orthonormal basis in the four dimensional space and get some $\W$. Then, one can certainly find an unitary $\U$ following the product rule. However, there are infinitely many such optimal unitary for a given optimal $\ket{X}, \ket{Y}$ and an initial state $\ket{\psi_0}$. Getting the simplest one out of them is not so obvious with that approach. Life is more difficult, when one considers some other measurement basis for Eve that leads to the optimal PIJSs $\ket{X}, \ket{Y}$ having all entries non-zero. In such case, basis completion is a challenging task.  
% The question is, whether we can avoid any trial and error approach and follow some proper way to complete the basis. 
% Here, we provide a simpler hack to find an optimal unitary leaving aside the burden of basis completion.

\subsection{A divide and conquer hack to get an optimal unitary}
We can think of the optimal unitary as the following partitioned matrix
	$\U = [\U_x~~\U_y]$.
Then, the optimal PIJSs can be written in the following way   
\begin{eqnarray*}  %\label{eq:post-interaction-joint-states_XY}
	\ket{X^{\star}} = \U_x \ket{e}, &~~~& \ket{Y^{\star}} = \U_y \ket{e}. \tab
\end{eqnarray*}
Therefore, for some initial state $\ket{e}$, if we can find two such submatrices $\U_x$ and $\U_y$, we can construct the optimal unitary from them.

For example, consider that Eve measures in computational basis. Then, one can write the optimal PIJSs in such an way that (see Sec.~\ref{sec:details}) 
\begin{eqnarray*}  %\label{eq:post-interaction-joint-states_XY}
	\U_{x} = (\ket{00}~~\ket{11})\tensorprod \id_2, 
	&~~~\text{and}~~~& 
	\U_{y} = (\ket{10}~~\ket{01}) \tensorprod \sigma_x, 
	\tab
\end{eqnarray*}
for the initial state $\ket{e} = \ket{\Delta^{\hada}}_{E} := \ket{\Delta_{xy}}_{E_1}\ket{\Delta_{uv}^{\hada}}_{E_2}$,
which consists of the following factored states 
\begin{eqnarray} \label{notations:del:del_hada}
\ket{\Delta_{\b}} &:=& \sqrt{F_{\b}}\ket{0} \!+\! \sqrt{D_{\b}}\ket{1}, 
\nonumber \\  %&\tab&
%		\ket{\overbar{\Delta}_{uv}}_{E_2} &=& \sigma_z\ket{\Delta_{uv}}_{E_2}. \tab 
\ket{\Delta_{\b}^{\hada}} &:=& \hada\ket{\Delta_{\b}} = \Dps_{\b}\ket{0} \!+\! \Dms_{\b}\ket{1}. 
%	\ket{\overbar{\Delta}_{uv}^{\hada}}_{E_2} &:=& \hada\ket{\overbar{\Delta}_{uv}}_{E_2} = \Dms_{uv}\ket{0}_{E_2} \!+\! \Dps_{uv}\ket{1}_{E_2}, \tab \\
\end{eqnarray}
%where $\hada = \invsqrttwo\left(\sigma_z+\sigma_x\right)$ is the Hadamard transform. 

An optimal unitary can be read from this as below.
\begin{eqnarray} \label{opt-unitary:IS_del-hada}
	\U_{\Delta^{\hada}}^{c} = \UDeltaHada.
\end{eqnarray} 
The superscript $E_2$ denotes which subsystem of Eve will it work on.

\subsection{Alternate solutions for optimal unitaries} \label{sec:alt-unitary}
Here we explain how to find alternate optimal unitaries for an IS by already knowing an optimal unitary for that IS. We completely characterize the two-level arbitration. % on the factors.
\begin{theorem}
For a given initial state $\ket{\psi_0}$, let an optimal unitary is known as $\U_{\psi_0}$. For the same initial state, a new optimal unitary $\U_{\psi_0}^{\prime}$ can be found in one of the following ways.
\begin{enumerate}
	\item A change in the basis spanning the orthogonal subspace of the IS $\ket{\psi_0}$ leads to an alternate optimal unitary
	\begin{eqnarray*} %\label{sol:unitary:wrt:local}
		\U_{\psi_0}^{\prime} &=&  \U_{\psi_0} (\id_2\tensorprod\Gamma_{\psi_0^{\perp}}).
	\end{eqnarray*} 
	The local unitary $\Gamma_{\psi_0^{\perp}} = 
	\begin{bmatrix}
	1 & \cdot  \\
	\cdot & T_{\psi_0^{\perp}}^{\dagger}  \\
	\end{bmatrix}$ 
	makes an alternate choice $\W_{\psi_0}^{\prime}$ for $\W_{\psi_0}$:
	\begin{eqnarray*} %\label{sol:unitary:wrt:local}
		\tab\W_{\psi_0}^{\prime} &=& \W_{\psi_0} \Gamma_{\psi_0^{\perp}} = \left[~\ket{\psi_0}~~\ket{\psi_1}^{\prime}~~\ket{\psi_2}^{\prime}~~\ket{\psi_3}^{\prime} ~\right]. 
	\end{eqnarray*}	
	The 3 dimensional unitary $T_{\psi_0^{\perp}}$ transforms the orthonormal basis $\ket{\psi_i}_{i=1,2,3}$ to a newer one, while $\Gamma_{\psi_0^{\perp}}$ leaves $\ket{\psi_0}$ intact. 
	% to choose a newer basis from the orthogonal subspace of $\ket{\psi_0}$.
	%------------------------------
	%============	NEW ITEM: U_XY	==============================
	%--------------------------------
	\item 
	A change in the basis spanning the orthogonal subspace of the PIJSs $\ket{X^{\star}}, \ket{Y^{\star}}$ leads to an alternate optimal unitary
	\begin{eqnarray*} %\label{sol:unitary:wrt:local}
		\tab \U_{\psi_0}^{\prime} &=&  \U_{XY}^{\prime}~\W_{\psi_0} ~=~ \U_{XY}~\Gamma_{X^{\perp}Y^{\perp}}~\W_{\psi_0}. %~~\ne~~  \U_{\psi_0}~\Gamma_{X^{\perp}Y^{\perp}}.
	\end{eqnarray*}
	The global unitary 
	\begin{eqnarray*} %\label{sol:unitary:wrt:local}
		\Gamma_{X^{\perp}Y^{\perp}} &=& \text{diag}~(\Gamma_{X^{\perp}}, \Gamma_{Y^{\perp}})
	\end{eqnarray*}
	%	\begin{eqnarray*} %\label{sol:unitary:wrt:local}
	%		\Gamma_{X^{\perp}Y^{\perp}} &=& 
	%		\begin{bmatrix}
	%			1 & \cdot & \cdot & \cdot  \\
	%			\cdot & \Gamma_{X^{\perp}}^{\dagger} & \cdot & \cdot  \\
	%			\cdot & \cdot & 1 & \cdot  \\
	%			\cdot & \cdot & \cdot & \Gamma_{Y^{\perp}}^{\dagger}  \\
	%		\end{bmatrix}
	%	\end{eqnarray*}
	transforms $\U_{XY}$ to a new one $\U_{XY}^{\prime} =  \U_{XY}~\Gamma_{X^{\perp}Y^{\perp}}$ having the following matrix representation:
	\begin{eqnarray*} 
		\tab \begin{bmatrix}
			~\ket{\mathbf{X}^{\star}} & \ket{X_{1}^{\prime}} & \ket{X_{2}^{\prime}} & \ket{X_{3}^{\prime}} & \ket{\mathbf{Y}^{\star}} & \ket{Y_{1}^{\prime}} & \ket{Y_{2}^{\prime}} & \ket{Y_{3}^{\prime}} ~
		\end{bmatrix}
	\end{eqnarray*}
	 by changing the basis-states $\{\ket{X_i}, \ket{Y_i}\} \mapsto \{\ket{X_i^{\prime}}, \ket{Y_i^{\prime}}\}$ for $i=1,2,3$ while leaving the optimal PIJSs $\ket{X^{\star}}, \ket{Y^{\star}}$ intact. 	
	%Thus, for the same optimal states $\ket{X}, \ket{Y}$, it introduces an alternate choice  for $\U_{XY}$ with the following matrix representation:
	\item due to a change in both of the above bases.
\end{enumerate}
\end{theorem}
Note that, the first rule doesn't require the factorization. Given an optimal unitary, an alternate solution can be found by post-multiplying the former by $\id_2\tensorprod\Gamma_{\psi_0^{\perp}}$. 
For instance, let's find an alternate optimal unitary for the initial state $\ket{\psi_0} = \ket{\Delta^{\hada}}_{E} := \ket{\Delta_{xy}}_{E_1}\ket{\Delta_{uv}^{\hada}}_{E_2}$. A solution $\U_{\Delta^{\hada}}$ is already found in Eq.~\eqref{opt-unitary:IS_del-hada}. We can simply post-multiply it by some $\id_2\tensorprod\Gamma_{\psi_0^{\perp}}$, where the local unitary is chosen as, say,
\begin{eqnarray*}
	\Gamma_{\psi_0^{\perp}} &=& %\left( ~\ket{00} ~~\ket{\bar{0}1} ~~\ket{10} ~~\ket{\bar{1}1} ~\right)
	%=
	\begin{pmatrix}	
		1 		& \cdot 		& \cdot & \cdot \\	
		\cdot 	& \invsqrttwo 	& \cdot & \invsqrttwo \\	
		\cdot 	& \cdot 		& 1 	& \cdot \\	
		\cdot 	& \invsqrttwo 	& \cdot & -\invsqrttwo \\		
	\end{pmatrix}.
\end{eqnarray*}
The post-multiplication will affect the (2nd, 4th) and (6th, 8th) columns of the unitary $\U_{\Delta^{\hada}}$ as follows
\renewcommand{\arraystretch}{1.6}
\begin{equation*}
	\begin{array}{ccccccccc}
		C_2 &\mapsto&  \frac{C_2+C_4}{\sqrt{2}},	
		&&
		C_4 &\mapsto&  \frac{C_2-C_4}{\sqrt{2}};	
		\\			
		C_6 &\mapsto&  \frac{C_6+C_8}{\sqrt{2}},	
		&&
		C_8 &\mapsto&  \frac{C_6-C_8}{\sqrt{2}}.
	\end{array}
\end{equation*}
%\begin{equation*}
%\begin{array}{ccccccccc}
%\ket{C_2} &\mapsto&  \frac{\ket{C_2}+\ket{C_4}}{\sqrt{2}},	
%&&
%\ket{C_4} &\mapsto&  \frac{\ket{C_2}-\ket{C_4}}{\sqrt{2}};	
%\\			
%\ket{C_6} &\mapsto&  \frac{\ket{C_6}+\ket{C_8}}{\sqrt{2}},	
%&&
%\ket{C_8} &\mapsto&  \frac{\ket{C_6}-\ket{C_8}}{\sqrt{2}}.
%\end{array}
%\end{equation*}
\renewcommand{\arraystretch}{1}
Any such alternate solution for the optimal unitary $\U_{\Delta^{\hada}}$ introduces more non-NULL entries than that in Eq.~\eqref{opt-unitary:IS_del-hada}. Thereby, the one in Eq.~\eqref{opt-unitary:IS_del-hada} is the simplest among all other alternatives. 

Although nothing can stop one to apply an arbitration on $\U_{XY}$ at that stage by getting the factors of $\U_{\Delta^{\hada}}$ guided by Eq.~\eqref{unitary:factorize:local} following painstaking calculations while completing a basis for the initial state $\ket{\Delta^{\hada}}_{E}$, we find it easier when we get an optimal unitary for the initial state $\ket{00}$. Because, for the later case, one can choose a basis for the orthogonal subspace to enforce $\W_{\Delta^{\hada}}$ to be the identity matrix, and $\U_{XY}$ becomes same as $\U_{\Delta^{\hada}}$.
%However, we have to move a few more steps to apply an arbitration on $\U_{XY}$, when we get a factorization easily. 
But to do so, we have to devise the methods to get an optimal unitary for a different IS using the knowledge of a given optimal unitary for some other IS.

%=== Listing opt-unitaries	==================
{ %\setlength{\tabcolsep}{0.5em} % for column spacing
	\renewcommand{\arraystretch}{1.6}% for row-spacing
\begin{table*}[htbp]
	\centering	
	\caption{An optimal unitary for each of the initial states $\ket{00}$, and $\ket{\phi^{+}_{xy}} = \frac{\ket{00}+\ket{11}}{\sqrt{2}}$ when Eve measures in four dimensional computational basis and Fuchs basis, respectively. Here $\Omat_2$ and $\vec{\mathbf{0}}_2$ are the two dimensional NULL matrix and NULL row-vector, respectively.}
	\label{tab:opt-unitaries}
	\begin{tabular}{C}
		\toprule 
		\U_{00}^{\mathcal{C}}$ = $\UzeroExpanded 
		\\ 
		\midrule 
		\U_{\phi^{+}}^{\mathcal{F}}$ = $\UBellFuchs 
		\\ 
		\bottomrule
	\end{tabular}
\end{table*}
}
%In Table~\ref{tab:opt-unitaries} 

%\clearpage

\subsection{Finding an optimal unitary when Eve's initial state changes}

%\clrV{Having an optimal unitary in hand for the initial state $\ket{e} = \ket{\Delta^{\hada}}_{E}$, one can certainly find various factors guided by Eq.~\eqref{unitary:factorize:local} following painstaking calculations, and thereby read off the orthonormal basis states in the orthogonal subspace of the optimal PIJSs and those in the orthogonal subspace of the initial state. However, the process is less calculation intensive if one can get an optimal unitary for the initial state $\ket{00}$. Because, in that case, one can choose the four dimensional canonical basis for $\W$ and $\U_{XY}$ becomes same as $\U$. Then, one can directly read the PIJS-basis from the optimal unitary itself. We defer the process and look for alternate solutions for the initial state $\ket{e} = \ket{\Delta^{\hada}}_{E}$.}

%Note, from the factorization, that the joint unitary $\U$ depends on two things: 1. the initial state $\IS$ of Eve, and 2. the measurement directions $\M \equiv \M_{xy}$ of Eve (it automatically fixes its $uv$ counterpart $\M_{uv}$). Thus, $\U = \U(\M, \IS) = \U^{\M}_{\IS}$. 

%So, once an optimal unitary is found for some initial state, one may wish to find an optimal unitary for some other initial state, or, for a different measurement basis with Eve, or, both. We resolve them one by one. 

%Fix a measurement basis $\M$ for Eve. 
%If an unitary $\U^{\M}_{\IS=e}$ is known for some initial state $\ket{e}$, one can find an unitary $\U^{\M}_{\IS=f}$ for some other initial state $\ket{f}$, just by knowing the local unitary that transforms $\ket{e}_{E} \rightarrow \ket{f}_{E}$. 
If an unitary $\U_{e}$ is known for some initial state $\ket{e}$, one can find an unitary $\U_{f}$ for some other IS $\ket{f}$, just by knowing the local unitary $T_{ef}$ that transforms $\ket{e} \mapsto \ket{f}$ [see Sec.~\ref{sec:altIS} for details]. 
%\begin{theorem} \label{thm:unitary:when-init-st-changes}
%	The description of the PIJS depends on the initial state $\IS$ of Eve and the joint unitary $\U$. Same PIJS can be produced by taking a different initial state and a different unitary. The later global unitary can then be found from the former global unitary, if one knows the local unitary transform between the initial states. 
%Let 
%\begin{eqnarray*}
%	T_{ef} : \ket{e} \rightarrow \ket{f} 
%\end{eqnarray*}
%be a local unitary that transforms the initial states. 
Then, the change in the global unitary is reflected as
\begin{eqnarray}\label{rule:ISchange}
\U_{f} = \U_{e}\left(\id_2^{A}\tensorprod T_{ef}^{E\dagger}\right).
\end{eqnarray}
%\end{theorem}

For instance, consider the task to find an optimal unitary for the initial state $\ket{\Delta}_{E} := \ket{\Delta_{xy}}_{E_1}\ket{\Delta_{uv}}_{E_2}$, which is a small tweak $T_{ef} = \id_2\tensorprod\hada ~:~ \ket{\Delta^{\hada}}_{E} \mapsto \ket{\Delta}_{E}$  of the earlier initial state $\ket{\Delta^{\hada}}_{E}$. Then, the global unitary is transformed as follows
\begin{equation*}
\begin{array}{lllllllll}
\U_{\Delta}^{\comp}  &=& \U_{\Delta^{\hada}}^{\comp}(\id_2^{A}\tensorprod\id_2^{E_1}\tensorprod\hada^{E_2}).	
\end{array}
\end{equation*}
The corresponding matrix is a tweak of the one in Eq.~\eqref{opt-unitary:IS_del-hada} while each inner sub-matrix $\id_2, \sigma_x$ gets post-multiplied by the Hadamard transformation $\hada$.

%We are just a step away to get the auxiliary basis states, as 
It's now easy to get an optimal unitary for the IS $=\ket{00}_{E}$. A local unitary $T_{ef} = A_{xy}^{E_1}\otimes A_{uv}^{E_2}$ maps $\ket{\Delta}_{E} := \ket{\Delta_{xy}}_{E_1}\ket{\Delta_{uv}}_{E_2} ~\mapsto~ \ket{00}$
for the two-dimensional unitary $A_{uv} = \sqrt{1-D_{uv}}\sigma_z + \sqrt{D_{uv}}\sigma_x$. 
The desired optimal unitary $\U_{00}^{\comp}$ is given in Table~\ref{tab:opt-unitaries}.
%Thus the desired optimal unitary $\U_{00}^{AE}$ is the one after post-multiplying $\U_{\Delta}^{AE}$ by $\id_2^{A}\tensorprod A_{xy}^{E_1}\otimes A_{uv}^{E_2}$ and is given in Table~\ref{tab:opt-unitaries}.

Now we can read the auxiliary basis states from the optimal unitary $\U_{00}$. The 2nd, 3rd, and 4th columns stand for the basis states $\ket{X_1}, \ket{X_2}, \ket{X_3}$ respectively, while the 6th, 7th and 8th columns stand for the basis states $\ket{Y_1}, \ket{Y_2}, \ket{Y_3}$ respectively. Now, the basis completion method works well without much trial and error calculations. At that stage, one may try getting alternate solutions by applying arbitration on $\U_{XY}$. 
% One can merge these intermediate stages easily to reach in a single step.

Can these complete information of the basis states provide any advantage to find an optimal unitary for any other initial state? One can certainly come up with a solution, but a simpler form is again a far cry without a trial and error in choosing a proper position for the basis states in $\U_{XY}$ and $\W_{\psi_0}$. For instance, one can try it for the IS $\ket{\phi_{xy}^{+}}: = \frac{\ket{00}+\ket{11}}{\sqrt{2}}$ to realize the difficulty. We avoid it by following the indirect approach: knowing an optimal unitary for the IS $\ket{00}$, find an optimal unitary for the IS $\ket{\phi_{xy}^{+}}$.	
Note that, a local unitary $T_{ef }= c\varA{-}\sigma_x\cdot(\hada\tensorprod\id_2) : \ket{00} \mapsto \ket{\phi_{xy}^{+}}$, which has the matrix form $\invsqrttwo\csigx\hadaTid = \invsqrttwo\Tef$, can lead to an optimal global unitary $\U_{\phi^{+}}^{\comp} = \U_{00}^{\comp}\left(\id_2^{A}\tensorprod (T_{ef}^{E})^{\dagger}\right)$. 

One can now extend the methods employed here to get an optimal unitary for an arbitrarily chosen initial state. Further, one can use the directives in Subsec.~\ref{sec:alt-unitary} to obtain as many optimal unitary as one may wish for a chosen initial state.

\subsection{Finding an optimal unitary when Eve's measurement setup changes.}
Note, from the factorization in Eq.~\eqref{unitary:factorize:local}, that the joint unitary $\U$ depends on two parameters: the initial state $\IS$ of Eve's ancilla, and Eve's measurement setup $\M \equiv \M_{xy}$. It is so because, the factors of the unitary $\W_{\psi_0}$ and $\U_{XY}$ depends on IS and $\M$, respectively. However, $\U = \U^{\M}_{\IS}$ represent an infinite collection of unitaries. 

So far we have explored the zoo of optimal unitaries when Eve measures in the computational basis. Now, we augment the hunting when a different measurement basis is used by Eve.
%dig here for an approach to get an optimal unitary for a different measurement basis used by Eve. 
Let's consider a different measurement basis $\{\ket{E_{\l}}\}$ which is a unitary transformation $\ket{E_{\l}} = \M_{xy}\ket{\l}$ of the computational basis chosen earlier.
%Let the measurement basis is changed from computational basis $\{\ket{\l}\}$ to some other basis $\{\ket{E_{\l}}\}$). It follows the mathematical rule $\ket{E_{\l}} = \M_{xy}\ket{\l}$, where the corresponding local unitary $\mathbf{M}_{xy} := \left[\ket{E_0}, \ket{E_1}, \ket{E_2}, \ket{E_3}\right]$ fixes the optimal POVM. 
%\begin{theorem}\label{thm:unitary:when-mes-basis-changes}
%	Suppose, Eve changes her measurement basis from the usual \textit{computational basis} to some other basis by the mathematical rule $\ket{E_{\l}} = \M_{xy}\ket{\l}$. Note that the corresponding local unitary $\mathbf{M}_{xy} := \left[\ket{E_0}, \ket{E_1}, \ket{E_2}, \ket{E_3}\right]$ fixes the optimal POVM. 
Then, the following retrospective effects could be observed on the optimal IVs, the optimal PIJSs, and the optimal global unitary.
\begin{enumerate} \label{rules:mes-basis-chng}
	\item The optimal IVs of Eve are changed as follows: 
	\begin{eqnarray}
	\ket{\IV_{xy}^{\star}}^{\M} &=& \M_{xy}\ket{\IV_{xy}^{\star}}^{\mathcal{C}}. 
	\end{eqnarray}
	\item The optimal PIJSs are transformed as follows: %($\ket{\PIJS}^{AE}_{xy}$)$^{\M} \rightarrow$ ($\ket{\PIJS}^{AE}_{xy}$)$^{\comp}$
	\begin{eqnarray}
	\ket{S_a}^{\M} &=& (\id_2\tensorprod \M_{xy})\ket{S_a}^{\mathcal{C}}, ~~a=x,y.
%	\ket{X}^{\M} \!=\! (\id_2\tensorprod \M_{xy})\ket{X}^{\mathcal{C}}, &~~&
%	\ket{Y}^{\M} \!=\! (\id_2\tensorprod \M_{xy})\ket{Y}^{\mathcal{C}}. \tab  
	\end{eqnarray}
	\item The global unitary gets tweaked as follows:
	\begin{eqnarray} \label{rules:mes-basis-chng:unitary}
	\U^{\M} &=& (\id_2\tensorprod \M_{xy})~\U^{\mathcal{C}}. \tab  
	\end{eqnarray}
\end{enumerate}
The first two claims are straight-forward, while the last claim is proved in Sec.~\ref{sec:details}.		
%\end{theorem}

To illustrate, consider the problem of finding an optimal unitary when Eve measures in Fuchs basis and chooses the IS as $\ket{\phi_{xy}^{+}}$. 
%Note that, those states correspond to the measurement basis $\{\ket{E_0},\ket{E_1},\ket{E_2},\ket{E_3}\} = \{\ket{00},\ket{11},\ket{10},\ket{01}\}$ for Eve. 
The optimal PIJSs, which are already enlisted in Sec.~\ref{sec:mes-basis-vs-optPIJS}, can also be found by the rule
$\ket{X}^{\mathcal{F}} = (\id_2 \otimes \M_{xy})\ket{X}^{\mathcal{C}}$, where $\M_{xy} = [\ket{00}, \ket{11}, \ket{10}, \ket{01}]$. 
To get an optimal unitary in the Fuchs basis, we exploit the already known structure of an optimal unitary $\U_{\phi^{+}}^{\mathcal{C}}$ which works on the same IS, but measures in computational basis. The transformation rule~\ref{rules:mes-basis-chng:unitary} leads to the optimal unitary $\U_{\phi^{+}}^{\mathcal{F}}$ as given in Table~\ref{tab:opt-unitaries}. 

%To get an optimal unitary for these states, one can follow the indirect approach. 
%We already have found an optimal unitary $\U_{\phi^{+}}^{\mathcal{C}}$ when Eve measures in the computational basis and uses the initial state $\ket{\phi_{xy}^{+}}$. From there, we can find an optimal unitary $\U_{\phi^{+}}^{\mathcal{F}}$ when Eve measures in the Fuchs basis and uses the same initial state $\ket{\phi_{xy}^{+}}$. All one needs is to apply the transformation rule~\eqref{rules:mes-basis-chng:unitary} 
%$\U_{\phi^{+}}^{\mathcal{F}} = (\id_2 \otimes \M_{xy})~\U_{\phi^{+}}^{\mathcal{C}}$ 
%capturing the change in the measurement basis to get an optimal unitary $\U_{\phi^{+}}^{\mathcal{F}}$ as given in Table~\ref{tab:opt-unitaries}.

%\clearpage     

\section{Proofs and calculations} \label{sec:details}

\subsection{Interrelation between optimal POVMs}  %-----  across the two MUBs

Since the conjugate relation for the encoding bases inherits to the PIJSs, the IVs in each of the encoding bases gets interrelated as follows. 
\begin{eqnarray} \label{eq:interaction-vecs_uv_to_xy}
\Scale[0.97]{
	2\sqrt{F_{uv}}\ket{\xi_u} \!=\! \sqrt{F_{xy}}(\ket{\xi_x}\!+\!\ket{\xi_y}) \!+\!%
	\sqrt{D_{xy}}(\ket{\zeta_x}\!+\!\ket{\zeta_y}),} \nonumber\bigskip\\
\Scale[0.97]{
	2\sqrt{F_{uv}}\ket{\xi_v} \!=\! \sqrt{F_{xy}}(\ket{\xi_x}\!+\!\ket{\xi_y}) \!-\!%
	\sqrt{D_{xy}}(\ket{\zeta_x}\!+\!\ket{\zeta_y}),} \nonumber\bigskip\\
\Scale[0.97]{
	2\sqrt{D_{uv}}\ket{\zeta_u} \!=\! \sqrt{F_{xy}}(\ket{\xi_x}\!-\!\ket{\xi_y}) \!+\!%
	\sqrt{D_{xy}}(\ket{\zeta_y}\!-\!\ket{\zeta_x}),} \nonumber\\
\Scale[0.97]{
	2\sqrt{D_{uv}}\ket{\zeta_v} \!=\! \sqrt{F_{xy}}(\ket{\xi_x}\!-\!\ket{\xi_y}) \!-\!%
	\sqrt{D_{xy}}(\ket{\zeta_y}\!-\!\ket{\zeta_x}).} \nonumber \\
\end{eqnarray}
%	
%		Then, for optimal interaction vectors~\mref{eq:interaction-vecs_opt_xy} in $\xy$ basis, one can find the following expressions for those in $\uv$ basis:	
%	\begin{eqnarray} \label{eq:interaction-vecs_uv_to_xy_opt}
%		\ket{\xi_u^{\star}} &=& \sqrt{F_{xy}}~\frac{\ket{E_0^{\star}} \!+\! \ket{E_1^{\star}}}{\sqrt{2}} + \sqrt{D_{xy}}~\frac{\ket{E_2^{\star}} \!+\! \ket{E_3^{\star}}}{\sqrt{2}}, \nonumber\\
%		\ket{\xi_v^{\star}} &=& \sqrt{F_{xy}}~\frac{\ket{E_0^{\star}} \!+\! \ket{E_1^{\star}}}{\sqrt{2}} - \sqrt{D_{xy}}~\frac{\ket{E_2^{\star}} \!+\! \ket{E_3^{\star}}}{\sqrt{2}},  \nonumber\\
%		\ket{\zeta_u^{\star}} &=& \sqrt{F_{xy}}~\frac{\ket{E_0^{\star}} \!-\! \ket{E_1^{\star}}}{\sqrt{2}} - \sqrt{D_{xy}}~\frac{\ket{E_2^{\star}} \!-\! \ket{E_3^{\star}}}{\sqrt{2}}, \nonumber\\
%		\ket{\zeta_v^{\star}} &=& \sqrt{F_{xy}}~\frac{\ket{E_0^{\star}} \!-\! \ket{E_1^{\star}}}{\sqrt{2}} + \sqrt{D_{xy}}~\frac{\ket{E_2^{\star}} \!-\! \ket{E_3^{\star}}}{\sqrt{2}}.  \nonumber \\
%	\end{eqnarray}
The sum and difference between the fidelity states (and similarly for the disturbed states) in $uv$ basis are written in terms of the Eve's states in $xy$ basis. 
\begin{subequations} \label{eq:IVs_uv-xy_grouped}
\begin{align}
	\sqrt{F_{uv}}\left(\ket{\xi_u}+\ket{\xi_v}\right) &= \sqrt{F_{xy}}\left(\ket{\xi_x}\!+\!\ket{\xi_y}\right), 
						\tag{\ref{eq:IVs_uv-xy_grouped}.F+} \label{eq:IVs_uv-xy_grouped.F+}	
	\bigskip\\
	\sqrt{F_{uv}}\left(\ket{\xi_u}-\ket{\xi_v}\right) &= \sqrt{D_{xy}}\left(\ket{\zeta_x}\!+\!\ket{\zeta_y}\right),  
						\tag{\ref{eq:IVs_uv-xy_grouped}.F--} \label{eq:IVs_uv-xy_grouped.F--}
	\bigskip\\
	\sqrt{D_{uv}}\left(\ket{\zeta_u}+\ket{\zeta_v}\right) &= \sqrt{F_{xy}}\left(\ket{\xi_x}\!-\!\ket{\xi_y}\right), \nonumber 
						\tag{\ref{eq:IVs_uv-xy_grouped}.D+} \label{eq:IVs_uv-xy_grouped.D+}
	\bigskip\\
	\sqrt{D_{uv}}\left(\ket{\zeta_u}-\ket{\zeta_v}\right) &= \sqrt{D_{xy}}\left(\ket{\zeta_y}\!-\!\ket{\zeta_x}\right). 
						\tag{\ref{eq:IVs_uv-xy_grouped}.D--} \label{eq:IVs_uv-xy_grouped.D--}
\end{align}
\end{subequations}
%\begin{eqnarray} \label{eq:IVs_uv-xy_grouped}
% 	\sqrt{F_{uv}}\left(\ket{\xi_u}+\ket{\xi_v}\right) &=& \sqrt{F_{xy}}\left(\ket{\xi_x}\!+\!\ket{\xi_y}\right) \nonumber\bigskip\\
% 	\sqrt{F_{uv}}\left(\ket{\xi_u}-\ket{\xi_v}\right) &=& \sqrt{D_{xy}}\left(\ket{\zeta_x}\!+\!\ket{\zeta_y}\right) \nonumber\bigskip\\
% 	\sqrt{D_{uv}}\left(\ket{\zeta_u}+\ket{\zeta_v}\right) &=& \sqrt{F_{xy}}\left(\ket{\xi_x}\!-\!\ket{\xi_y}\right) \nonumber\bigskip\\
% 	\sqrt{D_{uv}}\left(\ket{\zeta_u}-\ket{\zeta_v}\right) &=& \sqrt{D_{xy}}\left(\ket{\zeta_y}\!-\!\ket{\zeta_x}\right).
%\end{eqnarray}
Now, we use the optimal IVs for $xy$ and $uv$ basis as in Eqs.~\mref{eq:interaction-vecs_opt_xy, eq:interaction-vecs_opt_uv} to find the sum and difference of the parity IVs (disturbed or undisturbed) 
and feed them back into Eq.~\eqref{eq:IVs_uv-xy_grouped} to get the following relations: 
\begin{eqnarray*} %\label{eq:Ulu}
	\ket{F_0} \!+\! \ket{F_1} = \ket{E_0} \!+\! \ket{E_1}, &~~& \ket{F_2} \!+\! \ket{F_3} = \ket{E_0} \!-\! \ket{E_1}, \nonumber \\
	\ket{F_0} \!-\! \ket{F_1} = \ket{E_2} \!+\! \ket{E_3}, &~~& \ket{F_2} \!+\! \ket{F_3} = \ket{E_3} \!-\! \ket{E_2}. %\nonumber %\\
\end{eqnarray*} 	
Getting the relation between the optimal measurement directions in Eq.~\eqref{reln:optimalPOVMs} is now obvious.
%	\begin{eqnarray} \label{eq:Ulu}
%	2\ket{F_0^{\star}} &=& \ket{E_0^{\star}} + \ket{E_1^{\star}} + \ket{E_2^{\star}} + \ket{E_3^{\star}},  \nonumber \\ 
%	2\ket{F_1^{\star}} &=& \ket{E_0^{\star}} + \ket{E_1^{\star}} - \ket{E_2^{\star}} - \ket{E_3^{\star}},  \nonumber \\
%	2\ket{F_2^{\star}} &=& \ket{E_0^{\star}} - \ket{E_1^{\star}} - \ket{E_2^{\star}} + \ket{E_3^{\star}},  \nonumber \\
%	2\ket{F_3^{\star}} &=& \ket{E_0^{\star}} - \ket{E_1^{\star}} + \ket{E_2^{\star}} - \ket{E_3^{\star}}.
%	\end{eqnarray} 	

%\clearpage

\subsection{Proving the necessary and sufficient conditions}
Here we prove Thm.~\ref{thm:equivalent-iff-condns}.
	The following relations involving the amplitudes $\Dps_{uv}$ and $\Dms_{uv}$ defined in Eq.~\eqref{rel:DpDm} are heavily used in the derivations here.
	\begin{eqnarray*} %\label{rel:calD_uv}
		(\Dps_{uv})^2\!-\!(\Dms_{uv})^2 = 2\concave{D_{uv}},&~~ (\Dps_{uv})^2+(\Dms_{uv})^2 = 1, \tab \nonumber\\
		2\Dps_{uv}\Dms_{uv} = 1-2D_{uv}.& \tab
	\end{eqnarray*} 
%\textbf{\emph{Proof of Lemma~\ref{lem:overlapIV_uv-mes_xy}:}}
\begin{proof}[\textbf{Proof of the iff condition~\ref{iff:overlapIV_uv-mes_xy} of  Thm.~\ref{thm:equivalent-iff-condns}}]	 
	The catch here is to unfold the states in Eq.~\eqref{eq:Ulu} for the projectors $E_{\l}$ while using the Schmidt form of the PIJSs, and use them in Eq.~\eqref{eq:iff_Fuchs}. In Eq.~\eqref{eq:Ulu}, for $a=u$,
%	\begin{equation*} %\label{eq:condn_optG}
%	\begin{aligned} 
%	\sqrt{D_{uv}}\ket{U_{\l^{0} u}} = \varepsilon_{\l}^{0}~\sqrt{1-D_{uv}} \ket{V_{\l^{0} u}},
%	\end{aligned}	
%	\end{equation*}
%	the state vectors $\ket{U_{\l^{0} u}}$ and $\ket{V_{\l^{0} u}}$ can be written as follows	
	\begin{eqnarray*} %\label{eq:Ulu}
		\ket{U_{\l^{0} u}} &=& B_{u}\tensorprod E_{\l}~\ket{U} \nonumber \\
		\ &=& \sqrt{1-D_{uv}}~\braket{E_{\l}}{\xi_u}\left(\ket{u}\ket{E_{\l}}\right). \\
		\ket{V_{\l^{0} u}} &=& B_{u}\tensorprod E_{\l}~\ket{V} \nonumber \\
		\ &=& \sqrt{D_{uv}}~\braket{E_{\l}}{\zeta_v}\left(\ket{u}\ket{E_{\l}}\right).
	\end{eqnarray*}
	Feeding them back into Eq.~\eqref{eq:iff_Fuchs.u} leads to the first of the equations~\eqref{rel:ElXiuZetav_ElXivZetau}. The other relation can similarly be derived 
	from the \ifft~ condition~\eqref{eq:iff_Fuchs.v} while using the Schmidt form of the PIJSs and unfolding the states in Eq.~\eqref{eq:Ulu} for $a=v$.
\end{proof}

%------------- proof of thm 1.2 ----------------------------
\begin{proof}[\textbf{Proof of the iff~condition~\ref{iff:overlap-ratio} of  Thm.~\ref{thm:equivalent-iff-condns}}]
	The \ifft~conditions in Eq.~\eqref{rel:ElXiuZetav_ElXivZetau} can be grouped as follows:
	\begin{eqnarray*} %\label{eq:Ulu}
		\bra{E_{\l}}{(\ket{\xi_u}\pm\ket{\xi_v})} &=& \varepsilon_{\l}^{0}~\bra{E_{\l}}{(\ket{\zeta_v}\pm\ket{\zeta_u})}.
	\end{eqnarray*} 
%	Summing the first two of the equations~\eqref{eq:interaction-vecs_uv_to_xy}, we get,
%	\begin{eqnarray*} %\label{eq:Ulu}
%		\sqrt{1-D_{uv}}(\ket{\xi_u}+\ket{\xi_v}) &=& \sqrt{1-D_{xy}}(\ket{\xi_x}+\ket{\xi_y}).
%	\end{eqnarray*}
%	Similarly, summing the last two of the equations~\eqref{eq:interaction-vecs_uv_to_xy}, we get,
%	\begin{eqnarray*} %\label{eq:Ulu}
%		\sqrt{D_{uv}}(\ket{\zeta_u}+\ket{\zeta_v}) &=& \sqrt{1-D_{xy}}(\ket{\xi_x}-\ket{\xi_y}).
%	\end{eqnarray*}
Now, we look back to the interrelations between the IVs in $xy$ and $uv$ basis,  
viz., use Eqs.~\mref{eq:IVs_uv-xy_grouped.F+, eq:IVs_uv-xy_grouped.D+}. Taking the inner product of the IVs in each of these equations with the measurement direction $\ket{E_{\l}}$, and then taking the ratio of the like sides, we get,
%	\begin{eqnarray*}   %=== 2 liner
%		\frac{\braket{E_{\l}}{\xi_x}+\braket{E_{\l}}{\xi_y}} {\braket{E_{\l}}{\xi_x}-\braket{E_{\l}}{\xi_y}} 
%		&=& \frac{\sqrt{F_{uv}}}{\sqrt{D_{uv}}}\cdot \frac{\braket{E_{\l}}{\xi_u}+\braket{E_{\l}}{\xi_v}} {\braket{E_{\l}}{\zeta_u}+\braket{E_{\l}}{\zeta_v}} \\	 
%		\ &=& \frac{\sqrt{F_{uv}}}{\sqrt{D_{uv}}} \cdot \varepsilon_{\l}^{0}.
%	\end{eqnarray*}
	\begin{eqnarray*}    %=== 1 liner
		\frac{\braket{E_{\l}}{\xi_x}\!+\!\braket{E_{\l}}{\xi_y}} {\braket{E_{\l}}{\xi_x}\!-\!\braket{E_{\l}}{\xi_y}} 
		&=& \frac{\sqrt{F_{uv}}}{\sqrt{D_{uv}}} 
			\frac{\braket{E_{\l}}{\xi_u}\!+\!\braket{E_{\l}}{\xi_v}} 				{\braket{E_{\l}}{\zeta_u}\!+\!\braket{E_{\l}}{\zeta_v}} 
		= \frac{\sqrt{F_{uv}}}{\sqrt{D_{uv}}} \varepsilon_{\l}^{0}.
	\end{eqnarray*}
	By componendo and dividendo, we get, 
	\begin{eqnarray*} 
		\frac{\braket{E_{\l}}{\xi_x}}{\braket{E_{\l}}{\xi_y}} = \frac{\Dp_{uv}^{(+\varepsilon_{\l}^{0})}}{\Dp_{uv}^{(-\varepsilon_{\l}^{0})}} = \left(\frac{\Dp_{uv}}{\Dm_{uv}}\right)^{\varepsilon_{\l}^{0}}.
	\end{eqnarray*}
	We used here the improvised notation of Eq.~\eqref{notation:curly-D-epsa}.
%	Here, we improvise to the following notation
%	\begin{eqnarray*} 
%		\Dp_{uv}^{(\pm\varepsilon_{\l}^{0})} &=& \invsqrttwo\left(\sqrt{1-D_{uv}}\pm\varepsilon_{\l}^{0}\sqrt{D_{uv}}\right). %\nonumber \\
%	\end{eqnarray*}
	%Clearly, $\Dp_{uv}^{(-\varepsilon_{\l}^{0})} = \Dp_{uv}, \Dm_{uv}$ for $\varepsilon_{\l}^{0}=-1,+1$ respectively. Similarly, $\Dp_{uv}^{(+\varepsilon_{\l}^{0})} = \Dm_{uv}, \Dp_{uv}$ for $\varepsilon_{\l}^{0}=-1,+1$ respectively. 	
	The ratio $\nicefrac{\Dp_{uv}^{(+\varepsilon_{\l}^{0})}}{\Dp_{uv}^{(-\varepsilon_{\l}^{0})}}$ becomes $\nicefrac{\Dp_{uv}}{\Dm_{uv}}$ or its inverse depending on whether the sign $\varepsilon_{\l}^{0}$ of the eigenvalue assumes $+1$ or $-1$, respectively.
	
	Similarly, to establish the other ratio $\nicefrac{\braket{E_{\l}}{\zeta_x}}{\braket{E_{\l}}{\zeta_y}}$ of Eq.~\eqref{rel:ElXixXiy_ZetaxZetay}, we consider Eqs.~\mref{eq:IVs_uv-xy_grouped.F--, eq:IVs_uv-xy_grouped.D--} and follow the same procedure as above.
\end{proof}

%------------- proof of thm 1.3 ----------------------------
%---------- opt-IV-expr ---------------------
\begin{proof}[\textbf{Proof of the iff condition~\ref{iff:optIVs} of  Thm.~\ref{thm:equivalent-iff-condns}}]
	The proof follows from the \ifft~condition~\ref{iff:overlap-ratio}, viz., Eq.~\eqref{rel:ElXixXiy_ZetaxZetay}. The overlaps in the ratio $\nicefrac{\braket{E_{\l}}{\xi_x}}{\braket{E_{\l}}{\xi_y}}$ can be unfolded using some (complex) constant of proportion $r_{\l,\xi}$ as follows.
	\begin{eqnarray*} \label{eq:int-vec-coord}
		\braket{E_{\l}}{\xi_x} = r_{\l,\xi}~\Dp_{uv}^{(+\varepsilon_{\l}^{0})}, &~~~&
		\braket{E_{\l}}{\xi_y} = r_{\l,\xi}~\Dp_{uv}^{(-\varepsilon_{\l}^{0})}. \tab
	\end{eqnarray*}
	Note that, these overlaps constitute the components of the fidelity states %$\ket{{\xi_x}}$ and $\ket{{\xi_y}}$ 
	when expressed in the eigenbasis $\{\ket{E_{\l}}\}$.
	
	Similarly, in the ratio $\nicefrac{\braket{E_{\l}}{\zeta_x}}{\braket{E_{\l}}{\zeta_y}}$, the overlaps can be written, for some complex number $r_{\l,\zeta}$, in the following way.
	\begin{eqnarray*} 
		\braket{E_{\l}}{\zeta_x} = r_{\l,\zeta}~\Dp_{uv}^{(+\varepsilon_{\l}^{0})}, &~~~&
		\braket{E_{\l}}{\zeta_y} = r_{\l,\zeta}~\Dp_{uv}^{(-\varepsilon_{\l}^{0})}.
	\end{eqnarray*}
	These are the components of the disturbed states %$\ket{{\zeta_x}}$ and $\ket{{\zeta_y}}$ 
	when expressed in the eigenbasis $\{\ket{E_{\l}}\}$.
	
	Then we can write down the IVs with respect to the eigenbasis $\{\ket{E_{\l}}\}$ as follows.
	\begin{eqnarray*} %\label{rel:Dpmepsa}
		\ket{\xi_x} &=& \sum\limits_{\l} r_{\l,\xi}~\Dp_{uv}^{(+\varepsilon_{\l}^{0})}~\ket{E_{\l}}, \nonumber \\
		\ket{\xi_y} &=& \sum\limits_{\l} r_{\l,\xi}~\Dp_{uv}^{(-\varepsilon_{\l}^{0})}~\ket{E_{\l}}, \nonumber \\
		\ket{\zeta_x} &=& \sum\limits_{\l} r_{\l,\zeta}~\Dp_{uv}^{(+\varepsilon_{\l}^{0})}~\ket{E_{\l}}, \nonumber \\
		\ket{\zeta_y} &=& \sum\limits_{\l} r_{\l,\zeta}~\Dp_{uv}^{(-\varepsilon_{\l}^{0})}~\ket{E_{\l}}.
	\end{eqnarray*}
	
	But, we observe that, $\Dp_{uv}^{(+\varepsilon_{\l}^{0})} = \D^{+}_{uv}, \D^{-}_{uv}$ for $\varepsilon_{\l}^{0}=+1,-1$ respectively.
	Similarly, $\Dp_{uv}^{(-\varepsilon_{\l}^{0})} = \D^{-}_{uv}, \D^{+}_{uv}$ for $\varepsilon_{\l}^{0}=+1,-1$ respectively. Thereby,
	%\begin{eqnarray*} %\label{rel:Dpmepsa}
	%\frac{\braket{E_{\l}}{\xi_x}}{\braket{E_{\l}}{\xi_y}} &=& \frac{\braket{E_{\l}}{\zeta_x}}{\braket{E_{\l}}{\zeta_y}} \nonumber \\ 
	%~ &=& \frac{\Dp_{uv}^{(+\varepsilon_{\l}^{0})}}{\Dp_{uv}^{(-\varepsilon_{\l}^{0})}} = 
	%\begin{cases}
	%\nicefrac{\Dp_{uv}}{\Dm_{uv}}, \textbf{~~for~~} \varepsilon_{\l}^{0}=+1, \\
	%\nicefrac{\Dm_{uv}}{\Dp_{uv}}, \textbf{~~for~~} \varepsilon_{\l}^{0}=-1. \nonumber \\
	%\end{cases}
	%\end{eqnarray*}
	in the expression of the IVs, we can group the basis vectors $\ket{E_{\l}}$ according to the sign of the measurement outcome. For instance, each of the fidelity states get two groups: $\ket{E_{\l\xi}^{\pm}}$ groups the measurement directions for $\pm$ve outcomes. Similarly, the two groups for the disturbed states correspond to $\ket{E_{\l\zeta}^{\pm}}$. The following equation captures the grouping:
	% Eq.~\eqref{rel:basis-optIV_grp-mes-sign} 	grouping	 
	\begin{eqnarray*} \label{rel:basis-optIV_grp-mes-sign}
		\ket{E_{\l\xi}^{\pm}} \!:=\! \!\sum\limits_{\l:~\pm\sf ve~outcomes} r_{\l,\xi}\ket{E_{\l}}\!, \\
		%&~~~& 
		\ket{E_{\l\zeta}^{\pm}} \!:=\!	 \sum\limits_{\l:~\pm\sf ve~outcomes} r_{\l,\zeta}\ket{E_{\l}}. %\tab 
	\end{eqnarray*}	
	%		\begin{eqnarray} \label{rel:basis-optIV_grp-mes-sign}
	%			\ket{E_{\l\xi}^{\pm}} \!=\! \!\sum\limits_{\l: \varepsilon_{\l}^{0}=\pm 1} r_{\l,\xi}\ket{E_{\l}}\!, %\nonumber \\ 
	%		%	\ket{E_{\l\xi}^{-}} &=& \sum\limits_{\l: \varepsilon_{\l}^{0}=-1} r_{\l,\xi}\ket{E_{\l}}, \nonumber \\ 
	%			&~~~& \ket{E_{\l\zeta}^{\pm}} = \sum\limits_{\l: \varepsilon_{\l}^{0}=\pm 1} r_{\l,\zeta}\ket{E_{\l}}. 
	%		%	\ket{E_{\l\zeta}^{-}} &=& \sum\limits_{\l: \varepsilon_{\l}^{0}=-1} r_{\l,\zeta}\ket{E_{\l}}. \nonumber \\ 
	%		\end{eqnarray}	
	
	With these grouping, the IVs can be described as in Eq.~\eqref{eq:opt-IVs:new}. That the vectors $\{\ket{E_{\l\xi}^{+}}, \ket{E_{\l\xi}^{-}}, \ket{E_{\l\zeta}^{+}}, \ket{E_{\l\zeta}^{-}}\}$ form an orthonormal basis, can be argued as follows.
	As defined, the states in $E_\l^+ := \{\ket{E_{\l\xi}^{+}}, \ket{E_{\l\zeta}^{+}}\}$ are mutually orthogonal to the states in $E_\l^- := \{\ket{E_{\l\xi}^{-}}, \ket{E_{\l\zeta}^{-}}\}$.
	Then, the normalization constraint on the fidelity (or disturbed) states together induces the normalization constraint on the states in $E_\l^+$ (or $E_\l^-$). Moreover, the orthogonality between the fidelity states and the disturbed states inherits the orthogonality within the states in $E_\l^{+}$ as well the orthogonality within the states in $E_\l^{-}$.

	The last but not the least is the fact that each of the states $\{\ket{E_{\l\xi}^{+}}, \ket{E_{\l\xi}^{-}}, \ket{E_{\l\zeta}^{+}}, \ket{E_{\l\zeta}^{-}}\}$  can be expressed in terms of exactly two of the measurement directions $\{\ket{E_{\l}}\}$. It is so because, the sign of the measurement outcomes are evenly distributed for an optimal interaction: two +ve outcomes, and two -ve outcomes.
	%i.e., 	the sign parameter $\varepsilon_\l^0$ takes values $+1$ for two different $\l$ and $-1$ for the remaining two $\l$. Had it been not this way, then, w.l.o.g, let's assume that $\varepsilon_\l^0$ is +ve for only one outcome $\l$. 
	Had it not been this way, then, w.l.o.g, let's assume the possibility for only one +ve outcome. 
	Then, each of the states $\ket{E_{\l\xi}^{+}}, \ket{E_{\l\zeta}^{+}}$ should have only one of the measurement directions $\ket{E_\l}$ in their description. While the normalization constraint on these states indicate the coefficients $r_{\l,\xi}, r_{\l,\zeta}$ to be unimodular, their mutual orthogonality enforces one of these coefficients to be zero, leading to a contradiction.
\end{proof}

%----------- Better necc-suff condn 1.1a----------------
\begin{proof}[\textbf{Proof of Corollary~\ref{iff:IV-overlaps} of Thm.~\ref{thm:equivalent-iff-condns}}] 
	The proof follows from %Thm.~\ref{thm:equivalent-iff-condns}.\ref{iff:overlapIV_uv-mes_xy}
	condition~\ref{iff:overlapIV_uv-mes_xy} of the same theorem and Lem.~\ref{lem:interreln:parity-overlaps}.
	
%	Note that while we are considering the optimality in $xy$ basis, the \ifft~condition~\eqref{rel:iff} 	involves the IVs from the said basis only, whereas the \ifft~ condition~\eqref{rel:ElXiuZetav_ElXivZetau} deals with the IVs for $uv$ basis. 

	Clearly, an equality of the overlaps in Lem.~\ref{lem:interreln:parity-overlaps} lead to the desired result~\eqref{rel:iff}. To establish this equality, we consider the \ifft~ conditions~\eqref{rel:ElXiuZetav_ElXivZetau}, but for optimality in $uv$ basis, viz.
	\begin{eqnarray*} %\label{eq:Ulu}
		\braket{F_{\l}}{\xi_x} &=& \varepsilon_{\l}^{1}~\braket{F_{\l}}{\zeta_y}, \\
		\braket{F_{\l}}{\xi_y} &=& \varepsilon_{\l}^{1}~\braket{F_{\l}}{\zeta_x}.
	\end{eqnarray*}
	%\clrR{Note: used $F_{\l}$ instead of $F_{\l}$}.
	Multiplying the like sides of these two equations and adding over the measurement outcomes $\l$ in $uv$ basis, we get,
	\begin{eqnarray*} %\label{eq:Ulu}
		\ & \sum_{\l}\braket{\xi_x}{F_{\l}}\braket{F_{\l}}{\xi_y} = \sum_{\l}\braket{\zeta_x}{F_{\l}}\braket{F_{\l}}{\zeta_y}. 
	\end{eqnarray*}
	Since the projectors $F_{\l}$ consist a POVM, their completeness relation leads to the equality between the two overlaps $\braket{\xi_x}{\xi_y}$ and $\braket{\zeta_x}{\zeta_y}$, and consequently the desired result follows from Lem.~\ref{lem:interreln:parity-overlaps}. 
\end{proof}

\subsection{The two representations of the optimal IVs are unitarily equivalent} \label{sec:equivalence}
To establish the equivalence of the optimal IV$_{xy}$ in Eq.~\eqref{eq:interaction-vecs_opt_xy} and those in Eq.~\eqref{eq:opt-IVs:new} we make a matrix-vector representation of the IVs. We introduce a few notations for that in Table~\ref{tab:mat-vec-optIVs}.
%\begin{eqnarray*}
%	\vint_{xy} &:=& (\ket{\xi_x},\ket{\xi_y},\ket{\zeta_x},\ket{\zeta_y}), \\
%	(\POVM_{\xi\xi\zeta\zeta}^{+-+-})_{xy} &:=& (\ket{E_{\l\xi}^{+}}, \ket{E_{\l\xi}^{-}}, \ket{E_{\l\zeta}^{+}}, \ket{E_{\l\zeta}^{-}}), \\
%	(\POVM_{\xi\zeta\xi\zeta}^{++--})_{xy} &:=& (\ket{E_{\l\xi}^{+}}, \ket{E_{\l\zeta}^{+}}, \ket{E_{\l\xi}^{-}}, \ket{E_{\l\zeta}^{-}}),  \\
%	(\POVM_{\sigma_{\lambda}})_{xy} &:=& (\ket{E_{\l_0}^{+}},\ket{E_{\l_1}^{+}},\ket{E_{\l_2}^{-}},\ket{E_{\l_3}^{-}}), \\
%	\sigma_{\lambda} &:=& (\l_0^{+},\l_1^{+},\l_2^{-},\l_3^{-}), \\
%	\POVM_{xy} &:=& (\ket{E_0},\ket{E_1},\ket{E_2},\ket{E_3}), \\
%	\text{and,~~} 
%	\POVM_{uv} &:=& (\ket{F_0},\ket{F_1},\ket{F_2},\ket{F_3}).
%	%\text{and,~~} \vcan &:=& (\ket{00},\ket{01},\ket{10},\ket{11}).
%\end{eqnarray*} 
\begin{table}[htbp]
%	\begin{centre}
\caption{Notations: matrix-vector form of optimal IV}
\label{tab:mat-vec-optIVs}
		\begin{tabular}{CCRCL}
			\ &\ & \vint_{xy} &:=& (\ket{\xi_x},\ket{\xi_y},\ket{\zeta_x},\ket{\zeta_y}), 
			\\ 
			\POVM_{xy}^{\NEW} 	&\equiv& 	\POVM_{\xi,\xi,\zeta,\zeta}^{+-+-} &:=& (\ket{E_{\l\xi}^{+}}, \ket{E_{\l\xi}^{-}}, \ket{E_{\l\zeta}^{+}}, \ket{E_{\l\zeta}^{-}}), 
			\\
			\ &\ & \POVM_{\xi,\zeta,\xi,\zeta}^{++--} &:=& (\ket{E_{\l\xi}^{+}}, \ket{E_{\l\zeta}^{+}}, \ket{E_{\l\xi}^{-}}, \ket{E_{\l\zeta}^{-}}),  
			\\
			\ &\ & \POVM_{0,2,1,3}^{++--} &:=& (\ket{E_{0}^{+}},\ket{E_{2}^{+}},\ket{E_{1}^{-}},\ket{E_{3}^{-}}), 
			\\
			\POVM_{xy}^{\OLD} 	&\equiv& 	\POVM_{0,1,2,3}^{+-+-} 
			&:=& 	(\ket{E_0^{+}},\ket{E_1^{-}},\ket{E_2^{+}},\ket{E_3^{-}}), 
			\\ 
			%	\POVM_{uv} &:=& (\ket{F_0},\ket{F_1},\ket{F_2},\ket{F_3}).
			\ &\ & \matd_{uv} &:=& \id_2\tensorprod \left(\Dps_{uv}\id_2+\Dms_{uv}\sigma_x\right).
		\end{tabular}
%	\end{centre}
\end{table}
%\begin{eqnarray*}
%	\vint_{xy} &:=& (\ket{\xi_x},\ket{\xi_y},\ket{\zeta_x},\ket{\zeta_y}), 
%	\\
%	\POVM_{xy}^{\NEW} 	\equiv 	\POVM_{\xi,\xi,\zeta,\zeta}^{+-+-} &:=& (\ket{E_{\l\xi}^{+}}, \ket{E_{\l\xi}^{-}}, \ket{E_{\l\zeta}^{+}}, \ket{E_{\l\zeta}^{-}}), 
%	\\
%	\POVM_{\xi,\zeta,\xi,\zeta}^{++--} &:=& (\ket{E_{\l\xi}^{+}}, \ket{E_{\l\zeta}^{+}}, \ket{E_{\l\xi}^{-}}, \ket{E_{\l\zeta}^{-}}),  
%	\\
%	\POVM_{0,2,1,3}^{++--} &:=& (\ket{E_{0}^{+}},\ket{E_{2}^{+}},\ket{E_{1}^{-}},\ket{E_{3}^{-}}), 
%	\\
%	\POVM_{xy}^{\OLD} 	\equiv 	\POVM_{0,1,2,3}^{+-+-} 
%	&:=& 	(\ket{E_0^{+}},\ket{E_1^{-}},\ket{E_2^{+}},\ket{E_3^{-}}), 
%	\\ 
%	%	\POVM_{uv} &:=& (\ket{F_0},\ket{F_1},\ket{F_2},\ket{F_3}).
%	\matd_{uv} &:=& \id_2\tensorprod \left(\Dps_{uv}\id_2+\Dms_{uv}\sigma_x\right).
%\end{eqnarray*}
Here, OLD denotes IVs in Eq.~\eqref{eq:interaction-vecs_opt_xy} and NEW denotes IVs in Eq.~\eqref{eq:opt-IVs:new}. Those optimal IVs can be expressed in matrix-vector form as follows:
%Then, the optimal IV$_{xy}$ in Eq.~\eqref{eq:interaction-vecs_opt_xy} and Eq.~\eqref{eq:opt-IVs:new} can be written as 
\begin{eqnarray*} %\label{eq:interaction-vecs_opt_xy:mat}	 
(\vint^{\star}_{xy})^{\OLD} &=& \POVM_{xy}^{\OLD}~\matd_{uv}, \\	
(\vint^{\star}_{xy})^{\NEW} &=& \POVM_{xy}^{\NEW}~\matd_{uv}.
\end{eqnarray*} 
\newcommand{\invequiv}{\mathbin{\rotatebox[origin=c]{90}{$\equiv$}}}
To establish the equivalence, it's enough to show that $\POVM_{xy}^{\NEW}$ is unitarily equivalent to $\POVM_{xy}^{\OLD}$. The intermediate transformations are as follows:
\begin{equation*}
	\begin{array}{lclllllll}
	\stackrel{\stackrel{\POVM_{xy}^{\OLD}}{\invequiv}}
			 {\POVM_{0,1,2,3}^{+-+-}}					
			& \stackrel{S_w}{\longrightarrow} 
			& \POVM_{0,2,1,3}^{++--}	
			& \stackrel{\mathbf{R}}{\longrightarrow} 
			& \POVM_{\xi,\zeta,\xi,\zeta}^{++--}	
			& \stackrel{S_w}{\longrightarrow} 
			&  
			\stackrel{\stackrel{\POVM_{xy}^{\NEW}}{\invequiv}}
					 {\POVM_{\xi,\xi,\zeta,\zeta}^{+-+-}}
	\end{array}
\end{equation*}
%\begin{equation*}
%	\begin{array}{lclllllll}
%		\POVM_{xy}^{\OLD} &\equiv& (\POVM_{0,1,2,3}^{+-+-})_{xy} && \\	
%		\		& \stackrel{\Pi_{1324}}{\longrightarrow} & (\POVM_{0,2,1,3}^{++--})_{xy}	\\	
%		\		& \stackrel{\mathbf{K}}{\longrightarrow} & (\POVM_{\xi,\zeta,\xi,\zeta}^{++--})_{xy}	\\
%		\		& \stackrel{\Pi_{1324}}{\longrightarrow} & (\POVM_{\xi,\xi,\zeta,\zeta}^{+-+-})_{xy} ~~\equiv~~ \POVM_{xy}^{\NEW}.
%	\end{array}
%\end{equation*}
All the three maps are post-multiplication to transform the column-space, e.g., $\POVM_{0,2,1,3}^{++--} = \POVM_{0,1,2,3}^{+-+-}~S_w$ etc.
The swap operation %$S_w$  
\begin{eqnarray*}
	S_w &:=& 
	\begin{bmatrix} % bmatrix for [], pmatrix for ()
		1 & 0 & 0 & 0 \\
		0 & 0 & 1 & 0 \\
		0 & 1 & 0 & 0 \\
		0 & 0 & 0 & 1 
	\end{bmatrix}
\end{eqnarray*}
corresponds to the permutation $\Pi_{1324}$.

The unitary $\mathbf{R}:= {\sf diag} (\mathbf{R}^{+}, \mathbf{R}^{-})$ works on the measurement directions in order to affect unitarily the two subspaces, one for positive outcomes and the other for negative outcomes. To be specific, the measurement directions $\{\ket{E_{0}^{\pm}},\ket{E_{2}^{\pm}}\}$ go through an unitary transformation $\mathbf{R}^{\pm}$ in that subspace. 

%And, $\mathbf{R}:= {\sf diag} (\mathbf{R}^{+}, \mathbf{R}^{-})$ affects the two mutually orthogonal subspaces span$\{\ket{E_{0}^{+}},\ket{E_{2}^{+}}\}$ and span$\{\ket{E_{1}^{-}},\ket{E_{3}^{-}}\}$ unitarily by $\mathbf{R}^{+}$ and $\mathbf{R}^{-}$ respectively.
% is an unitary that creates an unitary transform  of the two states  in their subspace and an unitary transform $\mathbf{K}^{-}$ of the two states $\ket{E_{1}^{+}},\ket{E_{3}^{+}}$ in their subspace.

Therefore, we get the following interrelation between the POVMs associated with the \NEW~and \OLD~optimal IVs.
\begin{eqnarray*} 
	\POVM_{xy}^{\NEW} 
	&=& 
	% (\POVM_{\xi,\zeta,\xi,\zeta}^{++--})_{xy}~\Pi_{1324} \\ 	
	%\ &=& 
	\POVM_{xy}^{\OLD}~S_w~\mathbf{R}~S_w.
\end{eqnarray*}
%Note that, the optimal states in Eq.~\eqref{eq:opt-IVs:new} correspond to the following transformation
%\begin{eqnarray*} \label{eq:IVs-in-measurement-basis}
%	(\vint^{\NEW}_{xy})^{\dagger} &=& \mathbb{D}_{uv}~(\POVM_{\xi,\xi,\zeta,\zeta}^{+-+-})_{xy}^{\dagger} \\
%	\ &=& \mathbb{D}_{uv}~\Pi_{1324}~\mathbf{K}^{\dagger}~\Pi_{1324}~(\POVM^{+-+-}_{0,1,2,3})_{xy}^{\dagger}
%\end{eqnarray*} 
%Note that $\Pi_{1324}^{\dagger}=\Pi_{1324}$.
%Therefore, the optimal IV$_{xy}$ in Eq.~\eqref{eq:opt-IVs:new} can be written as
%\begin{eqnarray*} %\label{eq:IVs-in-measurement-basis}
%	(\vint^{\star}_{xy})^{\NEW} = \POVM_{xy}^{\OLD}~\Pi_{1324}~\mathbf{K}~\Pi_{1324}~\matd_{uv}.
%\end{eqnarray*}
Hence the equivalence follows.

\subsection{Getting an optimal unitary for some initial state when Eve measures in the computational basis}
%\subsubsection{Find the optimal $\PIJS$.}
First, we find the optimal IVs and the optimal PIJSs as Eve measures in computational basis.

Her optimal IVs can be expressed as follows:
\begin{eqnarray*}
	\ket{\xi_x^{\star}}^{\comp} \!=\! \ket{0}_{E_1}\ket{\Delta_{uv}^{\hada}}_{E_2}, 
	&\tab&
	\ket{\xi_y^{\star}}^{\comp} \!=\! (\id_2^{E_1}\tensorprod\sigma_x^{E_2})\ket{\xi_x^{\star}}^{\comp}, \tab \nonumber \\
	%	\ket{\zeta_x^{\star}}_{c} \!=\! \ket{1}_{E_1}\ket{\Delta_{uv}^{\hada}}_{E_2},  &\tab& 
	%	\ket{\zeta_y^{\star}}_{c} \!=\! \ket{1}_{E_1}\ket{\overbar{\Delta}_{uv}^{\hada}}_{E_2}. \tab 
	\ket{\zeta_x^{\star}}^{\comp} \!=\! \ket{1}_{E_1}\ket{\Delta_{uv}^{\hada}}_{E_2},  
	&\tab& 
	\ket{\zeta_y^{\star}}^{\comp} \!=\! (\id_2^{E_1}\tensorprod\sigma_x^{E_2})\ket{\zeta_x^{\star}}^{\comp}. \tab 
\end{eqnarray*}	
Here the state $\ket{\Delta_{uv}^{\hada}}$ is as defined in Eq.~\eqref{notations:del:del_hada}.
%\begin{eqnarray} \label{notations:del:del_hada}
%	\ket{\Delta_{uv}}_{E_2} &:=& \sqrt{1\!-\!D_{uv}}\ket{0}_{E_2} \!+\! \sqrt{D_{uv}}\ket{1}_{E_2}, 
%	\nonumber \\  %&\tab&
%	%		\ket{\overbar{\Delta}_{uv}}_{E_2} &=& \sigma_z\ket{\Delta_{uv}}_{E_2}. \tab 
%	\ket{\Delta_{uv}^{\hada}}_{E_2} &:=& \hada\ket{\Delta_{uv}}_{E_2} = \Dps_{uv}\ket{0}_{E_2} \!+\! \Dms_{uv}\ket{1}_{E_2}. 
%	%	\ket{\overbar{\Delta}_{uv}^{\hada}}_{E_2} &:=& \hada\ket{\overbar{\Delta}_{uv}}_{E_2} = \Dms_{uv}\ket{0}_{E_2} \!+\! \Dps_{uv}\ket{1}_{E_2}, \tab \\
%\end{eqnarray}
%Note that $\hada = \invsqrttwo\left(\sigma_z+\sigma_x\right)$ is the Hadamard transform. 

Therefore, the optimal PIJS$^{AE}_{xy}$ can be expressed as follows
%\begin{eqnarray*}  %\label{eq:optPostInteractionState_X}
%\ket{X^{\star}}_{c} &=& \sqrt{1\!-\!D_{xy}}~\ket{0}\ket{\xi_x^{\star}}_{c} +	 \sqrt{\!D_{xy}}~\ket{1}\ket{\zeta_x^{\star}}_{c} \\
%\ &=& \ket{\Phi_{D_{xy}}^{+}}\ket{\Delta_{uv}}.
%\end{eqnarray*}  
\begin{eqnarray*}  %\label{eq:optPostInteractionState_X}
	\ket{X^{\star}}^{\comp} &=& \ket{\Phi_{D_{xy}}^{+}}_{AE_1}\ket{\Delta_{uv}^{\hada}}_{E_2}, \\ %&~~&
	\ket{Y^{\star}}^{\comp} &=& \ket{\Psi_{D_{xy}}^{+}}_{AE_1}\tensorprod \sigma_x^{E_2}\ket{\Delta_{uv}^{\hada}}_{E_2}. \tab
\end{eqnarray*}  
where
\begin{eqnarray*} % \label{eq:optPostInteractionState_X}
	\ket{\Phi_{D_{xy}}^{+}}_{AE_1} &=& \sqrt{1\!-\!D_{xy}}\ket{00}_{AE_1} \!+\! \sqrt{D_{xy}}\ket{11}_{AE_1}, \\
	\ket{\Psi_{D_{xy}}^{+}}_{AE_1} &=& \sqrt{1\!-\!D_{xy}}\ket{10}_{AE_1} \!+\! \sqrt{D_{xy}}\ket{01}_{AE_1}.
\end{eqnarray*}
%\begin{note}
%Note that the PIJS$^{AE}_{xy}$ are each factored: Alice's qubit is entangled with Eve's first qubit, while the second qubit of Eve works like an ancilla. %\clrR{Both of Eve's qubit can be factored for a different unitary in terms of $k_{uv}^{\pm}$, as done earlier. Put a diagram if possible.}
%\end{note}

To get an optimal unitary, we need to rewrite the PIJSs in matrix-vector form.
%------------------	item1	---------------------
First, note that the entangled states from the subsystem AE$_1$ can be expressed in matrix-vector form as follows %for Alices qubit and Eve's first qubit can be expressed as
\begin{eqnarray*} % \label{eq:optPostInteractionState_X}
	\ket{\Phi_{D_{xy}}^{+}}_{AE_1} &=& W_{x}^{AE_1}\ket{\Delta_{xy}}_{E_1}, \\ 
	\ket{\Psi_{D_{xy}}^{+}}_{AE_1} &=& W_{y}^{AE_1}\ket{\Delta_{xy}}_{E_1},
\end{eqnarray*}
with the $4\times 2$ matrices %$\mathbf{W}_{x}$ and $\mathbf{W}_{y}$ can be written in the following way 
%\begin{eqnarray*} 
%\mathbf{W}_{x} &=& \ket{0_{A}0_{E_1}}\bra{0_{E_1}}+\ket{1_{A}1_{E_1}}\bra{1_{E_1}}, \nonumber \\  
%\mathbf{W}_{y} &=& \ket{1_{A}0_{E_1}}\bra{0_{E_1}}+\ket{0_{A}1_{E_1}}\bra{1_{E_1}},
%\end{eqnarray*}
\begin{eqnarray*} 
	\mathbf{W}_{x}^{AE_1} &=& \ket{00}_{AE_1}\bra{0_{E_1}}+\ket{11}_{AE_1}\bra{1_{E_1}}, \nonumber \\  
	\mathbf{W}_{y}^{AE_1} &=& \ket{10}_{AE_1}\bra{0_{E_1}}+\ket{01}_{AE_1}\bra{1_{E_1}}.
\end{eqnarray*}
%------------------	item2	---------------------
Thereby, the optimal PIJS$^{AE}_{xy}$ can be expressed in matrix-vector form as follows:
\begin{eqnarray*}  %\label{eq:optPostInteractionState_X}
	\ket{X^{\star}}^{\comp} &=& \U_x^{AE}~\ket{\Delta_{xy}}_{E_1}\ket{\Delta_{uv}^{\hada}}_{E_2}, 
	\\ 
	\ket{Y^{\star}}^{\comp} &=& \U_y^{AE}~\ket{\Delta_{xy}}_{E_1}\ket{\Delta_{uv}^{\hada}}_{E_2}, \tab
\end{eqnarray*} 
with the $8\times 4$ matrices
\begin{eqnarray*}  %\label{eq:optPostInteractionState_X}
	\U_x^{AE} = W_{x}^{AE_1}\tensorprod\id_2^{E_2}, 
	&~~& 
	\U_y^{AE} = W_{y}^{AE_1}\tensorprod\sigma_x^{E_2}. \tab
\end{eqnarray*}

%\noindent \textbf{Case-study: } given an initial state, derive opt-unitary.
%\subsubsection{Opt-unitary for initial state = $\ket{\Delta^{\hada}}_{E} := \ket{\Delta_{xy}}_{E_1}\ket{\Delta_{uv}^{\hada}}_{E_2}$.}

%------------------	the unitary	---------------------
Then, for an initial state
%Consider the initial state for Eve to be 
\begin{eqnarray*}  %\label{eq:optPostInteractionState_X}
	\ket{\Delta^{\hada}}_{E} &:=& \ket{\Delta_{xy}}_{E_1}\ket{\Delta_{uv}^{\hada}}_{E_2}, \tab
\end{eqnarray*} 
%where
%\begin{eqnarray*}
%	\ket{\Delta_{xy}}_{E_1} &:=& \sqrt{1\!-\!D_{xy}}\ket{0}_{E_1} \!+\! \sqrt{D_{xy}}\ket{1}_{E_1}. \tab 
%\end{eqnarray*}
%and with the similar notation for $\ket{k_{uv}}_{E_2}$.
%Then, the following statements hold true:
an optimal unitary can be given as 
\begin{eqnarray*} \label{sol:unitary:00_gen}
	\U_{\Delta^{\hada}}^{AE} &=& \U_x^{AE}\bra{0}_{A} + \U_y^{AE}\bra{1}_{A} \\
	\ &=& W_{x}^{AE_1}\bra{0}_{A}\tensorprod\id_2^{E_2} + W_{y}^{AE_1}\bra{1}_{A}\tensorprod\sigma_x^{E_2} \\
	\ &=& (\ket{00}_{AE_1}\bra{00}+\ket{11}_{AE_1}\bra{01})\tensorprod\id_2^{E_2} \nonumber \\
	\ && + (\ket{10}_{AE_1}\bra{10}+\ket{01}_{AE_1}\bra{11})\tensorprod\sigma_x^{E_2}. \nonumber  %\\
	%		\ &=& 
	%		\begin{bmatrix} 
	%			\id_2^{E_2} & \Omat_2 & \Omat_2 & \Omat_2 \\
	%			\Omat_2 & \Omat_2 & \Omat_2 & \sigma_x^{E_2} \\
	%			\Omat_2 & \Omat_2 & \sigma_x^{E_2} & \Omat_2 \\
	%			\Omat_2 & \id_2^{E_2} & \Omat_2 & \Omat_2 \\ 
	%		\end{bmatrix}.
\end{eqnarray*}
%The optimal unitary is expressed in \clrR{table??}.
%	Here $\Omat_2$ is the 2-dimensional null matrix.

\subsection{Factorization of an optimal unitary} \label{sec:factorization}
The optimal unitary in Eq.~\eqref{sol:unitary:gen} can be factored in the following way
%\begin{proof} 
	%[\clrR{---ADD---}]
	\begin{eqnarray*} %\label{sol:unitary:wrt:local}
		\U_{\psi_0} &=& \sum\limits_{a=0}^{1}\sum\limits_{i=0}^{3} \ket{S_a}\bra{a_{A}}\bra{\psi_i}_{E} 
		\\ 
		&=& \sum\limits_{a=0}^{1}\sum\limits_{i=0}^{3} \ket{S_a}\bra{a_{A}}\bra{i}_{E}\ket{i}\bra{\psi_i}_{E} 
		\\
		\ &=& 
		\sum\limits_{a=0}^{1}\sum\limits_{i=0}^{3} \ket{S_a}\bra{a_{A}}\bra{i_{E}} ~\times~
		\sum\limits_{i=0}^{3} \id_2\tensorprod \ket{i}_{E}\bra{\psi_i} .
	\end{eqnarray*}
%\end{proof}

%\subsection{Factorization of an optimal unitary} \label{sec:factorization}
%The optimal unitary in Eq.~\eqref{sol:unitary:gen} can be factored in the following way
%%\begin{proof} 
%%[\clrR{---ADD---}]
%\begin{eqnarray*} %\label{sol:unitary:wrt:local}
%	\U_{\psi_0} &=& \sum\limits_{a=0}^{1}\sum\limits_{\l=0}^{3} \ket{S_a}\bra{a_{A}}\bra{\psi_\l}_{E} 
%	\\ 
%	&=& \sum\limits_{a=0}^{1}\sum\limits_{\l=0}^{3} \ket{S_a}\bra{a_{A}}\bra{\l}_{E}\ket{\l}\bra{\psi_\l}_{E} 
%	\\
%	\ &=& 
%	\sum\limits_{a=0}^{1}\sum\limits_{\l=0}^{3} \ket{S_a}\bra{a_{A}}\bra{\l_{E}} ~\times~
%	\sum\limits_{\l=0}^{3} \id_2\tensorprod \ket{\l}_{E}\bra{\psi_\l} .
%\end{eqnarray*}
%%\end{proof}
%%\subsection{Alternate solutions for optimal unitaries for a given initial state and given \PIJSt s.}

\subsection{Change in initial state and measurement direction of Eve} \label{sec:altIS}
The global unitary evolves the joint system as follows:
\begin{eqnarray*}
	\U^{\M}_{\IS=e} ~\ket{a}_{A} \ket{e}_{E} &=& \ket{S_a}_{AE}^{\M}.
\end{eqnarray*}
For $a \in \{x,y\}$, the PIJSs $S_a \in \{X,Y\}$ gets fixed by fixing the measurement directions $\M$.
However, the same PIJS $\ket{S}_{AE}^{\M}$ can be produced for a different $\IS$ and a different unitary:
\begin{eqnarray*}
	\U^{\M}_{\IS=f} ~\ket{a}_{A} \ket{f}_{E} &=& \ket{S_a}_{AE}^{\M}.
\end{eqnarray*}   
Given an unitary $\U^{\M}_{\IS=e}$, one can find an unitary $\U^{\M}_{\IS=f}$ by knowing the local unitary that transforms $\ket{e} \rightarrow \ket{f}$.

\textbf{Deriving Eq.~\eqref{rule:ISchange}}: 
%$T_{ef}$ is an unitary that transforms $\ket{e} \rightarrow \ket{f}$, i.e., 
Since $\ket{f} = T_{ef}\ket{e}$, we get
\begin{eqnarray*}
	\U_{f} ~\ket{a}_{A} \ket{f}_{E} = \ket{S_a}_{AE} 	
	&=& \U_{e} ~\ket{a}_{A} \ket{e}_{E} \\
	\					&=& \U_{e} ~\ket{a}_{A} \tensorprod T_{ef}^{\dagger}\ket{f}_{E} \\
	\					&=& \U_{e}\left(\id_2^{A}\tensorprod T_{ef}^{E\dagger}\right) ~\ket{a}_{A} \ket{f}_{E} .
\end{eqnarray*}
%\qed

\textbf{Deriving Eq.~\eqref{rules:mes-basis-chng:unitary}}:
%The first two claims are straightforward. The last claim goes as follows:
\begin{eqnarray*}
	\U^{\M}~\ket{0}_{A}\ket{\psi_0}_{E} = \ket{X}^{\M} &=& (\id_2\tensorprod \M_{xy})~\ket{X}^{\mathcal{C}} \\
	&=&  (\id_2\tensorprod \M_{xy})~\U^{\mathcal{C}}~\ket{0}_{A}\ket{\psi_0}_{E}.
\end{eqnarray*}
%\qed

%\clearpage
%\clearpage

\subsection{Optimal eavesdropping and some connections}

\subsubsection{Fidelity of Eve's state discrimination}
An optimal attack on the \textit{p$\&$m} scheme leaves Eve with an optimal state-discriminate problem.
For a specific encoding basis, the four different post-interaction states of Eve's ancilla can be grouped into two mutually orthogonal sets: one with the two fidelity states, and the other with the two disturbed states. Since Eve can discriminate these orthogonal sets (whether disturbed or not), all she is left with is to distinguish the two states in a set, e.g., distinguishing $\ket{\xi_a}$ from $\ket{\xi_{\bar{a}}}$, or, distinguishing $\ket{\zeta_a}$ from $\ket{\zeta_{\bar{a}}}$. Following the optimal strategy, Eve can distinguish the two such parity states (fidelity or disturbed) with probability~\cite{hels76}
\begin{eqnarray*}
	F_E^{\b} &=& \half + \half\sqrt{1-|\braket{\xi_a^{\b}}{\xi_{\bar{a}}^{\b}}|^2} \\
	\ 	&=& \half + \half\sqrt{1-(1-2D_{\bar{\b}})^2} \\
	\ &=& \half + \sqrt{D_{\bar{\b}}(1-D_{\bar{\b}})}.
\end{eqnarray*}
%For symmetric attack Eve's fidelity =$\half + \sqrt{D(1-D)}$.

\subsubsection{Secret-key rate} %{Secure zone for classical post-processing}

%One may refer to~\cite[Sec.~IV]{fuchs97} for another relevant model for optimal eavesdropping, where a fixed average disturbance $D_a:=\half\left(D_{xy}+D_{uv}\right)$ is considered across the two bases. In this case, optimality of the averaged IG and the averaged MI is achieved only if $D_{xy} = D_{uv} = D_a$.

The \emph{secrecy capacity} $C_s$ %(aka \emph{secret-key rate}) 
of the quantum channel between Alice and Bob is defined~\cite{CK78} as the maximum rate at which Alice can reliably send information to Bob leaving Eve's information on that data arbitrarily small. A necessary and sufficient condition for a positive secret-key rate is not known, but a lower bound is known~\cite{CK78}. For a more general scenario, considering the knowledge gain of Eve over Bob's data ($I_{BE}$) due to public discussion over the supplementary classical channel, one can lower bound the secrecy capacity~\cite{ehpp94} by the following formula 
\begin{eqnarray*}
	C_s&\ge& \max\{I_{AB}-I_{AE},~ I_{AB}-I_{EB}\}.
\end{eqnarray*} 
Thus the legitimate parties should consider the channel unsafe and abort the transmission whenever
\begin{eqnarray*}
	I_{AB} &\le& \min\{I_{AE},I_{EB}\}.
\end{eqnarray*} 
On the other hand, %for a given distribution $P(A,E,B)$ of the key between three parties, 
the legitimate parties can establish a secret key following some one-way CPP, \emph{iff} $I_{AB} > I_{AE}$ or $I_{AB} > I_{EB}$. 
For an optimal symmetric attack, $I_{AE}=I_{EB}$. Therefore, Alice and Bob lives in the \textit{secure zone} whenever $I_{AB} > I_{AE}$. The difference $I_{AB} - I_{AE}$, that captures the \emph{secret-key rate}, remains same during the error correction and privacy amplification. Thus, the condition transcends in order to establish a shared secret between the two legitimate parties.

%\clrV{Due to symmetric eavesdropping, the quantum channel between Alice-Bob (or Alice-Eve) can be interpreted as a binary symmetric channel with data-flipping rate $D$ (or $\half - \sqrt{D(1-D)}$). Thus, the respective MI becomes $I_{AB} = \half~\phi\left(1-2D\right)$ and $I_{AE} = \half~\phi\left(2\concave{D}\right)$ which coincides when $1-2D = 2\concave{D}$,
%%Then, 
%%\begin{eqnarray*}
%%	I_{AB} &=& \ln 2 + D\ln D + (1-D)\ln (1-D) = \half~\phi\left(1-2D\right). %\nonumber \\
%%\end{eqnarray*}
%%On the other hand, for optimal symmetric attack, 
%%\begin{eqnarray*}
%%	I_{AE} &=& \half~\phi\left(2\concave{D}\right).
%%\end{eqnarray*}
%%Clearly, $I_{AE}=I_{AB}$ happens when  
%%\begin{eqnarray*}
%%	1-2D &=& 2\concave{D}, 
%%\end{eqnarray*}
%i.e., for the \emph{critical value} $D=D^{\star}$ of Eq.~\eqref{QBER_max}.
%}
%%Therefore, the maximum tolerable disturbance for the channel between Alice and Bob is given by 
%%\begin{eqnarray} \label{QBER_max}
%%D^{\star} &=& \half\left(1-\frac{1}{\sqrt{2}}\right) \approx 0.146. 
%%\end{eqnarray} 

\subsubsection{Optimal state-discrimination vs Bell-violation}
The \textit{p$\&$m} scheme has its equivalent \textit{eb} counterpart where Alice prepares a %two-qubit
maximally entangled state $\frac{\ket{aa}+\ket{\bar{a}\bar{a}}}{\sqrt{2}}$ and send one of the particles to Bob. Both the parties measure the observables $\sigma_z, \sigma_x$, chosen randomly. 
%Assume, Alice and Bob can share an ensemble of maximally entangled bipartite state as in the entanglement-based Ekert-91~\cite{e91} QKD protocol.

The security of the \textit{eb} scheme is linked to the tests of quantum nonlocality~\cite{fuchs97}. Presence of non-locality is a certificate for OW-CPP. The degree of non-locality depends on the estimated value of the CHSH polynomial for which the legitimate parties sacrifice a subset of their particles. Alice measures one of the observables $\sigma_z, \sigma_x$ chosen randomly,
%$\sigma_a \stackrel{\mathcal{R}}{\longleftarrow} \{\sigma_x, \sigma_y\}$ 
while Bob measures one of the observables $\frac{\sigma_z+\sigma_x}{\sqrt{2}}, \frac{\sigma_z-\sigma_x}{\sqrt{2}}$ chosen randomly.
%$\sigma_b \stackrel{\mathcal{R}}{\longleftarrow} \{\frac{\sigma_x+\sigma_y}{\sqrt{2}}, \frac{\sigma_x-\sigma_y}{\sqrt{2}}\}$. 
The binary measurement outcomes $a_i,b_j \in\{-1,+1\}$ are used to estimate the CHSH correlation-coefficient which in turn is the expected value of the product of the outcomes.
%(\clrR{Rand draw symbol:: for arbt distribution: $\stackrel{R}{\longleftarrow}$, OR, $\stackrel{R}{\hookleftarrow}$, OR, $\stackrel{\star}{\longleftarrow}$. For uniform distribution: $\stackrel{\$}{\longleftarrow}$} )
\begin{eqnarray*}
	S &:=& E(a_1,b_1) + E(a_1,b_2) + E(a_2,b_1) - E(a_2,b_2). %\le 2. %\nonumber \\
\end{eqnarray*}
Due to some channel error $D$, each of the correlations $E(a_i,b_j|D)$ get reduced from its error-free counterpart $E(a_i,b_j)$ by a factor of $1-2D$: %as evident from the following equation:
\begin{eqnarray*}
	E(a_i,b_j|D) &=& F\cdot E(a_i,b_j) - D\cdot E(a_i,b_j) \nonumber \\
	\ &=& (1-2D)\cdot E(a_i,b_j). 
\end{eqnarray*}
Consequently, $S_D = (1-2D)S_0$.
%Interestingly enough, the Bell-CHSH inequality (\clrR{cite}) is violated whenever the QBER falls below this critical value. This intriguing connection is not well understood yet and thereby leaving some scope for further research. In the E91 scheme (\clrR{cite}), eavesdropping can be detected by checking whether the Bell inequality is violated. In the CHSH version (\clrR{cite}), the correlation signature $S$ doesn't exceed 2 for a classical channel, while it can reach at most $2\sqrt{2}$ in a quantum channel without any eavesdropping (and any other error). The effect of eavesdropping in the quantum channel reduces this correlation to $2\sqrt{2}(1-2D)$, which lies above 2 whenever the disturbance lies below the critical value $D_0$.

%It is worth to mention here an appealing connecting between the security of quantum cryptography and tests of quantum nonlocality~\cite{fuchs97}. To be specific, the maximum allowed QBER in Eq.~\eqref{QBER_max} is the critical perturbation up to which a Bell-CHSH violation~\cite{chsh69,bell64} can be observed. 
The CHSH inequality forbids the correlation coefficient $S$ to exceed 2 for \textit{local operations and classical communication} (LOCC).
However, for an error-free quantum channel, this inequality is violated %by a factor $\sqrt{2}$,
and the correlation amount reaches the maximum of $2\sqrt{2}$. 
Then, in a quantum channel with error $D$, the maximum amount of violation becomes $S_D^{\star}=(1-2D)2\sqrt{2}$. In order to maintain quantum non-locality, this reduced sum must exceed 2, which happens precisely for $D < D^{\star}$ as in Eq.~\eqref{QBER_max}.

\subsubsection{Optimal attack contracts the Bloch vectors} 
%\textbf{Optimal eavesdropping vs shrinking of Bloch vector}: 
The state $\ket{a}^{\b}$ of a two-level quantum system (qubit) corresponds to a Bloch vector $\vec{a}_{\b}$ on the surface of the \Poincare~sphere. Alices' density operator $\rho_A = \ket{a}^{\b}\bra{a}$ is a  convex combination $\half \left(\id+\vec{a}_{\b}\cdot\vec{\sigma}\right)$ of the Pauli operators. 
For the BB84 protocol, the states in the $Z$ and the $X$ bases correspond to the Bloch vectors $(0,0,\pm1)$ and $(\pm1,0,0)$, respectively.
Therefore, Alice sends the density operators $\half \left(\id\pm\sigma_{s}\right)$ (for, $s\in\{z,x\}$) to Bob.
But, due to eavesdropping, Bob receives the density
\begin{eqnarray*}
	\rho_B &=& F\ket{a}^{\b}\bra{a} + D\ket{\bar{a}}^{\b}\bra{\bar{a}} \\
	\  &=& F\cdot\half \left(\id_2 + \vec{a}_{\b}\cdot\vec{\sigma}\right) 
	+ D\cdot\half \left(\id_2 - \vec{a}_{\b}\cdot\vec{\sigma}\right) \\
	\ &=& \half \left(\id_2 + (F-D)\vec{a}_{\b}\cdot\vec{\sigma}\right) 
\end{eqnarray*}
While Alice sends the density $\half \left(\id+\vec{a}\cdot\vec{\sigma}\right)$, Bob receives $\half \left(\id+\eta_D\vec{a}\cdot\vec{\sigma}\right)$ with $\eta_D = 1-2D$. To be specific, the density operators $\half \left(\id\pm\sigma_{s}\right)$ (for, $s\in\{z,x\}$) from Alice get perturbed to $\half \left(\id\pm\eta_D\sigma_{s}\right)$ when it reaches Bob. Thus, eavesdropping shrinks the Bloch vectors by a factor of $\eta_D = 1-2D$.

%\clearpage 

%\section{Listing opt-unitaries}
%\begin{eqnarray*}
%	\U_{00}^{\mathcal{C}} &=& \UzeroExpanded \medskip \\ 
%%	\ && \text{Here,~~} \Omat_2 \text{~~is a 2-dim null matrix.}
%	\\ \\
%	\U_{\phi^{+}}^{\mathcal{F}} &=& \UBellFuchs \\
%	\ && \text{Here,~~} \Omat_2 \text{~~is a 2-dim null matrix}, \text{~~and~~} \Ovecrow_2 \text{~~is a 2-dim null row-vector.}
%\end{eqnarray*}

\section{Conclusion}
We have characterized the optimal attacks on BB84 protocol exhaustively, where an attacker entangles a four dimensional probe per transmitted qubit. We have considered the generalized asymmetric error rates across the two MUBs in order to uncover all possible choices for an attacker, while a symmetric attack automatically becomes a special case.
A necessary and sufficient condition is derived here to testify the optimality of an interaction performed by an eavesdropper. As it unveils, an optimal attack corresponds to a specific configuration of the attacker's post-interaction states: that the overlap between the two disturbed states are same as the overlap between the two undisturbed states and is equal to the difference between the fidelity and the disturbance at the receiving end. Interestingly enough, the optimal overlap is same as the reduction in Bell violation in the equivalent entanglement-based scheme. We have shown explicitly that the optimal states of the joint system can also be obtained by an optimal phase-covariant cloning mechanism, and vice versa.

For practical purposes, all an eavesdropper requires is the optimal unitary to evolve the joint system and the corresponding measurement that she must perform to glean the optimal information. 
We have developed the methods to characterize the optimal unitaries and demonstrated via examples. Our method could figure out the simplest one out of the infinite family of optimal unitaries in a most natural fashion. A salient feature of an optimal unitary, as we have noticed here is that, it first  transforms Eve's initial state to $\ket{00}$, and then creates the required entanglement between this transformed state with Alice's qubit. For that particular IS ($\ket{00}$), one can read off the auxiliary basis states spanning the eight dimensional subspace orthonormal to the optimal PIJSs, directly from an optimal unitary, and vice versa.  
As an optimal unitary is parameterized by the error-rate, an attacker may first fix the QBER she wishes to introduce and choose an optimal unitary (not unique) for a specific choice of her measurement and the IS. An attacker would like to choose such unitaries which, if feasible, is easier to design than its siblings. As a further work, an interested reader may explore the design of the optimal unitaries in terms of universal quantum gates.
%Since her unitary is not unique, she would like to choose possibly the simplest ones from designing perspectives, and we leave it as a further work to explore. 
%\clrV{An interesting implication of our calculations is that an optimal unitary is parameterized by the QBER that the legitimate parties. Therefore, an attacker may first fix the amount of the data flipping rate that she wishes to introduce in the quantum channel between the legitimate parties and design the optimal unitary accordingly. Another aspect is that her optimal unitary fixes her optimal measurement and vice versa.} To be more specific, a chosen configuration for optimal measurement fixes her optimal unitary which in turn can have infinitely many different choices depending on the choice of her initial state. To summarize, the attacker first decides a QBER below the threshold and chooses an optimal measurement in the four dimensional Hilbert space. Further, a choice of her initial state fixes the optimal unitary interaction that she needs to perform. 

%The work here provides enormous freedom for a potential eavesdropper to choose an optimal unitary attack. As a further work, an interested reader may explore the design of the optimal unitaries in terms of universal quantum gates.

%\section{bib}


\begin{thebibliography}{0}
		%\bibitem{noclone}
		%W. K. Wootters,  W. H. Zurek.
		%%A Single Quantum cannot be Cloned.
		%{\it Nature} {\bf 299},  802--803, (1982).
		
		\bibitem{bb84}
		C. H. Bennett and G. Brassard,
		%Quantum Cryptography: Public key distribution and coin tossing.
		in {\it Proceedings of the IEEE International Conference on Computers, 
			Systems, and Signal Processing, Bangalore, India, December 9-12, 1984}
		(IEEE-1984), Vol. \textbf{1}, pp. 175--179.
		
		\bibitem{fuchs97}
		C. A. Fuchs, N. Gisin, R. B. Griffiths, C. S. Niu, and A. Peres,
		%Optimal eavesdropping in quantum cryptography. I. Information bound and optimal strategy.
		\hyperref{http://journals.aps.org/pra/abstract/10.1103/PhysRevA.56.1163}{category}{name}
				{Phys. Rev. A \textbf{56}, 1163 (1997)}.
		%{Phys. Rev. A} {\bf 56}, 1163--1172 (1997).
		
		\bibitem{fuchs96}
		C. A. Fuchs, in
		%Information Gain vs. State Disturbance in Quantum Theory.
		{\it Proceedings of the Fourth Workshop on Physics and Computation, Boston, November 22--24, 1996}, pp. 229--259, \hyperref{https://arxiv.org/abs/quant-ph/9611010}{category}{name}{arXiv:quant-ph/9611010}.
		
		
%		\bibitem{bell64}
%		J.S.Bell,
%		%On the Einstein Podolsky Rosen paradox.
%		%Physics Physique Fizika \textbf{1}, 195 (1964).
%		\hyperref{https://journals.aps.org/ppf/abstract/10.1103/PhysicsPhysiqueFizika.1.195}{category}{name}
%				{Physics Physique Fizika \textbf{1}, 195 (1964)}.		
		
		
		\bibitem{chsh69}		
		J.F. Clauser, M.A. Home, A. Shimony, and R.A. Holt, 
		%Proposed Experiment to Test Local Hidden-Variable Theories.
		%Phys. Rev. Lett. \textbf{23}, 880 (1969).
		\hyperref{https://journals.aps.org/prl/abstract/10.1103/PhysRevLett.23.880} {category}{name}
				{Phys. Rev. Lett. \textbf{23}, 880 (1969)}.		
		
		
		\bibitem{cirelson80}  
		B.S. Cirel'son, 
		%Quantum generalizations of Bell's inequality.
		%Lett. Math. Phys. \textbf{4}, 93 (1980).
		%1980, Volume 4, Issue 2, pp 93–100
		\hyperref{https://link.springer.com/article/10.1007/BF00417500}	{category}{name}
				{Lett. Math. Phys. \textbf{4}, 93 (1980)}.
		
		
		\bibitem{ach17}
		A. Acharyya, and G. Paul,
		%Revisiting optimal eavesdropping in quantum cryptography: Optimal interaction is unique up to rotation of the underlying basis
		\hyperref{https://journals.aps.org/pra/abstract/10.1103/PhysRevA.95.022326}{category}{name}
				{Phys. Rev. A \textbf{95}, 022326 (2017)}.
		%{Phys. Rev. A} {\bf 56}, 1163--1172 (1997).
		
		
		\bibitem{CK78} %-- Csiszar-Korner-1978
		I. Csisz\'ar, J. K\"orner:
		%Broadcast channels with confidential messages. 
		%IEEE Trans. Information Theory \textbf{24(3)}: 339-348 (1978)
		\hyperref{https://ieeexplore.ieee.org/document/1055892}
		{category}{name}{IEEE Trans. Information Theory \textbf{24(3)}, 339-348 (1978)}.
		
		
		\bibitem{ehpp94} %-- EHPP-94: ref 3 of Fuchs
		A. K. Ekert, B. Huttner, G. M. Palma, and A. Peres, 
		%Eavesdropping on quantum-cryptographical systems.
		%Phys. Rev. A \textbf{50}, 1047 (1994).
		\hyperref{https://journals.aps.org/pra/abstract/10.1103/PhysRevA.50.1047}
		{category}{name}{PhysRevA. \textbf{50}, 1047 (1994)}.
		
		
%		\bibitem{bbm92} %-- BBM-92: ref 16 of Fuchs
%		\clrR{TO CITE}. 
%		C. H. Bennett, G. Brassard, and N. D. Mermin, 
%		% Quantum cryptography without Bell's theorem
%		Phys. Rev. Lett. \textbf{68}, 557 (1992).
		
%		\bibitem{e91} %-- E-91: ref 17 of Fuchs
%		A. K. Ekert, 
%		%Quantum cryptography based on Bell's theorem.
%		%Phys. Rev. Lett. \textbf{67}, 661 (1991).
%		\hyperref{https://journals.aps.org/prl/abstract/10.1103/PhysRevLett.67.661}
%		{category}{name}{Phys. Rev. Lett. \textbf{67}, 661 (1991)}.
		
		
		%---Conference paper
%		\bibitem{gw2k} %-- Gisin-Wolf-2000  
%		N. Gisin, and S. Wolf, 
%		%Linking Classical and Quantum Key Agreement: Is There ``Bound Information”?.
%		In: \hyperref{https://link.springer.com/chapter/10.1007/3-540-44598-6_30}{category}{name}
%		{\it Proceedings of CRYPTO 2000. Lecture Notes in Computer Science, vol \textbf{1880}, pp. 482-500, Springer-Verlag}.
%		%{\it Proceedings of CRYPTO 2000. Advances in Cryptology -- CRYPTO 2000. Lecture Notes in Computer Science, vol \textbf{1880}, pp. 482-500, Springer-Verlag}.
		
		
%		\bibitem{gw99} %-- Gisin-Wolf-1999
%		N. Gisin, and S. Wolf, 
%		%Quantum Cryptography on Noisy Channels: Quantum versus Classical Key-Agreement Protocols.
%		%Phys.Rev.Lett. 83, 4200-4203 (1999).
%		\hyperref{https://journals.aps.org/prl/abstract/10.1103/PhysRevLett.83.4200}
%		{category}{name}{Phys.Rev.Lett. \textbf{83}, 4200-4203 (1999)}.
		
		
		
		% CLONING RELATED
		\bibitem{GH97} %-- Gisin-Huttner-1997
		N. Gisin, and B. Huttner,
		%Quantum cloning, eavesdropping and Bell's inequality
		\hyperref{https://doi.org/10.1016/S0375-9601(97)00083-2}
		{category}{name}{Phys. Lett. A \textbf{228}, 13-21 (1997)}.
		
		
%		\bibitem{BH96} %-- Buzek-illery-1996
%		V. Bu\u{z}ek, and M. Hillery,
%		%Quantum copying: Beyond the no-cloning theorem
%		Phys. Rev. A \textbf{54 (3)} (1996).
%		%VOLUME 54, NUMBER 3,  1996	
		
		
		\bibitem{qNC02}	% IF POSSIBLE, REMOVE____________________X
		M.~A.~Nielsen and I.~L.~Chuang,
		% Quantum Computation and Quantum Information.
		{\it Quantum Computation and Quantum Information} (Cambridge University Press, Cambridge,
		UK, 2002).
%		
%		\bibitem{cover91}	% IF POSSIBLE, REMOVE____________________X
%			\clrR{USED??}
%		T. Cover and J. Thomas,
%		%Elements of Information Theory.
%		{\em Elements of Information Theory}, 1st ed. (Wiley, New York, 1991), pp. 18--20.	
		
		
		%================= refer these = newly added
		\bibitem{BCAM2K}
		D. Bru\ss{}, M. Cinchetti, G. Mauro D' Ariano, and C. Macchiavello,
%		Phase-covariant quantum cloning.
		\hyperref{http://journals.aps.org/pra/abstract/10.1103/PhysRevA.62.012302}{category}{name}{Phys. Rev. A \textbf{62}, 012302 (2000)}.
		
		
		\bibitem{hels76}
		C. W. Helstrom, 
		{\it Quantum Detection and Estimation Theory} 
		(Academic Press, New York, 1976) pp. 80–84.
		
	\end{thebibliography}
\end{document}